# Twist Grain Boundary phases in proper ferroelectric liquid crystals realm

*Damian Pociecha, Jadwiga Szydlowska, Nataša Vaupotič, Katarzyna Kwiatkowska, Marijus Juodka, Julian Spiess, John MD Storey, Corrie T Imrie,† Rebecca Walker, Ewa Gorecka**

D. Pociecha, J. Szydlowska, K. Kwiatkowska, Ewa Gorecka

Department of Chemistry, University of Warsaw, Warsaw, Poland

e-mail: gorecka@chem.uw.edu.pl

N. Vaupotič

Faculty of Natural Sciences and Mathematics, University of Maribor, Maribor, Slovenia;

Jozef Stefan Institute, Ljubljana, Slovenia;

M. Juodka, J. Spiess, J.M.D. Storey, C.T. Imrie, R. Walker

School of Natural and Computing Sciences, University of Aberdeen, Aberdeen, Great Britain

†Deceased 14th January 2025

Funding: This research was supported by the National Science Centre (Poland) under the grant no. 2024/53/B/ST5/03275.

The twist-grain-boundary (TGB) phases, characterized by a periodic, helical arrangement of blocks made of polar smectic phases, $SmA_F$ and $SmC_F$, have been discovered. They have been observed for rod-like molecules with a strong longitudinal dipole moment, featuring an (S)-2-methylbutyl end group having only weak twisting power, and emerge below the antiferroelectric $SmA_{AF}$ phase, where the lamellar structure is already well established. It is suggested that the structure is governed by electrostatic interactions amplified by weak chiral forces, in striking contrast to the mechanism of TGB phase formation found in non-polar materials. The TGB phases exhibit light selective reflection in the visible range, while the value of electric polarization confirms an almost perfectly ordered dipole alignment.

## 1. Introduction

In the field of liquid crystals (LC) the concepts of ferroelectricity and chirality are closely related. Ferroelectric properties of LC phases were discovered in the 1970s[1] and, for a long time, were considered inherently linked to molecular chirality. The emergence of long-range dipole order in smectic layers was thought to result from the lack of inversion symmetry elements in the system built of chiral molecules. In the tilted SmC* phase molecular chirality not only induces spontaneous electric polarization but also causes the director, and thus polarization, to rotate between adjacent layers, producing the structural chirality of the phase. The discovery of ferroelectricity for achiral bent-core molecules[2] introduced a new perspective on the polarity/chirality relationship.[3] In these systems polarization appears due to steric interactions restricting molecular rotation. The resulting polarization vector, together with the tilt direction and layer normal, might define either a left- or right-handed coordination system, thus a structural chirality emerges spontaneously even though the individual molecules themselves are achiral. This finding demonstrated that molecular chirality is not a prerequisite for the emergence of ferroelectricity in liquid crystals, but ferroelectricity and structural chirality are still related. The breakthrough discovery of proper ferroelectricity in the least-ordered liquid crystalline phase - the nematic $N_F$ phase,[4-6] and later in smectic phases,[7-11] seemed to decouple ferroelectricity from chirality; dipole-dipole interactions alone were found sufficient to induce ferroelectric order. Molecular chirality still influences proper ferroelectric LC phases, e.g. transforming the $N_F$ phase into its helical analogue, $N_{F*}$ similarly as observed for the non-polar nematic phase. Importantly, the helical structure of the $N_{F*}$ phase does not affect the value of local electric polarization, and likewise polar order does not modify the helical pitch considerably.[12-15] For a while, in proper ferroelectric liquid crystals chirality and ferroelectricity appeared to be independent phenomena. However, in 2024, a pivotal discovery revealed that helicity can also emerge spontaneously as a mechanism to avoid bulk polarization, once again merging chirality and ferroelectricity in soft matter.[16-19]

Here we will illustrate the behavior of a material made of weakly chiral but strongly polar molecules - prone to forming proper ferroelectric LC phases - and show that such combination might lead to complex and unexpected structural arrangements: Twist Grain Boundary (TGB) phases. Mesogenic molecules tend to arrange themselves into well-defined layers, while the molecular chirality causes a natural tendency of the molecules to twist. In general, lamellar structures expel the twist, however in some systems these two tendencies compete and the resulting TGB structure develops as a compromise between both ordering principles.[20] The TGB structure consists of blocks of smectic layers that are separated by a periodic array of screw dislocations, mediating rotation of the smectic blocks. The axis of the resulting helix is perpendicular to the smectic layer normal, and often the pitch is of the order of visible light wavelength. The structure resembles that of type-II superconductors, where magnetic flux lines penetrate the material in a periodic fashion.[21] The TGB phase typically appears in strongly chiral materials and is usually found within a temperature range between the isotropic liquid or N* and SmA or SmC phases, in which layer interactions are still weak.[22-24] It was only rarely observed below a SmA phase.[24,25] Until now, no TGB phase has been reported in proper ferroelectric liquid crystals, although recently a twisted organization of discrete polar smectic blocks, inherited from the twisted state of the $N_F$ phase, has been described.[26]

## 2. Structure properties

The studied compound, **RW4***, has a long, rigid mesogenic core with a substantial longitudinal dipole moment (~12D), and a single terminal chain containing an asymmetric carbon atom (**Figure 1**). The material is a modification of the previously studied compound **JK104**,[17] with a non-branched terminal chain, which showed a sequence of non-polar and proper ferroelectric phases (denoted by subscript F): N - SmA - $SmA_F$ - $SmC_F$. Notably, shorter **JK10*n*** homologues formed the spontaneously helical ferroelectric nematic phase, $N_{TBF}$. Both the optically pure enantiomer (S-**RW4***) and racemic mixture (rac-**RW4***) were studied, and they showed similar phase transition temperatures.

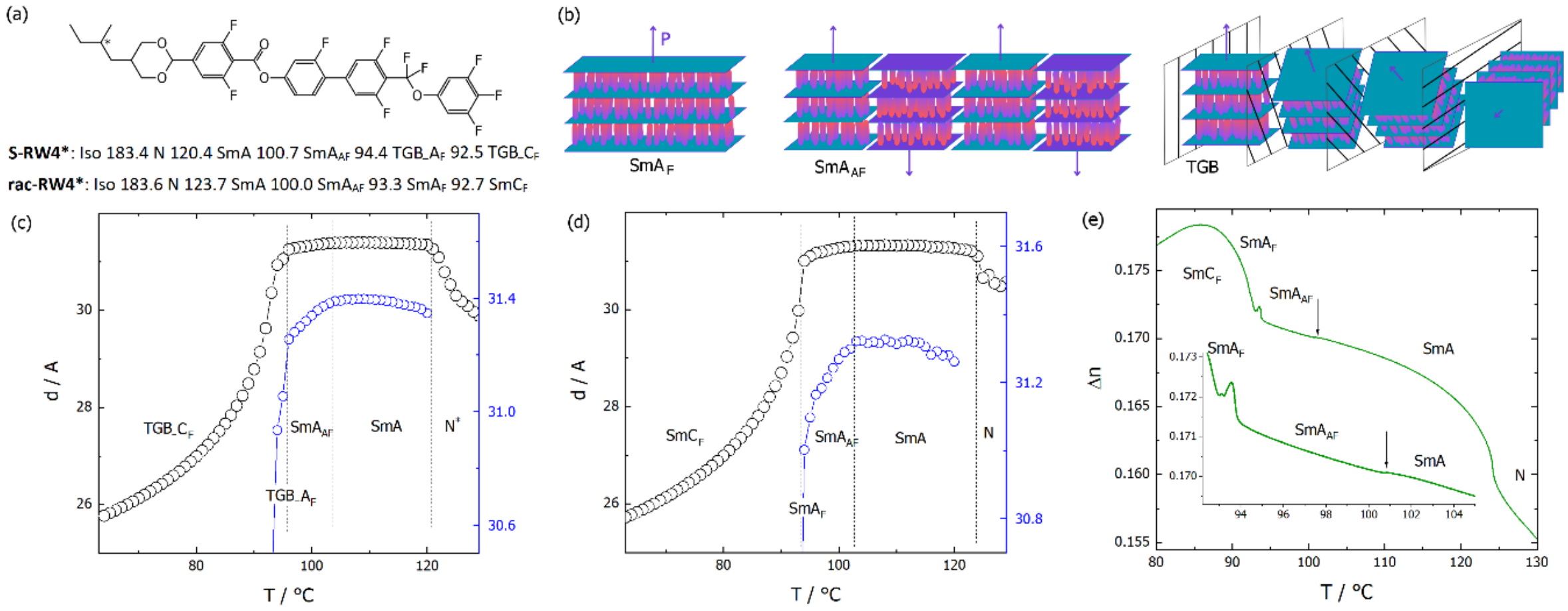

**Figure 1.** (a) Molecular structure and phase transition temperatures determined by DSC method for S-enatiomer and racemic mixture of **RW4*** compound. (b) Models of ferroelectric $SmA_F$, antiferroelectric $SmA_{AF}$ and TGB phases, polar molecules are represented by ellipsoids with different color ends, arrows show the spontaneous polarization vector. Layer spacing vs. temperature for (c) S-**RW4*** and (d) rac-**RW4***. (e) Optical birefringence vs. temperature for rac-**RW4***, measured in 1.6-µm-thick cell with planar anchoring. In the inset enlarged temperature range with SmA-$SmA_{AF}$- $SmA_F$ phases. Note, that in the SmC phase measured apparent optical retardance does not reflect actual birefringence changes, as the sample loses alignment with the formation of tilted domains (see **Figure S51**).

For both materials X-ray diffraction (XRD) studies revealed a nematic phase and a sequence of three orthogonal smectic phases (SmA-type with liquid like in-plane order, **Figure S49**), with transitions between them marked as slight changes in the slope of the layer spacing – temperature dependence **(Figure 1c, d**). On further cooling a strong decrease of layer thickness was observed, confirming formation of the tilted smectic C phase. There is no difference in layer spacing between the enantiomer and the racemic mixture. For rac-**RW4***, for which uniformly aligned samples could be obtained between glass plates treated for planar anchoring, the phase transitions were also tracked by optical birefringence, $\Delta n$, changes. At the N-SmA phase transition there is a step-like increase of $\Delta n$, and subsequent transitions between orthogonal smectic phases are marked by weak changes of birefringence, pointing to only small variations of the orientational order (**Figure 1e**).

The two highest-temperature SmA-type phases of rac-**RW4*** give a homeotropic texture when observed in free-standing films or in cells with homeotropic anchoring, with no changes visible at the phase transition between them. Apparently, in both phases the layers are oriented with the layer normal perpendicular to the sample surface. In the lowest temperature orthogonal smectic phase the samples start to lose the homeotropic texture and small wrinkle-like defects develop (**Figure S50**). On further cooling towards the SmC phase, the texture rebuilds completely, a strongly birefringent texture develops evidencing a book-shelf geometry of layers. Such a behavior can be explained assuming that the lowest temperature SmA phase becomes ferroelectric, with the polarization vector along the layer normal. For an axially polar structure there is a strong tendency to escape from homeotropic orientation of layers, even in the free suspended film samples, to avoid the charges at the film surface.[27] Thus, based on results of XRD and optical studies, one can speculate that the smectic phases in rac-**RW4*** appear on cooling in a sequence: non-polar SmA, antiferroelectric $SmA_{AF}$ with polarization compensated by the formation of separate blocks with antiparallel orientation (**Figure 1b**), ferroelectric $SmA_F$ and $SmC_F$, with the onset of polar order in $SmA_{AF}$ phase. It should be noticed that birefringence, measured in planar cell, at SmA- $SmA_{AF}$ phase transition **(Figure 1e)** shows the small step down (of order $10^{-4}$), which can be attributed to small splay of director thus polarization, by no more than 1-2 degree at the boundaries of the antiferroelectric grains (see SI).

In cells with planar anchoring, the enantiomeric S-**RW4*** material shows a Grandjean texture in N* phase (helical axis perpendicular to the cell surface) with the helical pitch in visible range (**Figure S52**) and fan texture in the SmA and $SmA_{AF}$ phases (**Figure 2** and **S53**). The sample undergoes a complete reorganization in the lower temperature phases, typical for the formation of TGB-type phase (**Figure 2e**). In planar geometry, in the TGB phase the Grandjean texture re-appears, and the helical structure gives selective reflection of light in the visible range. On lowering the temperature, judging from the sequence of colors, the helix winds, and the phase transition between TGB_$A_F$ and TGB_$C_F$ phases is marked by appearance of two reflection bands in the visible range (**Figure 2**). The selective reflection bands in the $TGB_C$ phase are asymmetric and become less intense with lowering temperature.

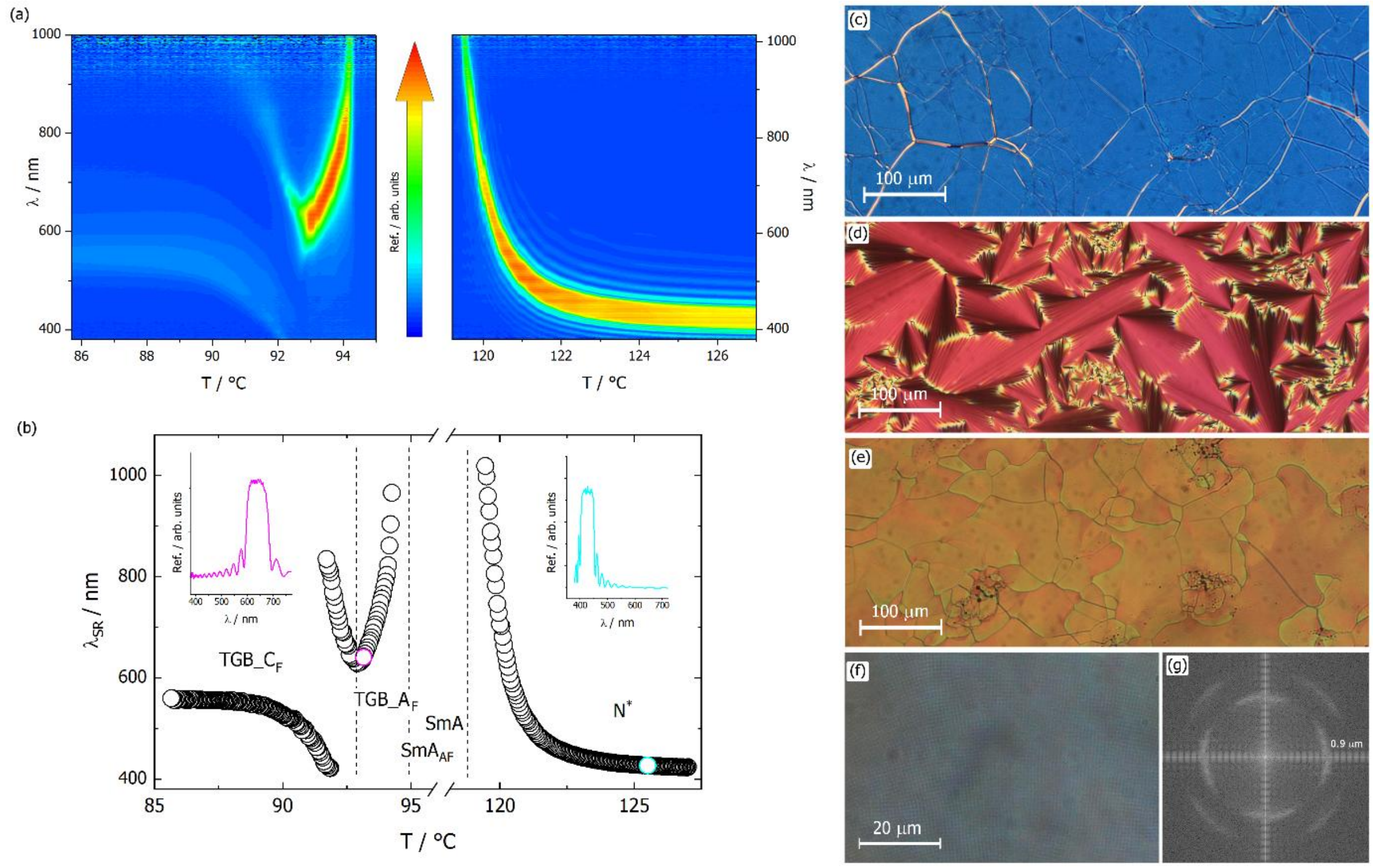

**Figure 2.** (a) 2D plot showing temperature evolution of selective reflection bands in N* phase (right) and TGB phases (left) for S-**RW4*** compound. (b) Selective reflection wavelength, $\lambda_{SW}$, vs. temperature; in the insets reflection vs. wavelength for chosen temperatures in N* (right) and TBG_$A_F$ (left) phases. Optical textures of (c) N* (d) $SmA_{AF}$, (e) TGB_$A_F$ and (f) TGB_$C_F$ phases. In TGB_$C_F$ square pattern texture is observed, which Fourier transform is presented in (g).

A contact cell, in which the enantiomer and racemic mixture form the diffused region with gradual change of optical purity, shows that the TGB_$A_F$ phase in S-**RW4*** corresponds to the temperature range of the $SmA_F$ phase in rac-**RW4***, and the TGB_$C_F$ phase to the $SmC_F$ phase (**Figure 3**).

In the temperature range corresponding to the TGB_$C_F$ phase a square grid pattern develops[28]. This is usually taken as evidence for the TGB_$C^*$ structure (proposed by Renn [29]), in which, in addition to the helical superstructure of layer blocks, each block exhibits the director helix of $SmC^*$ phase, the helical axes being mutually perpendicular.

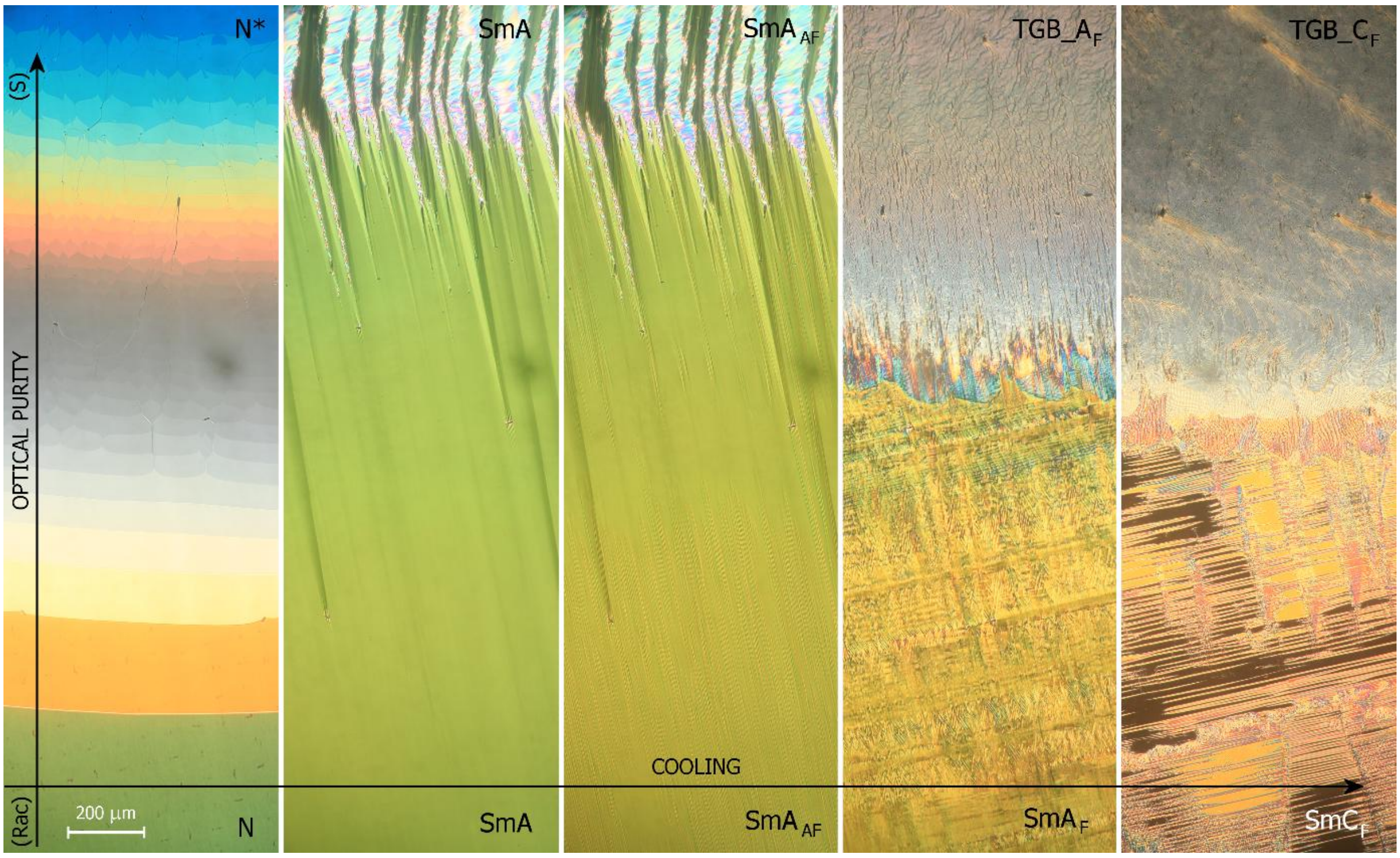

**Figure 3.** Temperature evolution of the optical textures in 3-μm-thick cell with planar anchoring, the area is presented in which optically pure S-**RW4*** (top) is in contact with racemic mixture rac-**RW4*** (bottom).

## 3. Polar properties

To confirm the polar nature of the LC phases, the Second Harmonic Generation (SHG) activity was monitored, which is inherent to materials with a non-centrosymmetric structure and is often used to prove the polar nature of a phase.[30-32] The experiments were conducted in planar cells: in this geometry, a strong SHG signal is expected as the spontaneous polarization is perpendicular to the light propagation direction. In the racemic mixture, an SHG signal appears in the $SmA_F$ and $SmC_F$ phases. The enantiomeric material was SHG silent in all phases because, as expected, the global polarization in the $TGB_A_F$ and $TGB_C_F$ phases is canceled by the helical structure. However, under an electric field, above some critical value (the cell with in-plane electric field was used) the SHG signal is observed already in the $SmA_{AF}$ phase, clearly showing the switching from an antiferroelectric non-SHG active to a ferroelectric, SHG active state (**Figure 3** and **S55**). In the $SmA_{AF}$, $TGB_A_F$ and $TGB_C_F$ phases, a strong SHG signal appears upon application of electric field and its intensity increases with lowering temperature (**Figure S56**).

The electric polarization was determined by measuring the switching current resulting from application of ac electric field. In the $SmA_F$ (or $TGB_A_F$) phase, a single current peak per half of the electric field cycle is found, while in the $SmA_{AF}$ phase a symmetric double peak is registered - consistent with the antiferroelectric nature of the phase (**Figure 4**). The modified electric field cycle (with two consecutive triangular wave functions of the same polarity) confirmed this assumption; the repolarization current peaks appear at each rise/fall of electric field, with the antiferroelectric ground state restored at zero field (**Figure S57**). In the $SmC_F$ (and TGB_$C_F$) phase, the current peak again becomes double, but asymmetric, due to the complex nature of the switching in this phase that involves both, repolarization and changes of the tilt and induction of orthogonal structure under the electric field.[10,11] The spontaneous electric polarization calculated from the current peak gradually increases and reaches ~4.5 μC cm$^2$ (**Figure 4a**), showing nearly perfect ordering of the dipole moments.

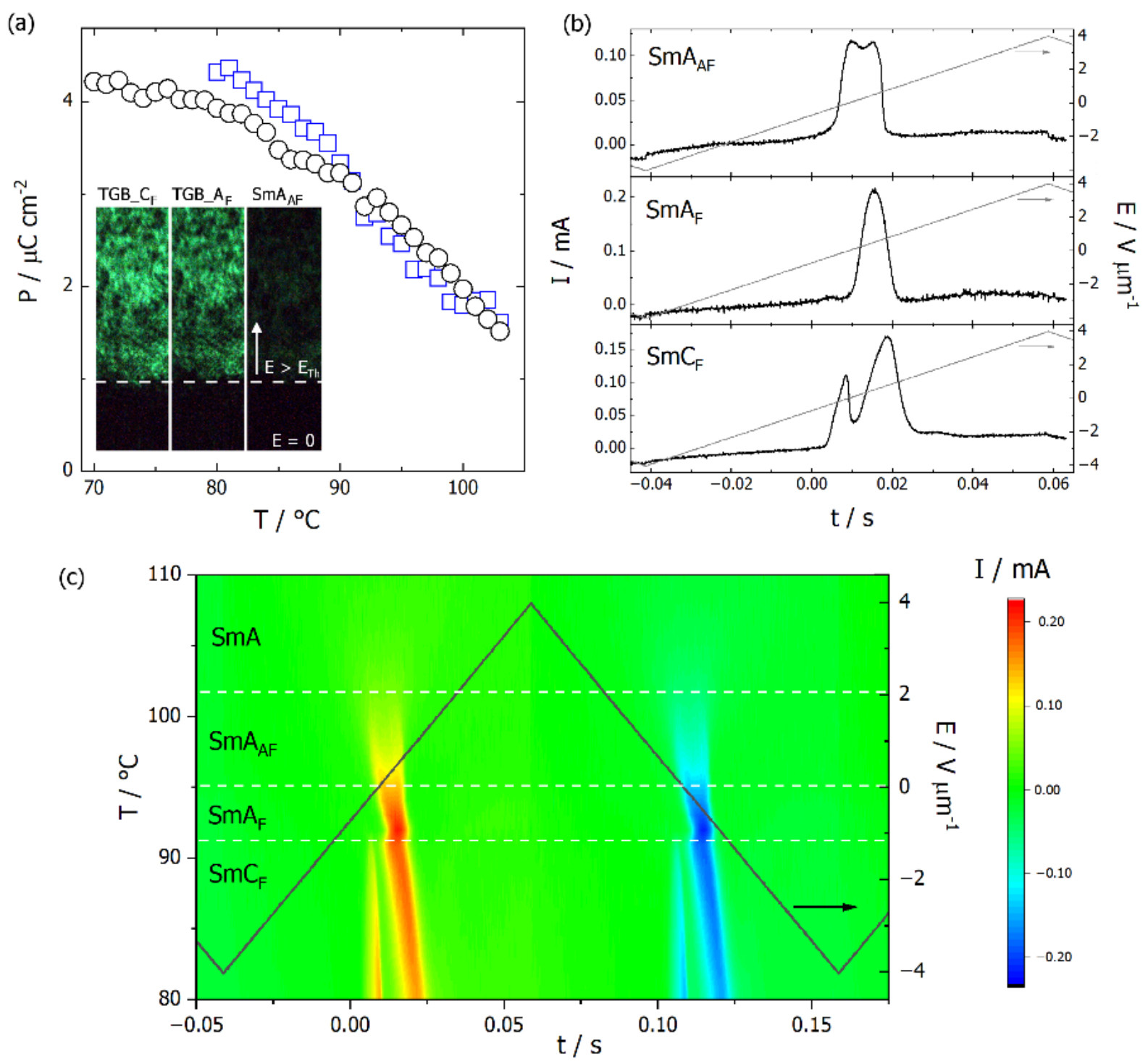

**Figure 4.** (a) Electric polarization vs. temperature for rac-**RW4*** (blue squares) and S-**RW4*** (black circles). In the inset the SHG-microscopy images taken in $SmA_{AF}$, TGB_$A_F$ and TGB_$C_F$ phases in the cell with in-plane electrodes, the SHG active areas (top part) are between electrodes, thus are exposed to electric field, the intensity of SHG signal reflects differences in electric polarization, SHG silent areas (bottom part) are on the electrode (b) switching current recorded under application of triangular-wave ac electric field in $SmA_{AF}$, $SmA_F$ and $SmC_F$ phases of rac-**RW4***. (c) Temperature evolution of repolarization current peaks in smectic phases of rac-**RW4***, evidencing very slight changes in the threshold field for polarization switching.

Dielectric spectroscopy measurements revealed a weak relaxation mode in the SmA and $SmA_{AF}$ phases. This mode softens with decreasing temperature, as evidenced by a decrease in relaxation frequency and an increase in mode strength, indicating the onset of polar order in the $SmA_F$ ($TGB_A_F$) phase. In the $SmA_{AF}$ phase, a second relaxation mode starts to appear at higher frequency, though the dielectric response remains relatively weak compared to that in the lower-temperature phases, possibly hinting at tilt fluctuations starting to develop on approaching the $SmC_F$ ($TGB_C_F$) phase (**Figure S58**). In tilted phases, the dielectric response is substantial, characteristic of proper ferroelectric liquid crystalline phases with strong fluctuations of polar order direction.[10]

## 4. Conclusion

Summarizing the results, for the studied material with strong longitudinal dipole moment, regardless of its optical purity, the higher-temperature phases are non-polar nematic (N), non-polar orthogonal SmA, and antiferroelectric $SmA_{AF}$. In the lower temperature range, the $SmA_F$ and $SmC_F$ phases, which appear in the racemic mixture, are in the enantiomerically pure material transformed into a twist grain boundary superstructure. This phase sequence, with a large temperature range of smectic phases preceding the TGB structure, is in a strong contrast to non-polar materials, where TGB phases typically emerge in a temperature range directly below the nematic phase, when the layer structure is weak. Moreover, for the system studied here, TGB phases emerge for mesogens featuring an (S)-2-methylbutyl end group, which is considered to have only weak helical twisting power,[33] while for non-polar materials a significantly stronger twisting power is required to induce a TGB phase formation. This suggests that the mechanism driving the TGB structure in the studied material is fundamentally different and is influenced by polar interactions. The observed transitions can be accounted for by a continuous phenomenological model. The free energy includes elastic contributions and Landau terms describing the phase transitions from the nematic to the smectic A phase (at temperature $T_{NS}$), from smectic A to smectic C phase (at $T_{AC}$) and the transition to the polar phase (at $T_P$). Details of the model are given in the Supporting Information. The most important terms in the elastic energy are due to the splay and twist deformation of the nematic director. In a racemic mixture there is no tendency for a spontaneous twist, while it appears in an enantiomer. In both, the racemic mixture and optically pure enantiomer, a spontaneous splay is favourable in polar phases due to the flexoelectric effect. If $T_{NS} > T_P > T_{AC}$, the model describes transitions from the apolar nematic phase to apolar smectic A phase, then to the polar smectic A phase and finally to the polar smectic C phase upon the reduction of temperature. At the onset of the polar order ($T < T_P$), polarization splay becomes favourable, which is directly related to the splay of the nematic director. In the SmA phase, splay of the director can be achieved by undulation of smectic layers which comes with no energy penalty, because constant splay does not affect the smectic layer thickness. However, a favourable splay cannot be achieved everywhere, thus regions of favourable splay with the up and down polarization interchange, being separated by regions (walls) without polar order (see Figure S59 in Supporting Information). A phase transition from the SmA to antiferroelectric $SmA_{AF}$ phase is thus observed. From the birefringence measurements for studied material the splay angle is estimated to be ±1.5 degree along the polar block. At temperatures close to $T_P$, melting of the polar order is not energetically costly, but the energy cost increases with decreasing temperature as $(T_P - T)^2$. Thus, at some temperature $T_F$ the cost of the wall with no polar order becomes the same as the cost of the wall with polar order but unfavourable splay. This leads to a phase transition to the $SmA_F$ phase where the neighbouring blocks of favourable splay have the same direction of polarization. With a further reduction of temperature, a transition to the ferroelectric $SmC_F$ phase is obtained. The reason that the $SmA_F$ and $SmC_F$ phases are observed in racemic mixture but not in enentiomer lies in the fact that a chiral material prefers also a spontaneous twist, which is impossible to accommodate along the layer without changing its thickness. In a chiral material, the transition from $SmA_{AF}$ to the polar TGB_$A_F$ phase is observed, because the walls between smectic blocks in the TGB structure can accommodate a favourable twist and unfavourable splay; the energy price of the latter being compensated by the energy gain due to the former. By assuming a constant splay of polarization within a block, the amplitude of the splay angle being small (as only a very weak change of optical birefringence was detected at the SmA-$SmA_{AF}$ phase transition), we estimated the order of magnitude of the blocks' size in polar phases to be of the order of 10 nm (see SI), which is consistent with the widths of the blocks measured in the ordinary TGB_A phase.[34] The width of the block is below the light diffraction limit, resembling blocks in the $SmZ_A$ ($N_x$) phase,[35,36] which are built of ferroelectric nematic domains. However, experimental confirmation of the blocks' size and how they are connected remains an open question. The TGB and $SmA_{AF}$ structures develop in a temperature range where the lamellar order is already strong, therefore we can disregard the possibility that the blocks' interfaces are molten.[37]

**Supporting Information**

Supporting Information (synthetic procedures and structural characterization of materials, experimental details, additional results and phenomenological model)

## References

1. Meyer, R.B., Liebert, L., Strzelecki, L., Keller, P., Ferroelectric liquid crystals, 1975, J. Physique Lett., 36, 69, 10.1051/jphyslet:0197500360306900.

2. Niori, T., Sekine, T., Watanabe, J., Furukawa, T., Takezoe, H., Distinct ferroelectric smectic liquid crystals consisting of banana shaped achiral molecules, 1996, Journal of Materials Chemistry, 6, 1231, https://doi.org/10.1039/JM9960601231.

3. Link, DR., Natale, G., Shao, R., Maclennan, JE., Clark, NA., Korblova, E., Walba, DM., Spontaneous formation of macroscopic chiral domains in a fluid smectic phase of achiral molecules, 1997, Science. 278, 1924, https://doi.org/10.1126/science.278.5345.1924. PMID: 9395390.

4. Nishikawa, H., Shiroshita, K., Higuchi, H., Okumura, Y. , Haseba, Y., Yamamoto, SI., Sago, K., Kikuchi, H., A Fluid Liquid-Crystal Material with Highly Polar Order, 2017, *Adv. Mater.*, 29, 1702354, https://doi.org/10.1002/adma.201702354

5. Mandle, R. J., Cowling, S. J., Goodby, J. W., A nematic to nematic transformation exhibited by a rod-like liquid crystal, 2017, Phys. Chem. Chem. Phys., 19, 11429, https://doi.org/10.1039/C7CP00456G.

6. Mertelj, A., Cmok, L., Sebastián, N., Mandle, R. J., Parker, R. R., Whitwood, A. C., Goodby, J. W., Čopič, M., Splay Nematic Phase, 2018, Phys. Rev. X, 8, 041025, https://doi.org/10.1103/PhysRevX.8.041025.

7. Chen, X.; Martinez, V.; Nacke, P.; Korblova, E.; Manabe, A.;Klasen-Memmer, M.; Freychet, G.; Zhernenkov, M.; Glaser, M. A.;Radzihovsky, L.; Maclennan, J. E.; Walba, D. M.; Bremer, M.; Giesselmann, F.; Clark, N. A. Observation of a uniaxial ferroelectric smectic A phase, 2022, Proc. Natl. Acad. Sci. U.S.A. 119, e2210062119, https://doi.org/10.1073/pnas.2210062119.

8. Kikuchi, H.; Matsukizono, H.; Iwamatsu, K.; Endo, S.; Anan, S.;Okumura, Y. Fluid Layered Ferroelectrics with Global $C_{\infty v}$ Symmetry, 2022, Adv. Sci. 9, 2202048, https://doi.org/10.1002/advs.202202048

9. Kikuchi, H., Nishikawa, H., Matsukizono, H., Iino, S., Sugiyama, T., Ishioka, T., Okumura, Y., Ferroelectric Smectic C Liquid Crystal Phase with Spontaneous Polarization in the Direction of the Director., 2024, Adv. Sci. 2024, 11, 2409827, https://doi.org/10.1002/advs.202409827

10. Strachan, G. J.; Górecka, E.; Szydłowska, J.; Makal, A.; Pociecha, D., Nematic and Smectic Phases with Proper Ferroelectric Order, 2025, *Adv. Sci.*, *12*, e2409754 DOI: 10.1002/advs.202409754.

11. Hobbs, J., Gibb, C. J., Pociecha, D., Szydłowska, J., Górecka, E., Mandle, R. J., Polar Order in a Fluid Like Ferroelectric with a Tilted Lamellar Structure − Observation of a Polar Smectic C (SmC P ) Phase, 2025, Angew. Chem. Int. Ed., 64, e202416545, https://doi.org/10.1002/anie.202416545.

12. Nishikawa, H., Araoka, F., A New Class of Chiral Nematic Phase with Helical Polar Order, 2021, Adv. Mater., 33, 2101305, https://doi.org/10.1002/adma.202101305.

13. Feng, C., Saha, R., Korblova, E., Walba, D., Sprunt, S. N., Jákli, A., Electrically Tunable Reflection Color of Chiral Ferroelectric Nematic Liquid Crystals, 2021, Adv. Opt. Mater., 9, 2101230, https://doi.org/10.1002/adom.202101230.

14. Zhao, X., Zhou, J., Li, J., Kougo, J., Wan, Z., Huang, M., Aya, S., Spontaneous helielectric nematic liquid crystals: Electric analog to helimagnets, 2021, Proc. Natl. Acad. Sci. U.S.A., 118, e2111101118, https://doi.org/10.1073/pnas.2111101118.

15. Ortega, J., Folcia, C. L., Etxebarria. J., Sierra, T., Ferroelectric chiral nematic liquid crystals: new photonic materials with multiple bandgaps controllable by low electric fields, 2022, Liq. Cryst., 49, 2128, https://doi.org/10.1080/02678292.2022.2104949.

16. Kumari, P., Basnet, B., Lavrentovich M. O., Lavrentovich O. D., Chiral ground states of ferroelectric liquid crystals, 2024, Science, 383,1364, DOI:10.1126/science.adl0834.

17. Karcz, J., Herman, J., Rychłowicz, N., Kula, P., Górecka, E., Szydlowska, J., Majewski, P. W., Pociecha, D., Spontaneous chiral symmetry breaking in polar fluid–heliconical ferroelectric nematic phase, 2024, Science 384, 1096, doi:10.1126/science.adn6812.

18. Gibb, C.J., Hobbs, J., Nikolova, D.I., Raistrick, T., Berrow, S. R., Mertelj, A., Osterman, N., Sebastián, N., Gleeson, H. F., Mandle, R. J., Spontaneous symmetry breaking in polar fluids, 2024, *Nat. Commun.*, **15**, 5845, https://doi.org/10.1038/s41467-024-50230-2.

19. Nishikawa, H., Okada, D., Kwaria, D., Nihonyanagi, A., Kuwayama, M., Hoshino, M., Araoka, F., Emergent Ferroelectric Nematic and Heliconical Ferroelectric Nematic States in an Achiral “Straight” Polar Rod Mesogen, 2024, Adv. Sci., 11, 2405718, https://doi.org/10.1002/advs.202405718.

20. Goodby, J. W., Waugh, M. A., Stein, S. M., Chin, E., Pindak, R., Patel J. S., Characterization of a new helical smectic liquid crystal, 1989, Nature, 337, 449, https://doi.org/10.1038/337449a0.

21. Renn, S., Lubensky. T., Abrikosov dislocation lattice in a model of the cholesteric to smectic-A transition, 1988, Phys. Rev. A, 38, 2132, https://doi.org/10.1103/PhysRevA.38.2132.

22. Goodby, J. W., Twist grain boundary and frustrated liquid crystal phases, 2002, Current Opinion in Colloid and Interface Science, 7, 326, https://doi.org/10.1016/S1359-0294(02)00090-0.

23. Yelamaggad, C.V., Achalkumar, A. S. , Bonde, N. L., Prajapati, A. K., Liquid Crystal Abrikosov Flux Phase: The Exclusive Wide Thermal Range Enantiotropic Occurrence, 2006, Chem. Mater., 18, 1076, https://doi.org/10.1021/cm052570+.

24. Glogarová, M., Novotná, V., Frustrated smectic liquid crystalline phases in lactic acid derivatives, 2016, Ph. Transit., 89, 829, https://doi.org/10.1080/01411594.2016.1180519.

25. Huang, H., Li, J., Jia, Y., Du, A., Zhang, B., TGBs below SmA*: Mesophase behaviours and optical properties, 2024, J. Mol. Liq., 393, 123692, https://doi.org/10.1016/j.molliq.2023.123692

26. Nishikawa, H., Okumura, Y., Kwaria, D., Nihonyanagi, A., Araoka, F., Spontaneous Twist of Ferroelectric Smectic Blocks in Polar Fluids, 2025, Adv. Mater., 2501946, https://doi.org/10.1002/adma.202501946.

27. Hedlund, K. G., Martinez, V., Chen, X., Park, C. S., Maclennan, J. E. , Glaser, M.A., Clark, N. A., Freely suspended nematic and smectic films and free-standing smectic filaments in the ferroelectric nematic realm, 2025, Phys. Chem. Chem. Phys., 27, 119, 10.1039/D4CP03425B.

28. Kuczynski, W., Stegemeyer, H., Polymorphism of twist grain boundary phases near the chiral NAC point, 1997, Proc. SPIE, 3318, 90, https://doi.org/10.1117/12.299944.

29. Renn, S., Multicritical behavior of Abrikosov vortex lattices near the cholesteric–smectic-A–smectic-$C$* point, 1992, Phys. Rev. A 45, 953, https://doi.org/10.1103/PhysRevA.45.953

30. Miyajima, D., Araoka, F., Takezoe, H., Kim, J., Kato, K., Takata, M., Aida T. , Columnar Liquid Crystal with a Spontaneous Polarization along the Columnar, 2010, J. Am. Chem. Soc., 132, 25, 8530, 10.1021/ja101866e.

31. Folcia, C.L., Ortega, J., Vidal, R. , Sierra, T., Etxebarria J., The ferroelectric nematic phase: an optimum liquid crystal candidate for nonlinear optics, 2022, Liq. Cryst., 49, 899, https://doi.org/10.1080/02678292.2022.2056927.

32. Lovšin, M., Petelin, A., Berteloot, B. , Osterman, N., Aya, S., Huang, M., Drevenšek-Olenik, I., Mandle, R.J., Neyts, K., Mertelj, A., Sebastian, N., Patterning of 2D second harmonic generation active arrays in ferroelectric nematic fluids, 2024, Giant, 19, 100315, https://doi.org/10.1016/j.giant.2024.100315.

33. Ocak, H., Bilgin-Eran, B., Guzeller, D., Prehm, M., Tschierske C., Twist grain boundary (TGB) states of chiral liquid crystalline bent-core mesogens, 2015, Chem. Commun., 51, 7512, https://doi.org/10.1039/C5CC01592H.

34. Navailles, L.; Pansu, B.; Gorre-Talini, L.; Nguyen, H. T.; Structural Study of a Commensurate TGB A Phase and of a Presumed Chiral Line Liquid Phase, 1998, Phys. Rev. Lett. 81, 4168. https://doi.org/10.1103/PhysRevLett.81.4168

35. Chen, X., Martinez, V. , Korblova, E., Freychet, G., Zhernenkov, M., Glaser, M. A., Wang, C., Zhu, C., Radzihovsky, L., Maclennan, J. E., Walba, D. M., Clark, N. A. , The smectic $Z_A$ phase: Antiferroelectric smectic order as a prelude to the ferroelectric nematic, 2023, *Proc. Natl. Acad. Sci. U.S.A.*, 120, e2217150120, https://doi.org/10.1073/pnas.221715012.

36. Cruickshank, E., Rybak, P., Majewska, M. M., Ramsay, S., Wang, C., Zhu, C., Walker, R., Storey, J. M. D., Imrie, C. T., Górecka, E., Pociecha, D., To Be or Not To Be Polar: The Ferroelectric and Antiferroelectric Nematic Phases, 2023, *ACS Omega*, 8, 36562, https://doi.org/10.1021/acsomega.3c05884.

37. Fernsler, J., Hough, L., Shao, R.-F., Maclennan, J. E., Navailles, L., Brunet, M., Madhusudana, N. V., Mondain-Monval, O., Boyer, C., Zasadzinski, J., Rego, J. A., Walba, D. M., Clark N. A., 2005, Proc. Natl. Acad. Sci. U.S.A., 102, 14191, https://doi.org/10.1073/pnas.050066410

*Supporting Information for:*

**Twist Grain Boundary phases in proper ferroelectric liquid crystals realm**

*Damian Pociecha, Jadwiga Szydlowska, Nataša Vaupotič, Katarzyna Kwiatkowska, Marijus Juodka, Julian Spiess, John MD Storey, Corrie T Imrie, Rebecca Walker, Ewa Gorecka**

## Synthetic Procedures and Structural Characterisation

**Reagents** All reagents and solvents that were available commercially were purchased from Sigma Aldrich, Fisher Scientific or Fluorochem and were used without further purification unless otherwise stated.

**Thin Layer Chromatography** Reactions were monitored using thin layer chromatography, and the appropriate solvent system, using aluminium-backed plates with a coating of Merck Kieselgel 60 F254 silica which were purchased from Merck KGaA. The spots on the plate were visualised by UV light (254 nm) or by oxidation using either a potassium permanganate stain, iodine or p-anisaldehyde dip.

**Column Chromatography** For normal phase column chromatography, the separations were carried out using silica gel grade 60 Å, 40-63 μm particle size, purchased from Fluorochem and using an appropriate solvent system.

**Structure Characterisation** All final products and intermediates that were synthesised were characterised using $^{1}$H NMR, $^{13}$C NMR and infrared spectroscopies. The NMR spectra were recorded on a 400 MHz Bruker Avance III HD NMR spectrometer. The infrared spectra were recorded on a Perkin Elmer Spectrum Two FTIR spectrometer with an ATR diamond cell.

The synthetic routes to the (S)-enantiomer and racemic versions of **RW4*** were identical, apart from utilisation of (S)-1-bromo-2-methylbutane and 1-bromo-2-methylbutane, respectively, in Step 1, and summarised in ***Scheme 1***.

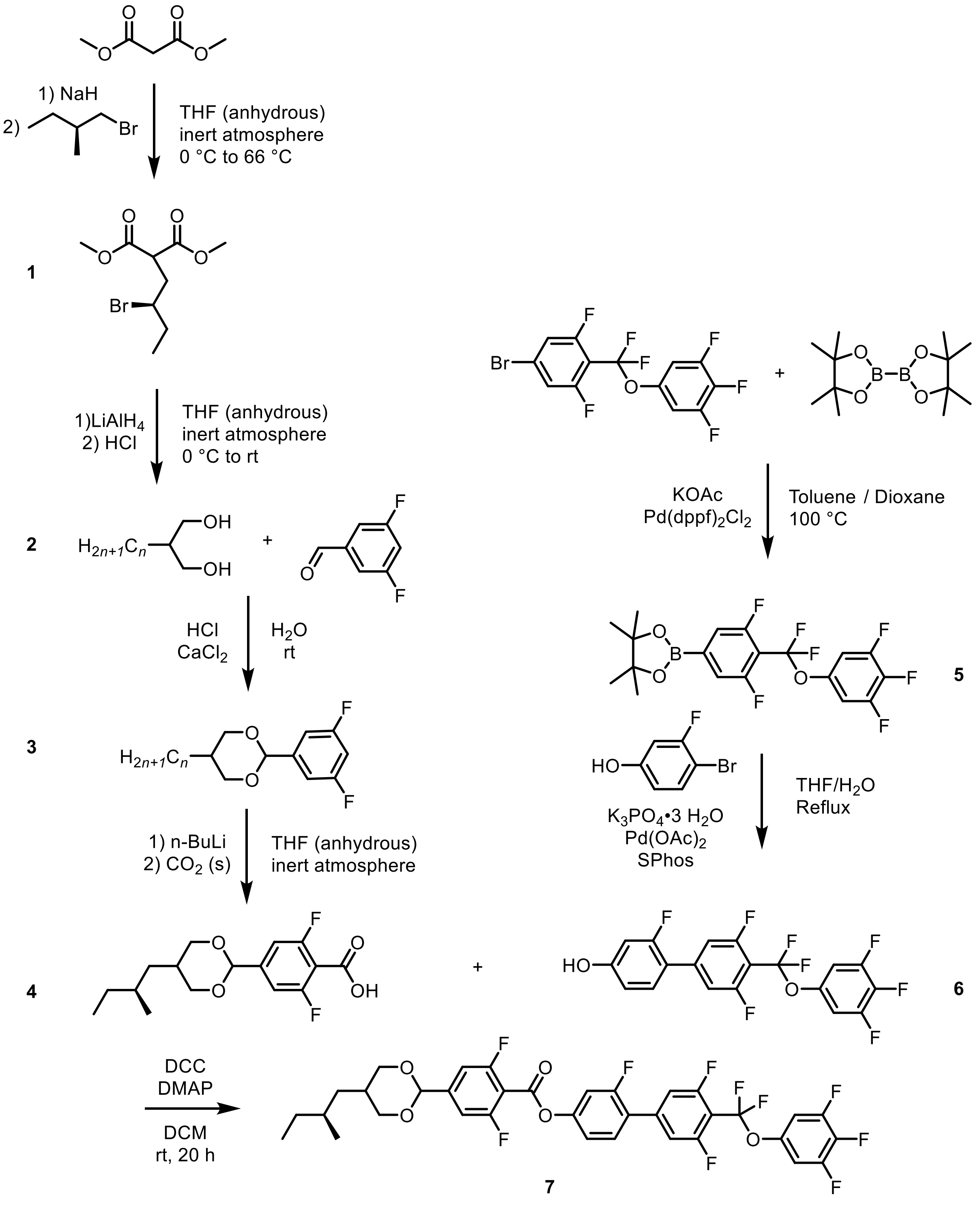

*Scheme S1. Synthetic route to new materials reported. Final product 7 was obtained as S-enantiomer (**S-RW4***) and as racemic mixture (**rac-RW4***)*

**Dimethyl 2-[(2S)-2-methylbutyl]propanedioate (1(S)) /**

**Dimethyl 2-(2-methylbutyl)propanedioate (1(rac))**

NaH (1.1 eq) was placed in a 2-necked RBF with a magnetic stirrer under an argon atmosphere. Hexane (3 x 4 mL) was added with a syringe, the mixture was stirred, followed by a removal of the hexane - mineral oil solution using a syringe. Then, anhydrous THF (15 mL) was added to the RBF using a syringe, and the reaction mixture cooled down to 0 $^oC$ using an ice bath. Dimethylmalonate (1.1 eq) was then added dropwise using a syringe and the reaction was left to stir for 30 minutes until all hydrogen evolution had ceased. Using the same technique, (S)-1-bromo-2-methylbutane* (1 eq) was added to the reaction mixture. The ice bath was then removed, and the reaction mixture stirred at reflux for 8h. The reaction was quenched using HCl (1 N), and the product was extracted using of $Et_2O$ (3 x 10 mL). Combined organic layers were washed with water (2 x 30 mL), brine (2 x 30 mL) and dried over $MgSO_4$. The solvent was removed *in vacuo* and flash chromatography (8:2 PE/EtOAc, Rf = 0.48) was performed to yield the pure compound as a clear liquid.

*Procedure for **1(rac)** used 1-bromo-2-methylbutane.

**1(S):**

**Yield:** 62.8 %

**$^1$H NMR** (400 MHz, $CDCl_3$): δ ppm = 3.68 – 3.59 (6H, m, -(O-$\underline{CH_3}$)$_2$), 3.42 – 3.32 (1H, m, -CO-$\underline{CH}$(-$CH_2$)-CO-), 1.91 – 1.82 (1H, m, -CH-$\underline{CH}$(-H)-CH-), 1.63 – 1.54 (1H, m, -CH-CH($\underline{-H}$)-CH-), 1.33 – 1.18 (2H, m, -CH-$\underline{CH_2}$-$CH_3$), 1.12 – 1.02 (1H, m, -CH(-H)-$\underline{CH}$(-$CH_3$)-$CH_2$-), 0.81 – 0.73 (6H, m, -CH(-$\underline{CH_3}$); -$CH_2$-$\underline{CH_3}$).

**$^{13}$C NMR** (101 MHz, $CDCl_3$): δ ppm = 170.09, 169.96, 52.33, 52.28, 49.69, 35.50, 32.37, 29.14, 18.57, 10.96.

**IR** ($\nu_{max}$/cm$^{-1}$): 2957 ($sp^3$ C-H stretch), 2933 ($sp^3$ C-H stretch), 2877 ($sp^3$ C-H stretch), 1734 (C=O stretch, ester).

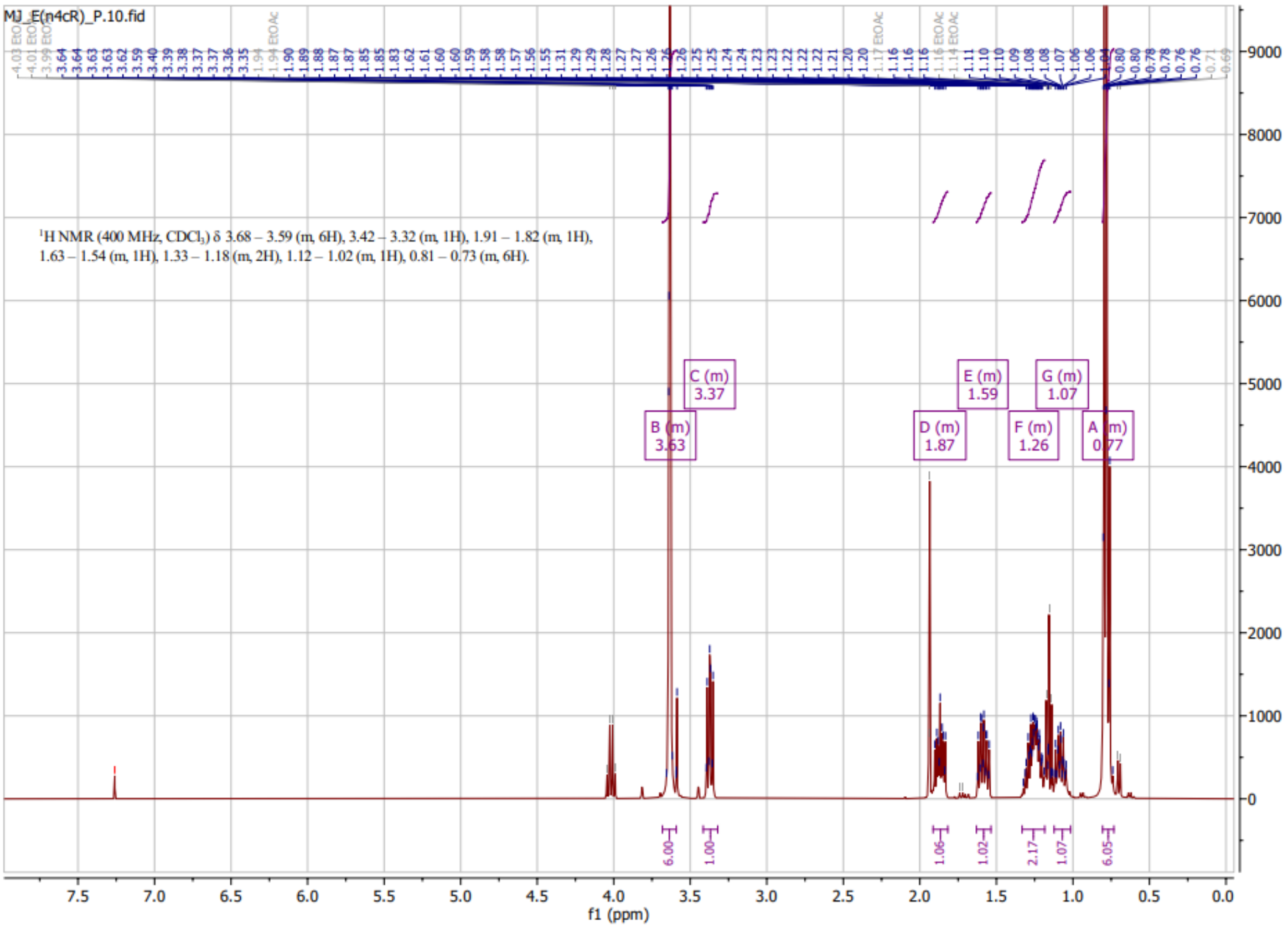

**Figure S1**: $^{1}$H NMR spectrum of **1(S)** in $CDCl_3$.

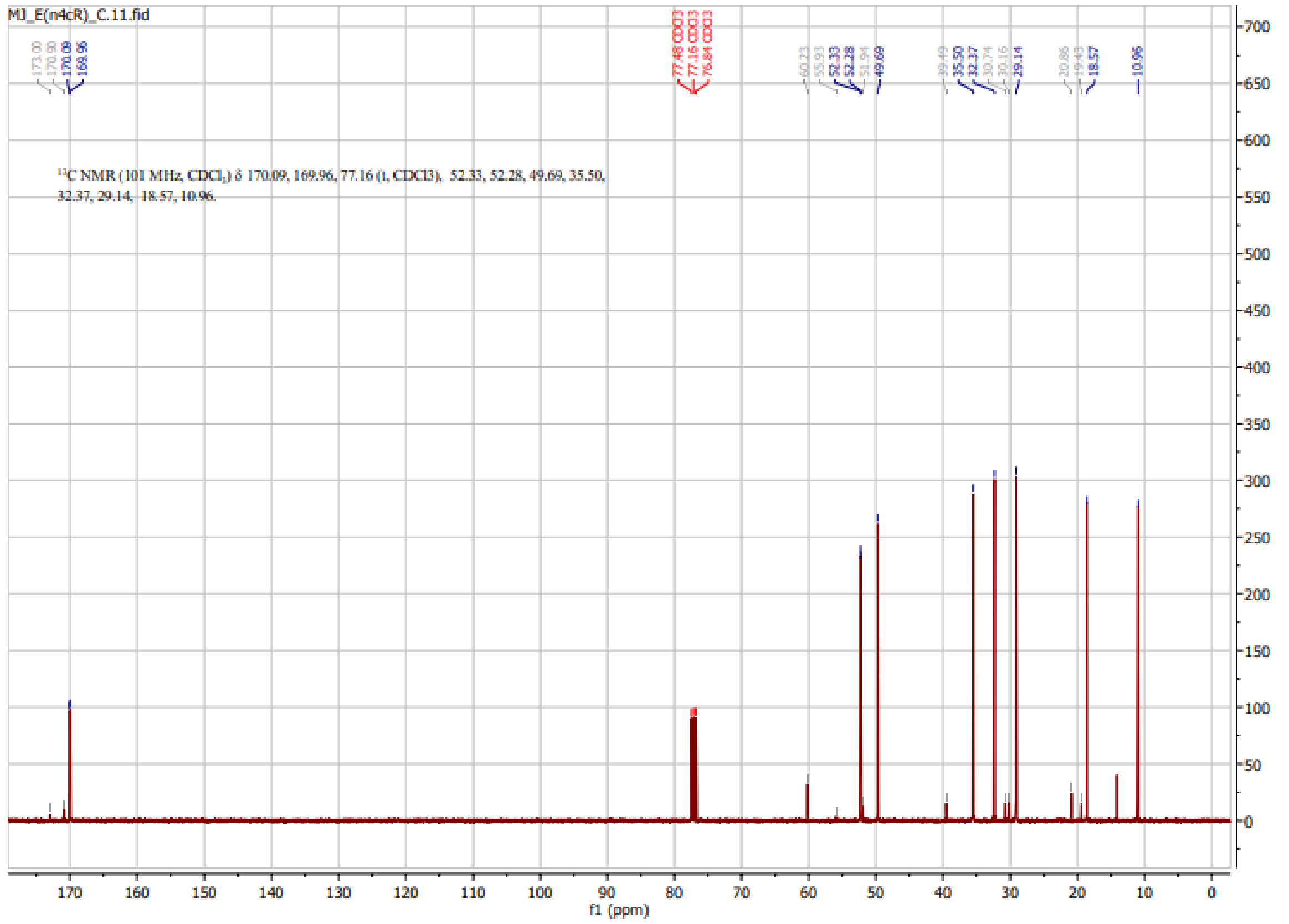

**Figure S2**: $^{13}$C NMR spectrum of **1(S)** in $CDCl_3$.

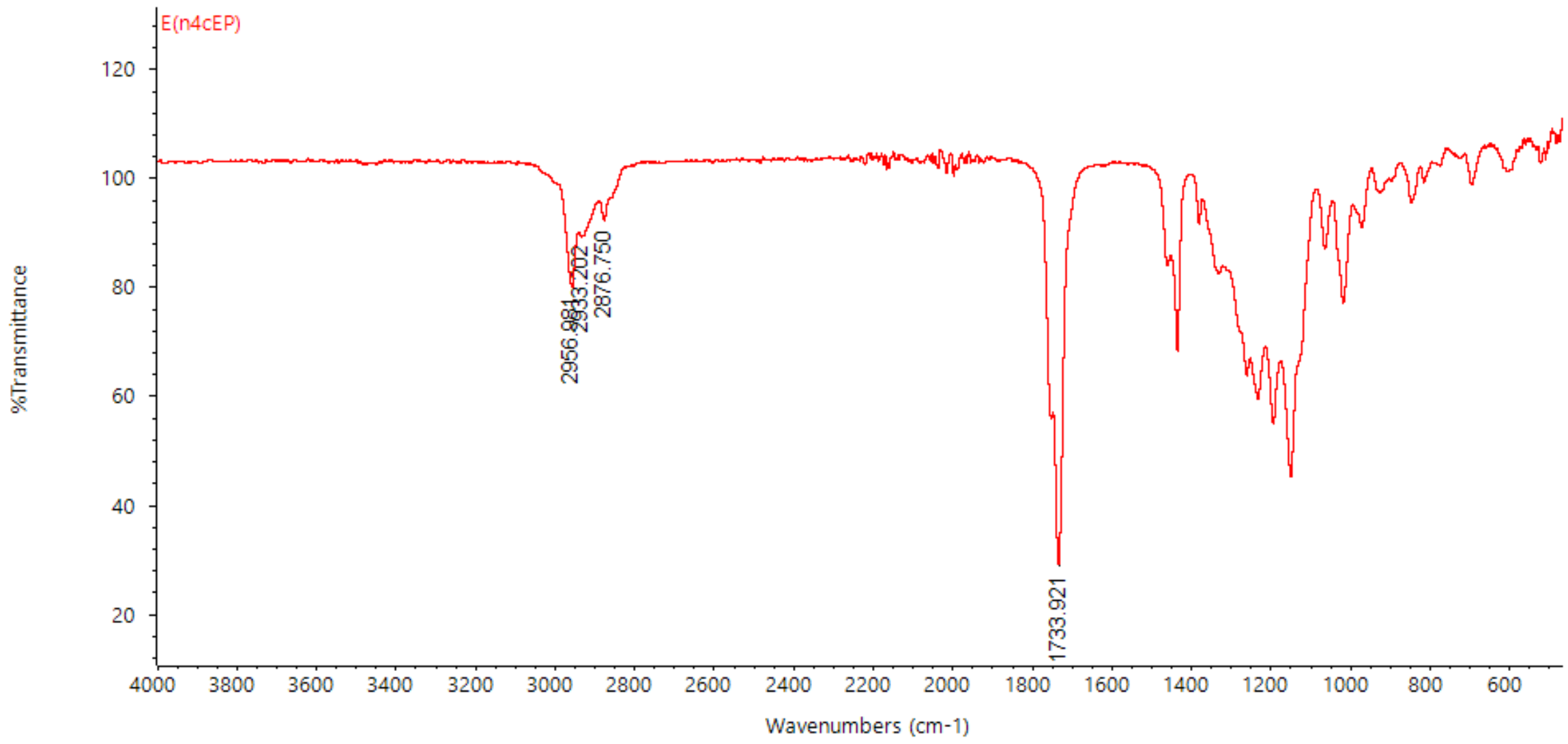

**Figure S3**: IR spectrum of **1(S)**.

**1(rac):**

**Yield:** 37.6 %

**$^1$H NMR** (400 MHz, $CDCl_3$): δ ppm = 3.70 (6H, d, *J* 1.6 Hz, -($OCH_3$)$_2$), 3.44 (1H dd, *J* 8.8 Hz, 6.6 Hz, -(CO)$_2$-<u>CH</u>-$CH_2$-), 1.99 – 1.89 (1H, m, -CH-C<u>H</u>(-H)-CH-), 1.71 – 1.59 (1H m, -CH-CH(-<u>H</u>)-CH-), 1.38 – 1.25 (2H, m, -CH-<u>$CH_2$</u>-$CH_3$), 1.21 – 1.09 (1H, m,-CH-<u>CH</u>-($CH_3$)-$CH_2$-), 0.91 – 0.76 (6H, m, -$CH_2$-<u>$CH_3$</u>, -CH-<u>$CH_3$</u> ).

**$^{13}$C NMR** (101 MHz, $CDCl_3$): δ ppm = 170.31, 170.17, 52.55, 52.50, 49.87, 49.85, 35.65, 35.63, 32.51, 32.49, 29.28, 29.26, 18.73, 18.71, 11.14, 11.12.

**IR** ($\nu_{max}$/cm$^{-1}$): 2958 (sp$^3$ C-H stretch), 2933 (sp$^3$ C-H stretch), 2877 (sp$^3$ C-H stretch), 1733 (C=O stretch, ester).

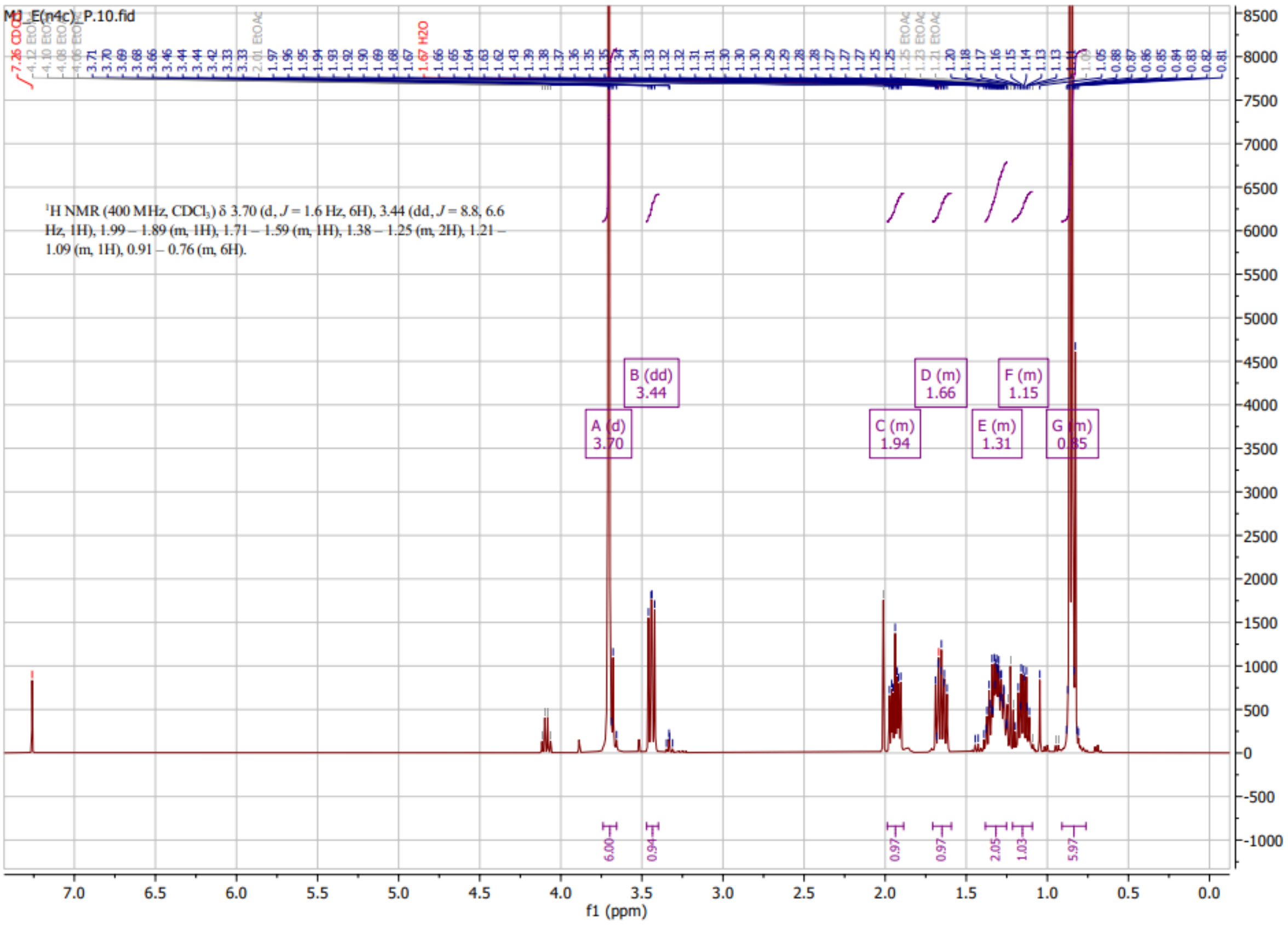

**Figure S4**: $^1$H NMR spectrum of **1(rac)** in $CDCl_3$.

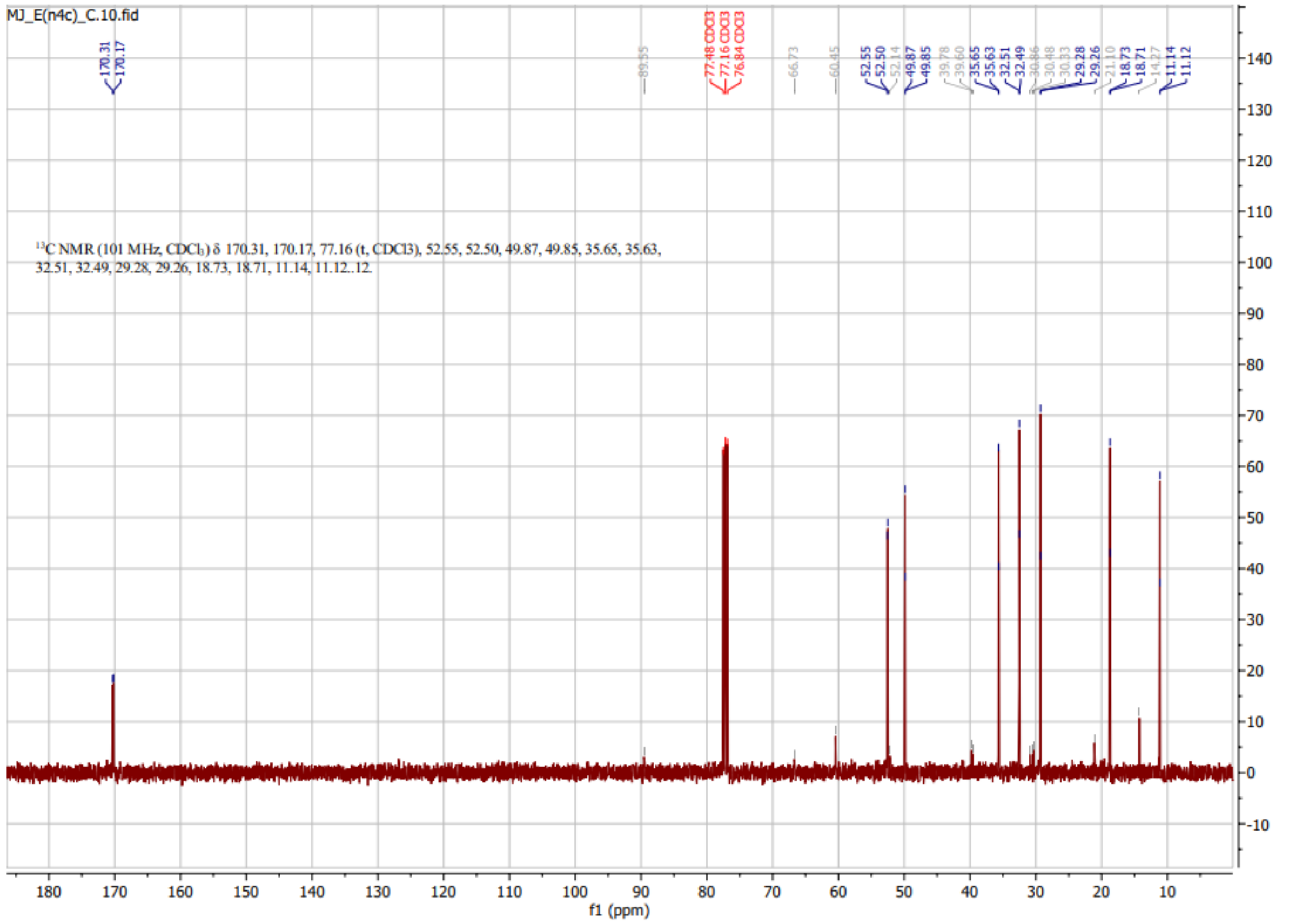

**Figure S5**: $^{13}$C NMR spectrum of **1(rac)** in $CDCl_3$.

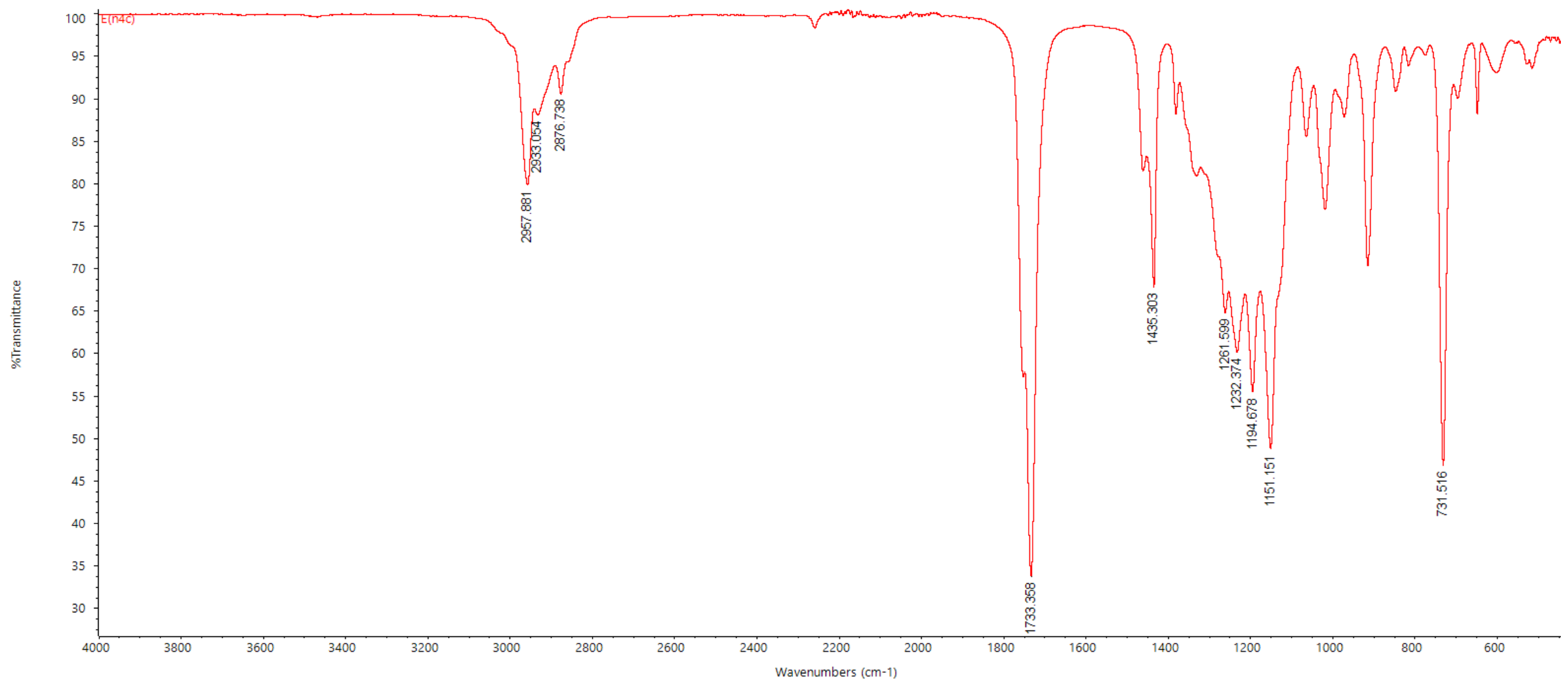

**Figure S6**: IR spectrum of **1(rac)**.

**2-[(2S)-2-methylbutyl]propane-1,3-diol (2(S)) /**

**2-[2-methylbutyl]propane-1,3-diol (2(rac))**

Anhydrous THF (20 mL) was added to a 2-necked RBF under argon atmosphere. To the stirred solvent, $LiAlH_4$ solution (2 M, 4.1 eq) was added using a syringe. The solution was cooled to 0 °C using an ice bath, and the 2-alkyl-dimethylmalonate (1 eq) was added dropwise using a syringe. The reaction mixture was allowed to slowly warm to room temperature and stirred for 8h. The reaction mixture was then cooled down to 0 °C once again and acidified with HCl (2 N) followed by the addition of dilute acetic acid solution (1 M, 10 mL) to quench the reaction. The product was then extracted with $Et_2O$ (3 x 10 mL), combined organic layers were washed with water (2 x 20 mL), brine (40 mL), dried over $MgSO_4$ and purified with flash chromatography (250 mL of 8:2, 100 mL of 1:1, 200 mL of 3:7, PE/EtOAc) to yield a colourless oil (Rf = 0.15 (1:1 PE/EtOAc)).

**2(S):**

**Yield:** 45.8 %

**$^1$H NMR** (400 MHz, $CDCl_3$): δ ppm = 3.85 – 3.74 (2H, m, -CH-$\underline{CH_2}$-O-), 3.68 – 3.55 (2H, m, -CH-$\underline{CH_2}$-O-), 2.47 (1H, s,-OH), 2.15 (1H, s, -OH), 1.94 – 1.83 (1H, m, -$CH_2$-$\underline{CH}$-($CH_2$-)$_2$), 1.43 – 1.29 (2H, m, -CH-$\underline{CH(-H)}$-CH-), 1.28 – 1.07 (2H, m, -CH-$\underline{CH(-H)}$-$CH_3$), 1.00 – 0.91 (1H, m,-$CH_2$-$\underline{CH}$(-$CH_3$)-$CH_2$-), 0.91 – 0.80 (6H, m,-CH-$\underline{CH_3}$; -$CH_2$-$\underline{CH_3}$).

**$^{13}$C NMR** (101 MHz, $CDCl_3$): δ ppm = 66.43, 65.62, 39.44, 34.78, 31.81, 29.78, 19.34, 11.30.

**IR** ($\nu_{max}$/cm$^{-1}$): 3600-3000 (broad, -OH stretch), 2958 (sp$^3$ C-H stretch), 2917 (sp$^3$ C-H stretch), 2875 (sp$^3$ C-H stretch).

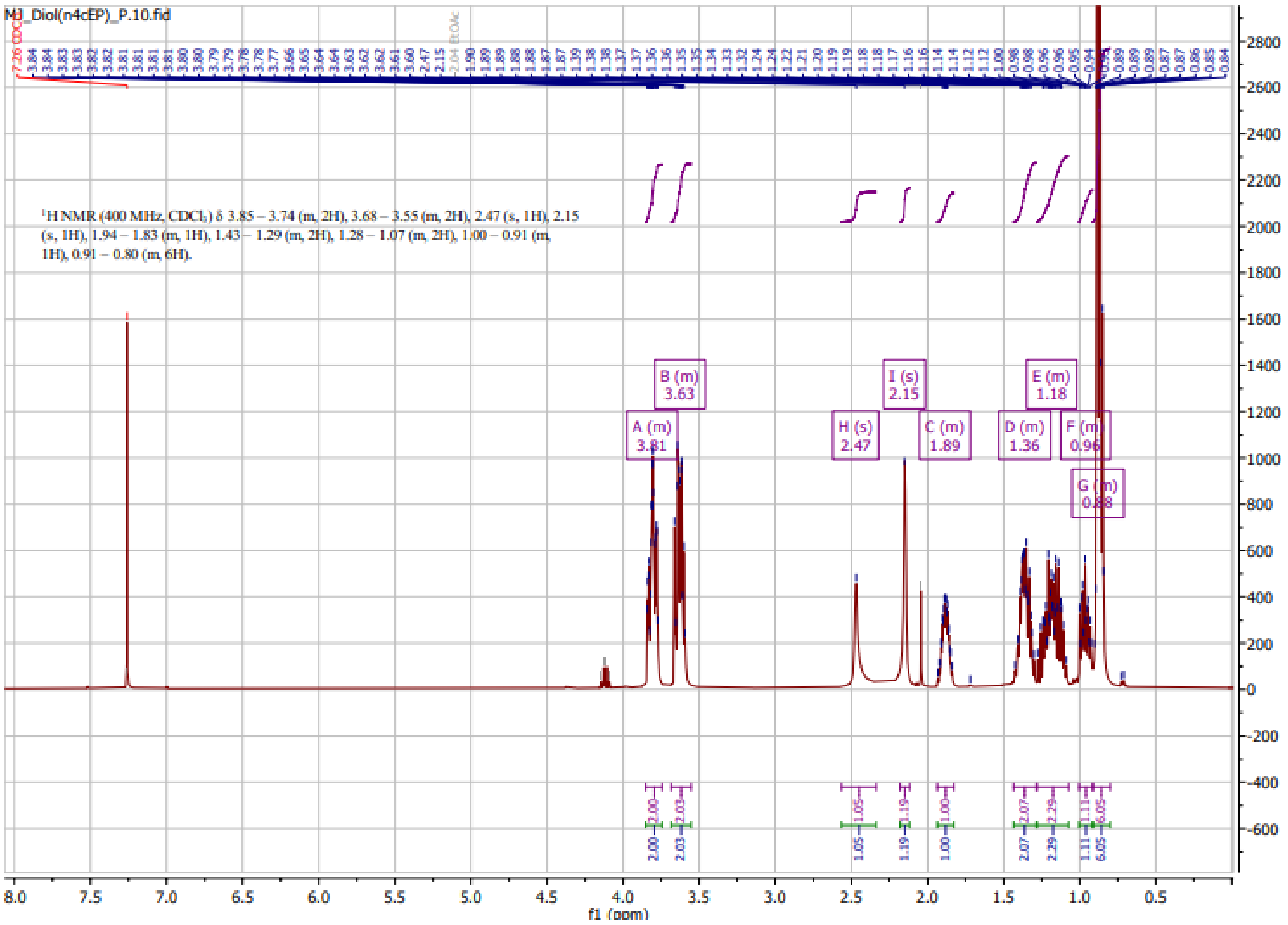

**Figure S7**: $^1$H NMR spectrum of **2(S)** in $CDCl_3$.

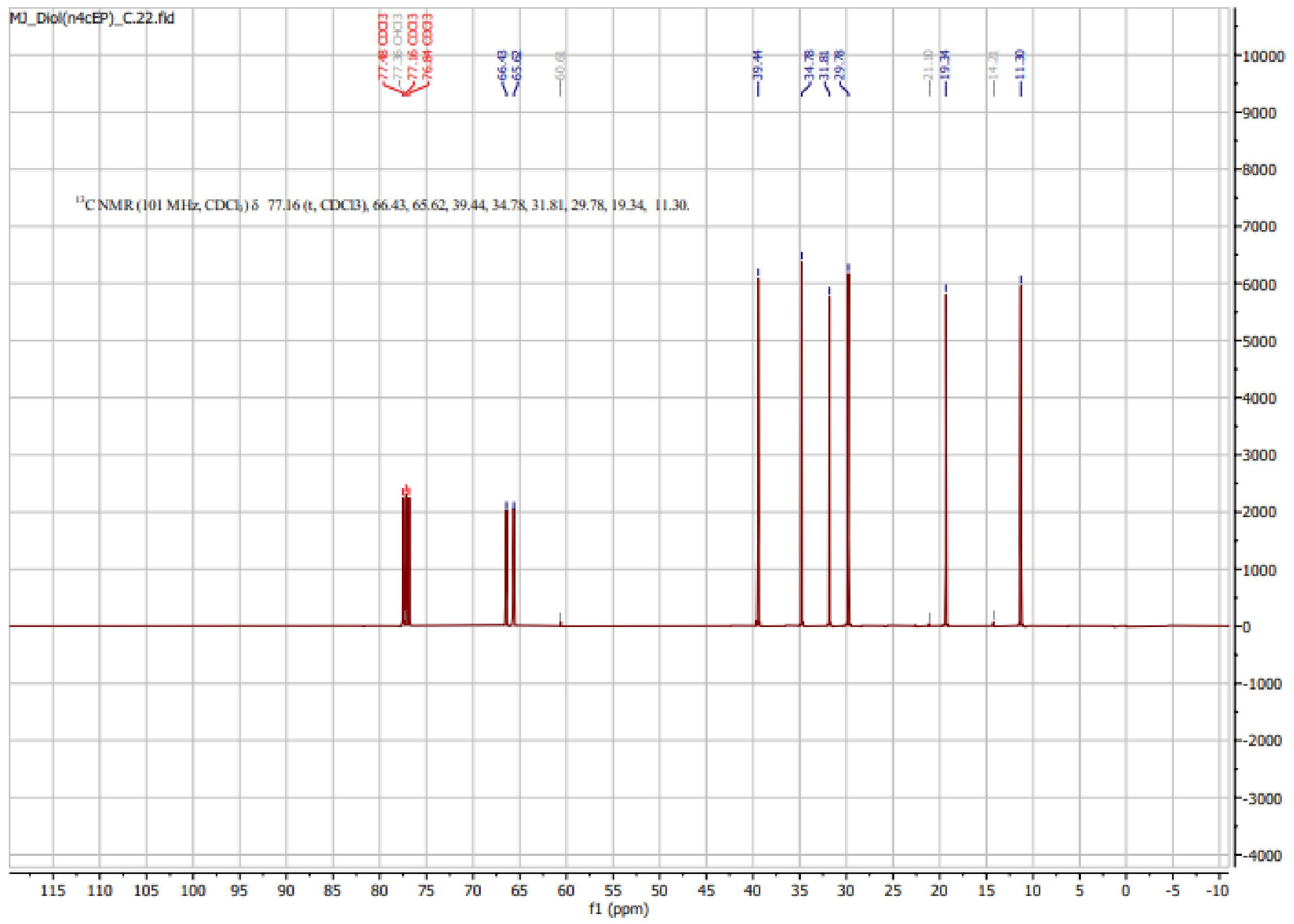

**Figure S8**: $^{13}$C NMR spectrum of **2(S)** in $CDCl_3$.

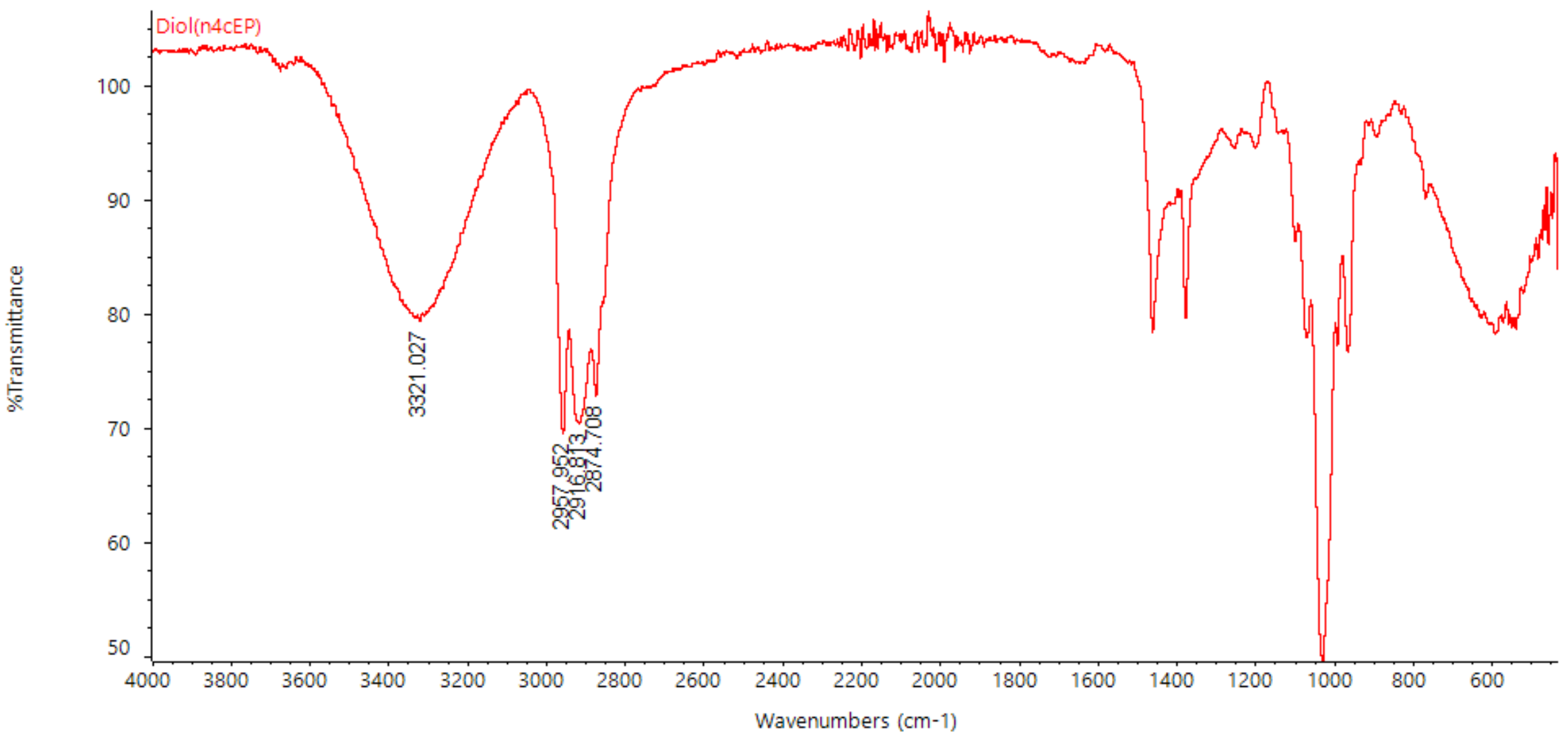

**Figure S9**: IR *s*pectrum of **2(S)**.

**2(rac):**

**Yield:** 31.3 %

**$^1$H NMR** (400 MHz, CDCl$_3$): δ ppm = 3.83 – 3.72 (2H, m, -CH-CH$_2$-O-), 3.65 – 3.52 (2H, m, -CH-CH$_2$-O-), 2.91 (2H, s, (-OH)$_2$), 1.91 – 1.81 (1H, m, ,-CH$_2$-CH-(CH$_2$-O-)$_2$), 1.43 – 1.27 (2H, m, -CH-CH$_2$-CH(-CH$_3$)- ), 1.22 – 1.07 (2H, m, -CH(-CH$_3$)-CH$_2$-CH$_3$), 0.98 – 0.90 (1H, m, -CH$_2$-CH(-CH$_3$)-CH$_2$-), 0.89 – 0.80 (6H, m, -CH-CH$_3$; -CH$_2$-CH$_3$).

**$^{13}$C NMR** (101 MHz, CDCl$_3$): δ ppm = 67.44, 66.63, 39.48, 34.78, 31.91, 29.84, 19.42, 11.37.

**IR** ($\nu_{max}$/cm$^{-1}$): 3600-3000 (broad, -OH stretch), 2958 (sp$^3$ C-H stretch), 2920- (sp$^3$ C-H stretch), 2874 (sp$^3$ C-H stretch).

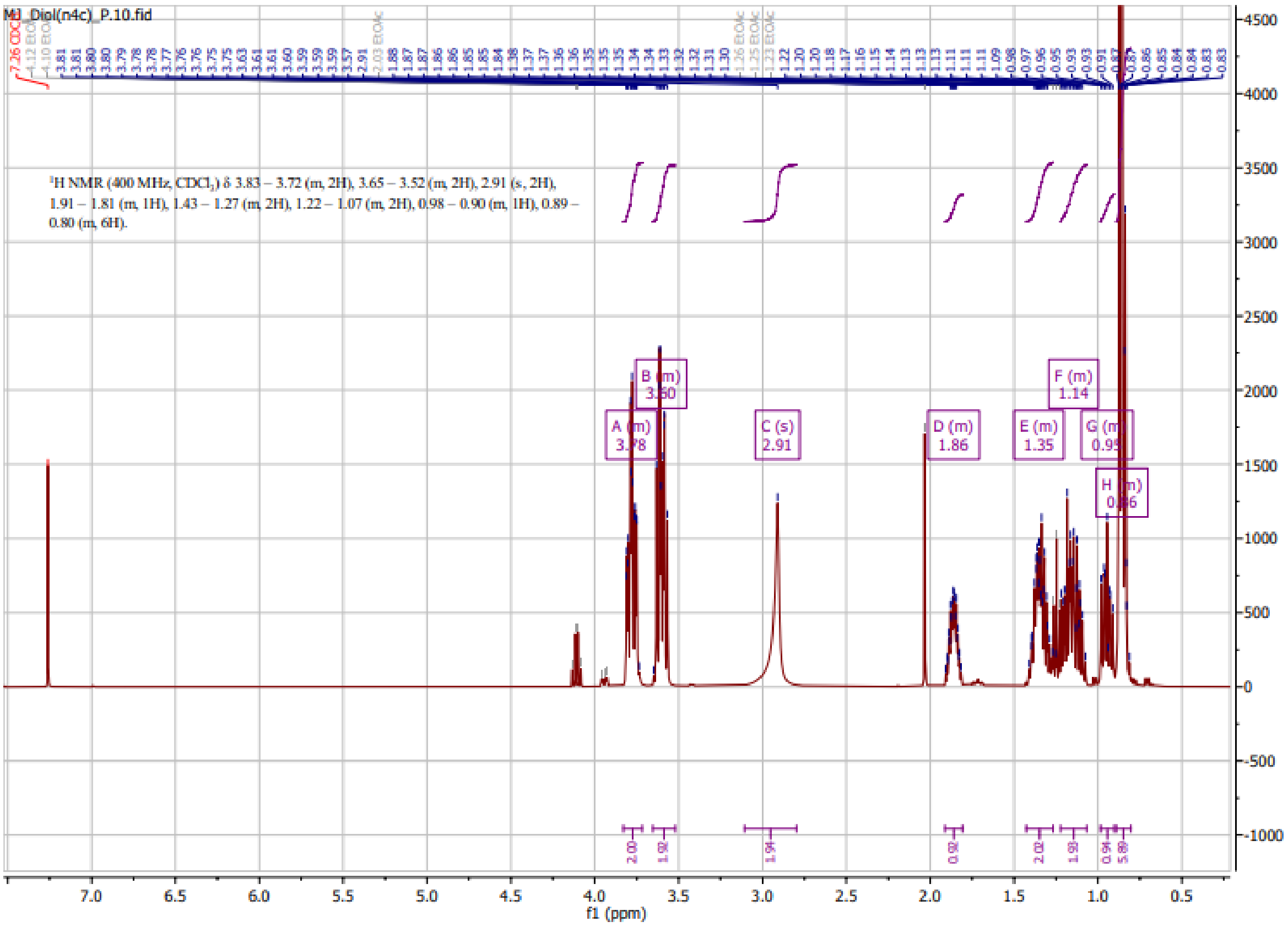

**Figure S10**: $^1$H NMR spectrum of **2(rac)** in CDCl$_3$.

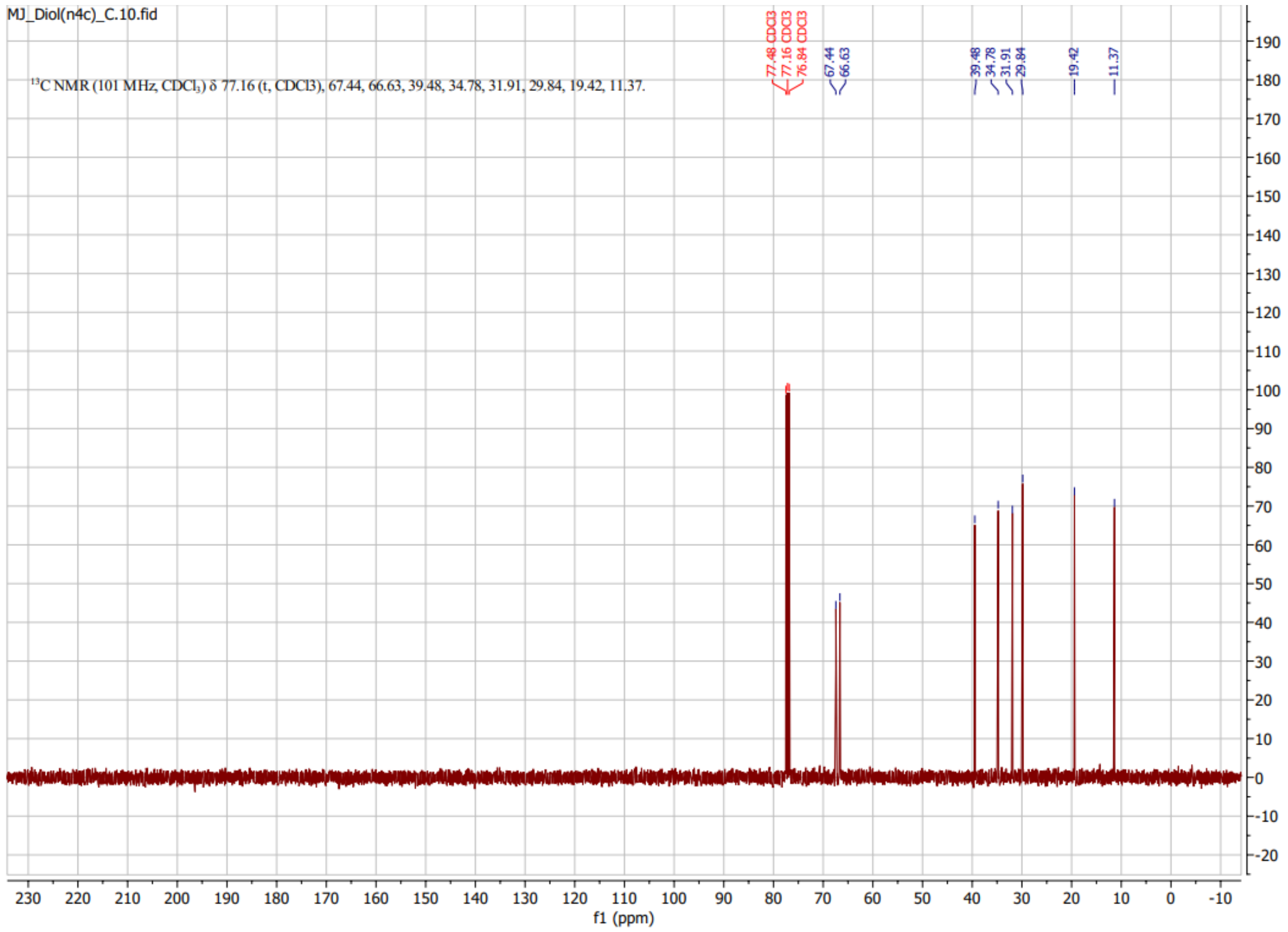

**Figure S11**: $^{13}$C NMR spectrum of **2(rac)** in $CDCl_3$.

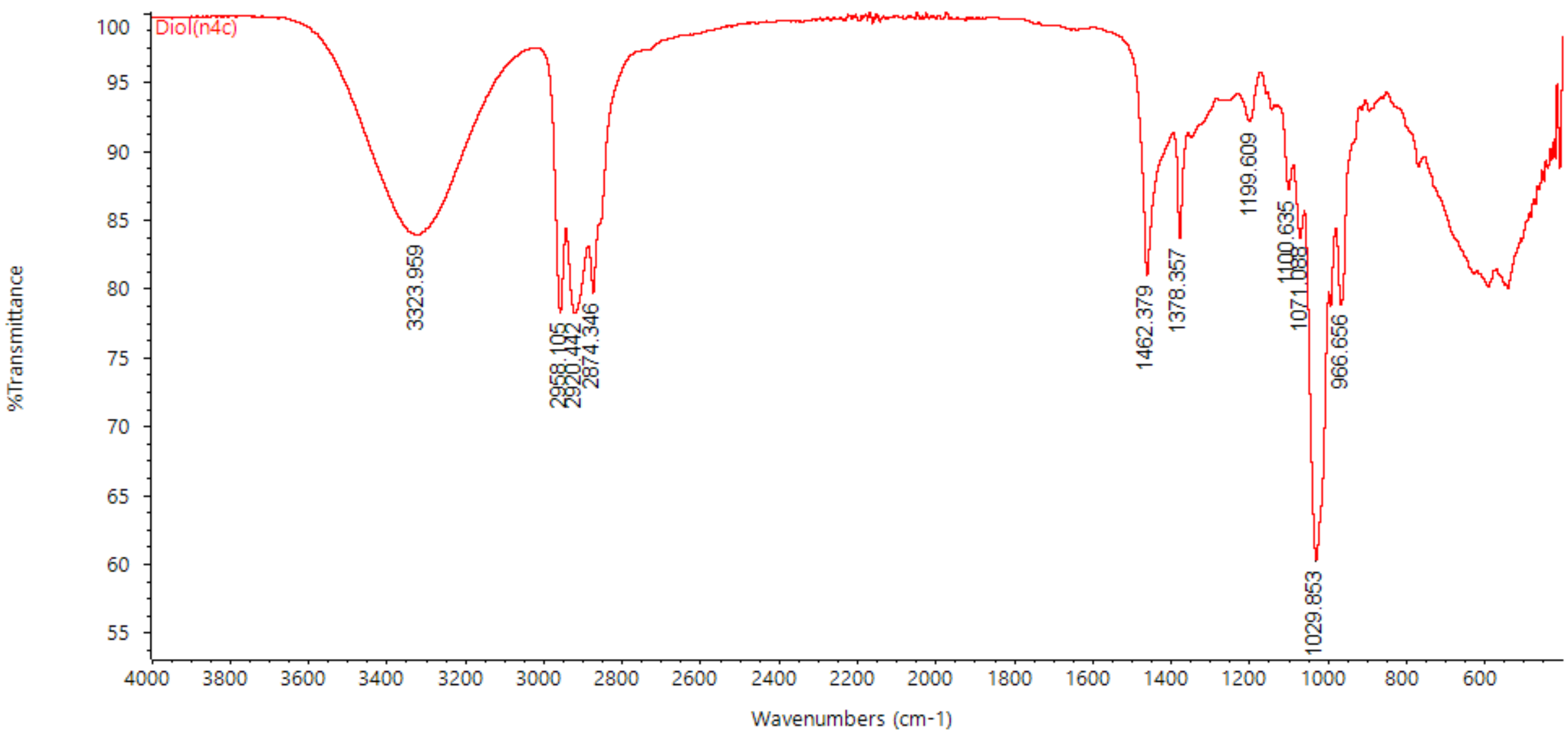

**Figure S12**: IR spectrum of **2(rac)**.

**2-(3,5-difluorophenyl)-5-[(2S)-2-methylbutyl]-1,3-dioxane (3(S)) /**

**2-(3,5-difluorophenyl)-5-[2-methylbutyl]-1,3-dioxane (3(rac))[1]**

In a RBF, equipped with magnetic stirrer, 2-alkylpropane-1,3-diol (1.1 eq) and 3,5-difluorobenzaldehyde (1 eq) are added into 1M HCl saturated aqueous $CaCl_2$ solution. The stirring is then turned on and the reaction is monitored with TLC (96:4, PE/EtOAc) until a full conversion is seen. Then, the reaction mixture is diluted with $H_2O$ (50 mL) and crude product is extracted with DCM (3 x 50 mL). Collected organic extracts are combined and washed with brine (80 mL), dried with anhydrous $MgSO_4$. $^1$H NMR ($CDCl_3$) is then performed and, in case of some unreacted aldehyde remaining, the crude is purified using Biotage® Selekt (PE/EtOAc gradient) to yield a colourless oil. (Rf (*trans*- isomer)= 0.55 (96/4 PE/EtOAc))

**3(S):**

**Yield:** 95.5 %

***$^1$H NMR** (400 MHz, $CDCl_3$): δ ppm = 7.07 – 6.97 (2H, m, Ar-H), 6.76 (1H, tt, *J* 8.9 Hz, 2.4 Hz, Ar-H), 5.36 (1H, s, Ar-CH-), 4.20 (2H, dddd, *J* 15.7 Hz, 11.2 Hz, 4.6 Hz, 2.3 Hz, -CH-$\underline{CH_2}$-O-), 3.50 (2H, td, *J* 11.1 Hz, 7.2 Hz, -CH-$\underline{CH_2}$-O-), 2.27 – 2.14 (1H, m, -CH(-H)-$\underline{CH}$-($CH_2$-O-)$_2$ ), 1.39 – 1.30 (2H, m, -CH-$\underline{CH_2}$-CH(-$CH_3$)-), 1.20 – 1.12 (1H, m, -CH(-$CH_3$)-$\underline{CH(-H)}$-$CH_2$-), 1.10 – 1.02 (1H, m, -CH(-$CH_3$)-$\underline{CH(-H)}$-$CH_2$-), 0.90 – 0.84 (7H, m, -$CH_2$-$\underline{CH_3}$, -CH-$\underline{CH(-CH_3)}$-$CH_2$-).

**$^{13}$C NMR** (101 MHz, $CDCl_3$): δ ppm = 162.98 (dd, $J_{C\text{-}F\text{-}i}$ = 248.1 Hz, $J_{C\text{-}F\text{-}m}$ = 12.4 Hz), 142.27 (t, $J_{C\text{-}F\text{-}o}$ = 9.2 Hz), 109.96 – 108.87 (m), 104.09 (t, $J_{C\text{-}F\text{-}m}$ = 25.4 Hz), 99.75 (t, $J_{C\text{-}F\text{-}l.r.c.}$ = 2.6 Hz), 72.98, 72.72, 35.08, 31.91, 31.30, 29.77, 19.37, 11.28. (*i* – ipso coupling; *o* – ortho coupling; *m* – meta coupling; l.r.c. – long range coupling)

**IR** ($\nu_{max}$/cm$^{-1}$): 3082 (sp$^2$ hybridised C-H stretching), 2962 (sp$^3$ hybridised C-H stretching), 2926 (sp$^3$ hybridised C-H stretching), 2876 (sp$^3$ hybridised C-H stretching), 1599 (-C=C- stretching, aromatic).

*Only *trans*-isomer peaks elucidated. *Cis*-/*trans*- ratio: 0.17/1

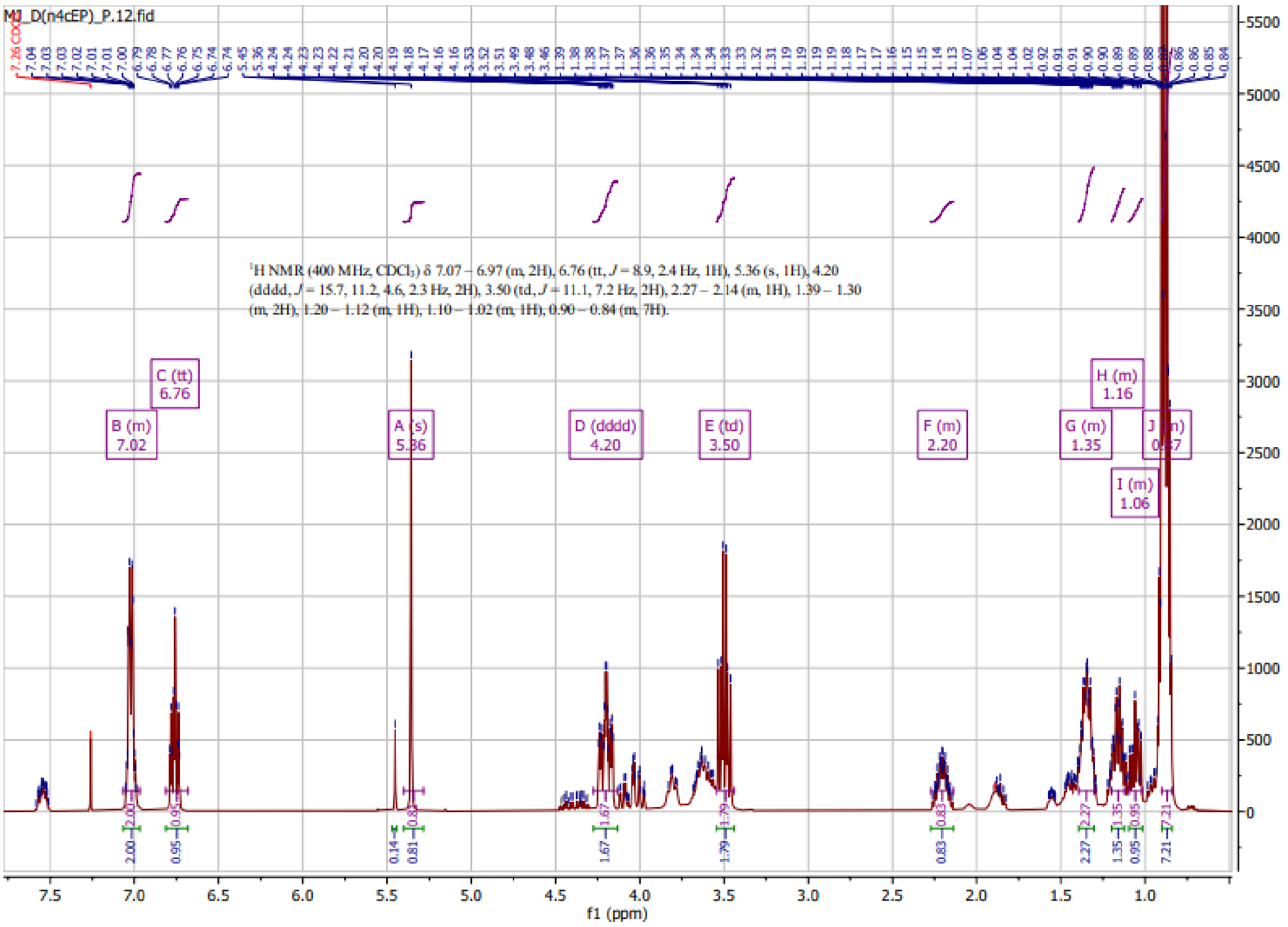

**Figure S13**: $^1$H NMR spectrum of **3(S)** in $CDCl_3$.

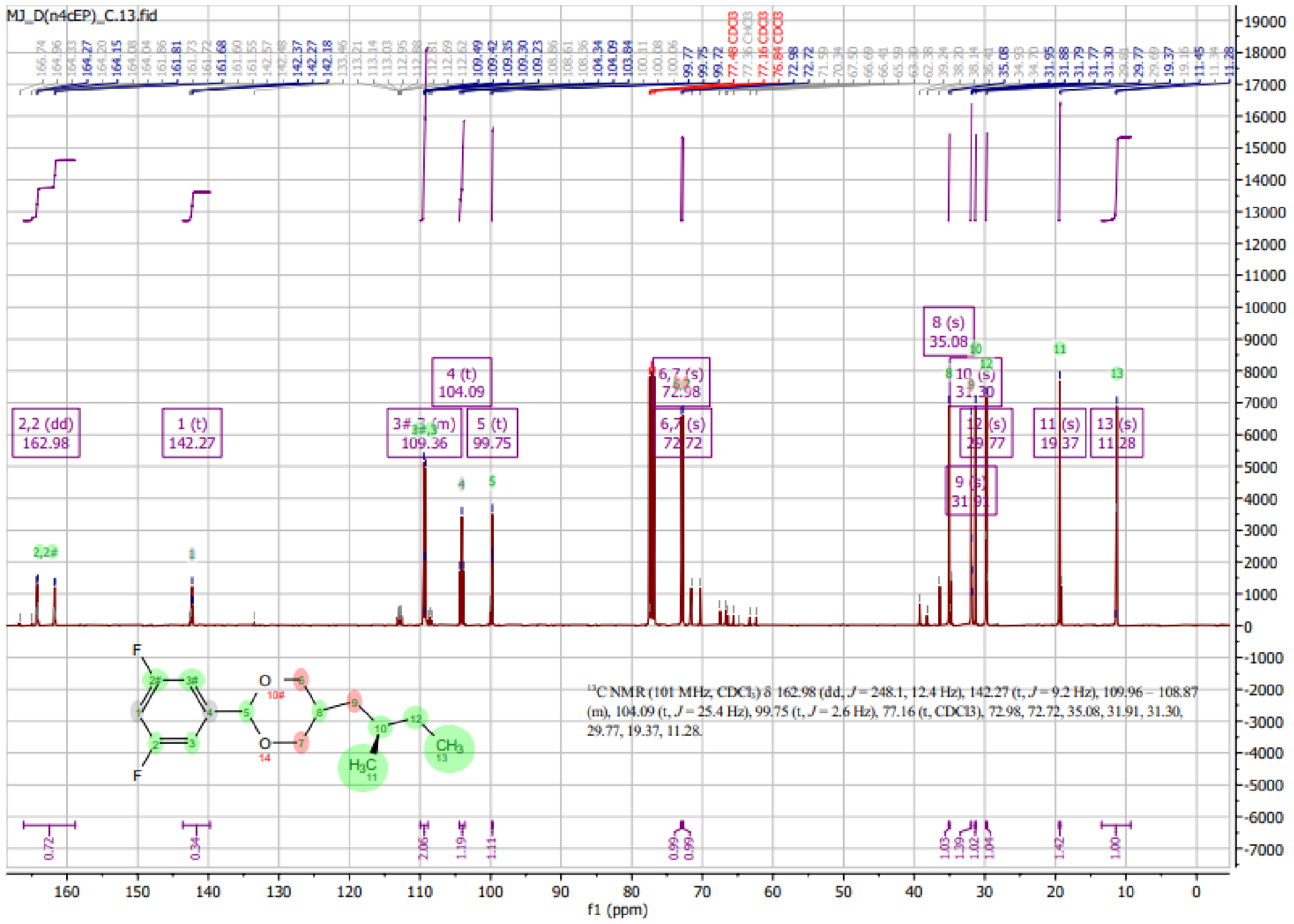

**Figure S14**: $^{13}$C NMR spectrum of **3(S)** in $CDCl_3$.

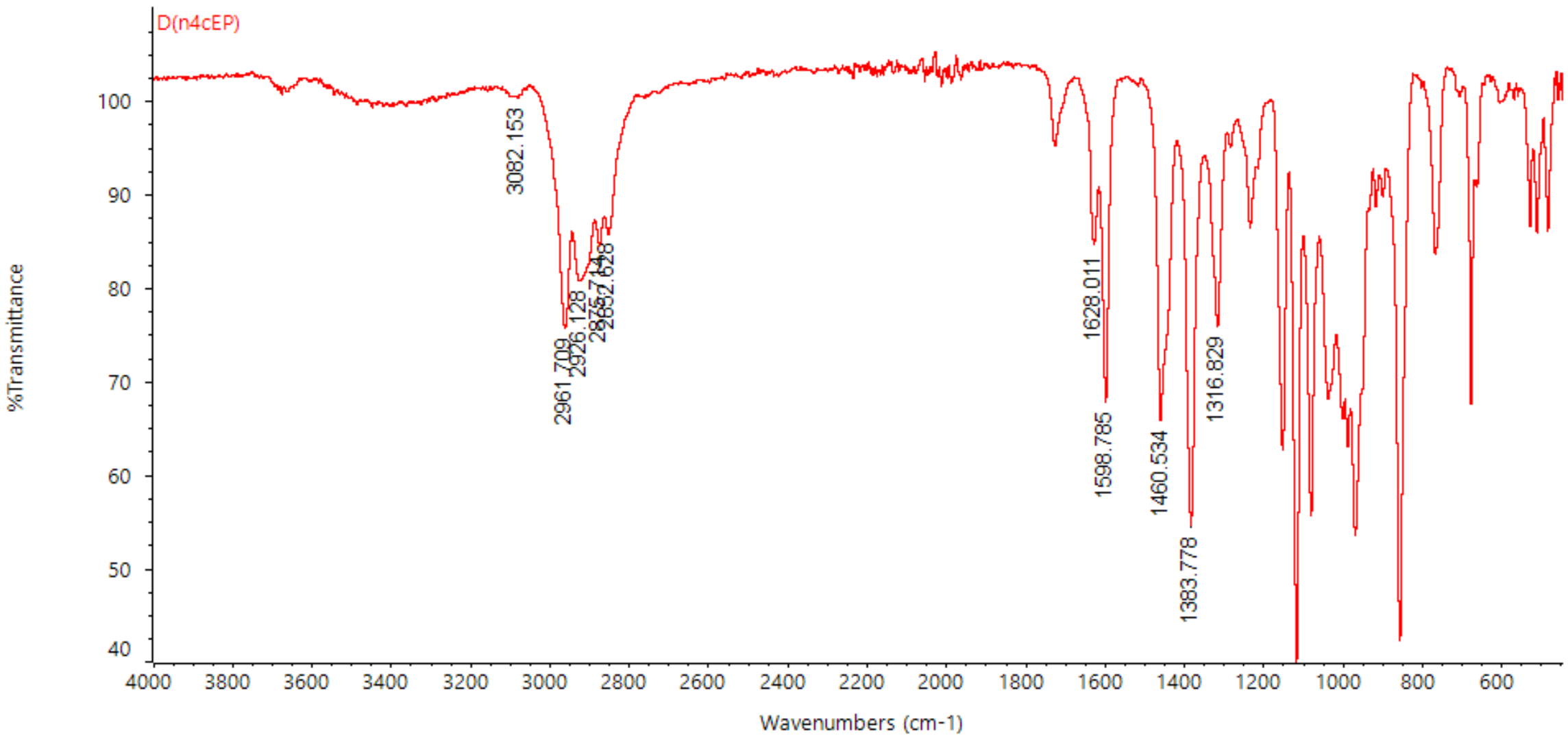

**Figure S15**: IR spectrum of **3(S)**.

**3(rac):**

**Yield:** 78.6 %

***[1]H NMR** (400 MHz, $CDCl_3$): δ ppm = 7.07 – 6.98 (2H, m, Ar-H), 6.77 (1H, tt, *J* 8.9 Hz, 2.5 Hz, Ar-H), 5.36 (1H, s, Ar-CH-), 4.21 (2H, dddd, *J* 15.8 Hz, 11.2 Hz, 4.6 Hz, 2.3 Hz, -CH-$CH_2$-O-), 3.49 (2H, td, *J* 11.1 Hz, 7.3 Hz, -CH-$CH_2$-O-), 2.28 – 2.14 (1H, m, -$CH_2$-CH-($CH_2$-O-)$_2$), 1.41 – 1.29 (2H, m, -CH-$CH_2$-CH(-$CH_3$)-), 1.23 – 1.11 (1H, m, -CH(-$CH_3$)- CH(-H)-$CH_2$-), 1.11 – 1.00 (1H, m, -CH(-$CH_3$)-CH(-H)-$CH_2$-), 0.91 – 0.86 (7H, m, -$CH_2$-$CH_3$, -CH-CH(-$CH_3$)-$CH_2$-).

**[13]C NMR** (101 MHz, $CDCl_3$): δ ppm = 162.96 (dd, $J_{C\text{-}F\text{-}i}$ = 248.1 Hz, $J_{C\text{-}F\text{-}m}$ = 12.4 Hz), 142.37 (t, $J_{C\text{-}F\text{-}o}$ = 9.2 Hz), 109.75 – 108.89 (m), 104.00 (t, $J_{C\text{-}F\text{-}m}$ = 25.3 Hz), 99.69 (t, $J_{C\text{-}F\text{-}l.r.c.}$ = 2.6 Hz), 72.93, 72.68, 35.06, 31.94, 31.29, 29.76, 19.33, 11.24. (*i* – ipso coupling; *o* – ortho coupling; *m* – meta coupling; l.r.c. – long range coupling)

**IR** ($\nu_{max}$/$cm^{-1}$): 3092 ($sp^2$ hybridised C-H stretching), 2962 ($sp^3$ hybridised C-H stretching), 2924 ($sp^3$ hybridised C-H stretching), 2849 ($sp^3$ hybridised C-H stretching), 2008 (C-H bending, aromatic overtone), 1599 (-C=C- stretching, aromatic).

*Only *trans*-isomer peaks elucidated. *Cis*-/*trans*- ratio: 0.14/1

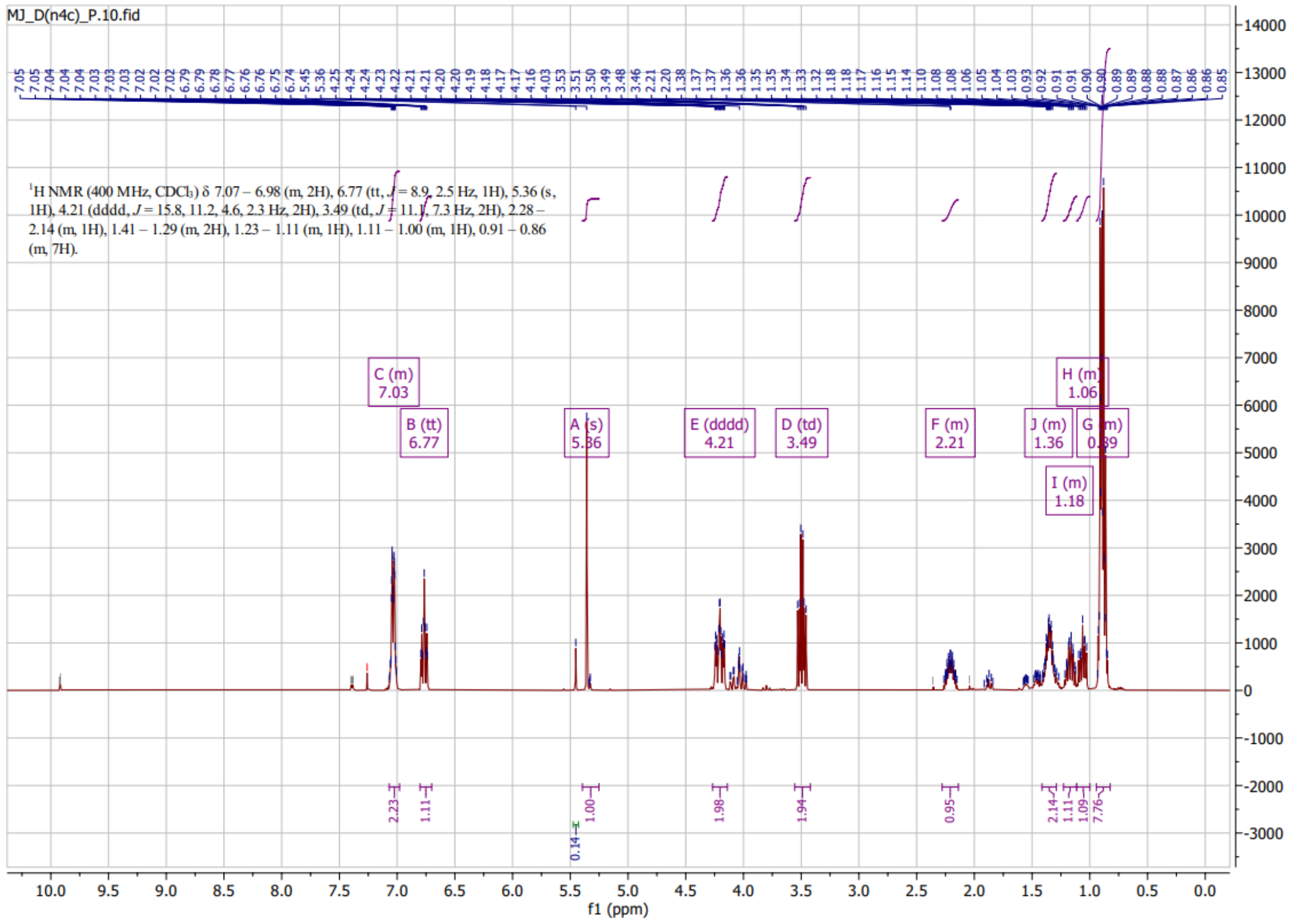

**Figure S16**: $^{1}$H NMR spectrum of **3(rac)** in $CDCl_3$.

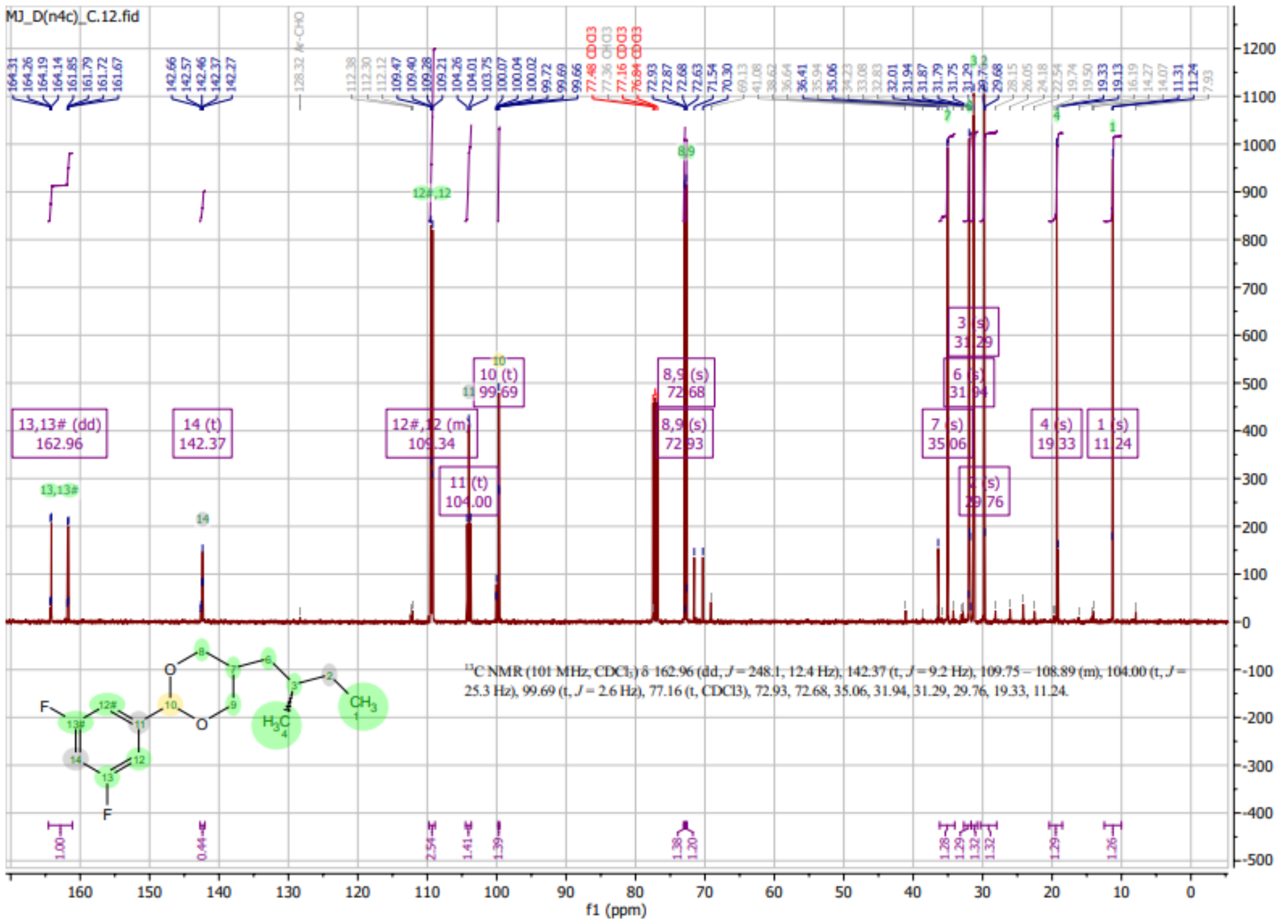

**Figure S17**: $^{13}$C NMR spectrum of **3(rac)** in $CDCl_3$.

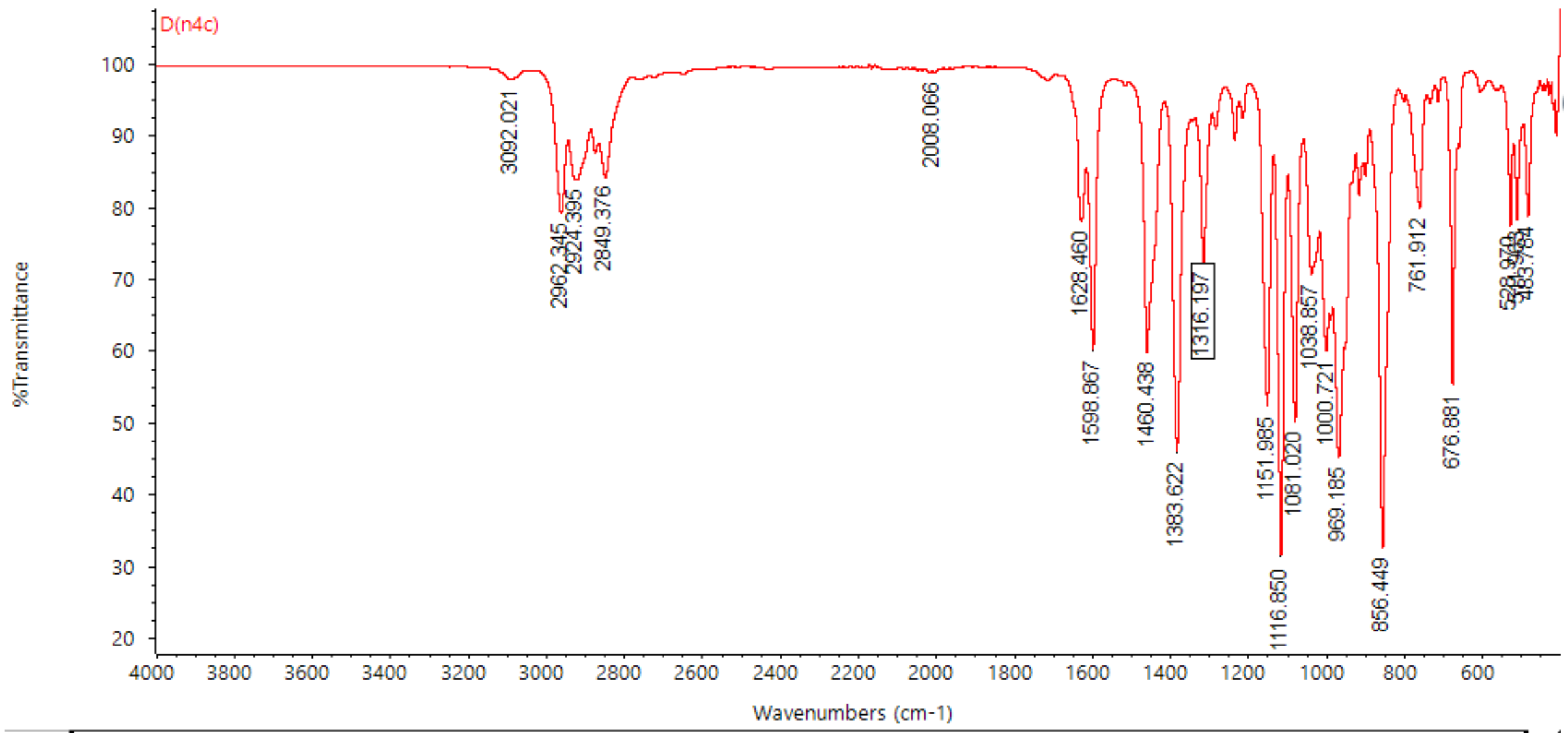

**Figure S18**: IR spectrum of **3(rac)**.

**2,6-difluoro-4-[5-[(2S)-2-methylbutyl]-1,3-dioxan-2-yl]benzoic acid (4(S)) / 2,6-difluoro-4-[5-[2-methylbutyl]-1,3-dioxan-2-yl]benzoic acid (4(rac))**[2]

Anhydrous 2-(3,5-difluorophenyl)-5-alkyl-1,3-dioxane (1 eq) was added to a RBF under an argon atmosphere. Anhydrous THF (25 mL) was added, the reaction vessel was cooled down to -78 $^{o}$C using acetone/dry ice bath. Then, *n*-BuLi (2.5 M, 1.1 eq) was added, and the reaction mixture was stirred under inert environment for at least 1 h. An excess of solid $CO_2$ was then added directly into the reaction vessel which was left to slowly heat up to room temperature. The reaction mixture was then acidified with HCl (1 M) to a pH=4, THF was removed *in vacuo*. The concentrated crude was diluted with $Et_2O$ (30 mL) and washed with $H_2O$ (2 x 20 mL). Then, diisopropylamine (1N) was added to form a water-soluble salt with the reaction product, and two additional washes with $H_2O$ were carried out (2 x 20 mL). The aqueous layers were combined and acidified until a white precipitate formation was observed. The crystals were then filtered off under vacuum, washed with $H_2O$ (15 mL) and, to ensure the highest possible purity whilst minimising product loss, recrystallised in hexane/toluene mixture (1/4) to yield large white needle-like crystals.

**4(S):**

**Yield:** 82.5 %

**m.p.** = 113.4 °C

**$^1$H NMR** (400 MHz, $CDCl_3$): δ ppm = 7.18 – 7.09 (2H, m, Ar-H), 5.37 (1H, s, Ar-CH-), 4.22 (2H, dddd, *J* 15.6 Hz, 11.4 Hz, 4.6 Hz, 2.2 Hz, -CH-<u>CH$_2$</u>-O-), 3.51 (2H, td, *J* 11.1 Hz, 7.2 Hz, -CH-<u>CH$_2$</u>-O-), 2.26 – 2.15 (1H, m, -CH$_2$-<u>CH</u>-(CH$_2$-O-)$_2$), 1.40 – 1.30 (2H, m, -CH-<u>CH2</u>-CH(-

$CH_3$)-), 1.21 – 1.13 (1H, m, -CH(-$CH_3$)-CH(-H)-$CH_3$), 1.10 – 1.03 (1H, m, -CH(-$CH_3$)-CH(-H)-$CH_3$), (7H, m, -$CH_2$-$CH_3$, -$CH_2$-CH(-$CH_3$)-CH(-H)-).

**$^{13}$C NMR** (101 MHz, $CDCl_3$): δ ppm = 166.14 (t, $J_{C\text{-}F\text{-}l.r.c.}$ = 1.8 Hz), 161.23 (dd, $J_{C\text{-}F\text{-}i}$ = 259.4 Hz, $J_{C\text{-}F\text{-}m}$ = 5.7 Hz), 145.39 (t, $J_{C\text{-}F\text{-}o}$ = 10.1 Hz), 110.44 – 109.99 (m), 109.28 (t, $J_{C\text{-}F\text{-}m}$ = 16.4 Hz), 98.85 (t, $J_{C\text{-}F\text{-}l.r.c.}$ = 2.3 Hz), 72.89, 72.64, 34.94, 31.82, 31.20, 29.67, 19.28, 11.19. (*i* – ipso coupling; *o* – ortho coupling; *m* – meta coupling; l.r.c. – long range coupling)

**IR** ($\nu_{max}$/cm$^{-1}$): 3400-2500 (OH stretch, carboxylic acid), 3080 (sp$^2$ hybridised -C-H stretch), 2967 (sp$^3$ hybridised -C-H stretch), 2899 (sp$^3$ hybridised -C-H stretch) 2001 (aromatic overtone), 1699 (C=O stretching, carboxylic acid), 1636 (C=C stretch, aromatic).

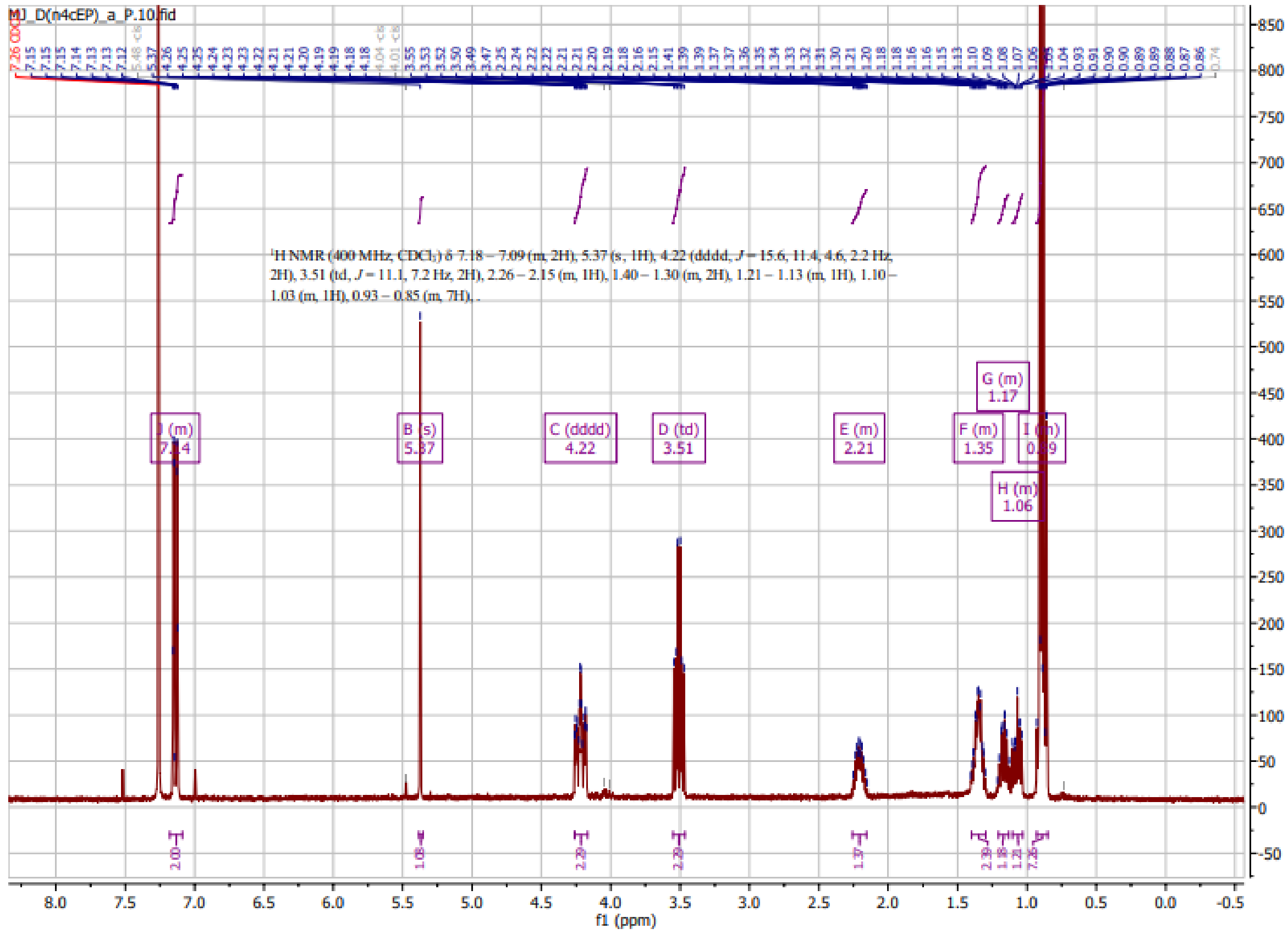

**Figure S19**: $^1$H NMR spectrum of **4(S)** in $CDCl_3$.

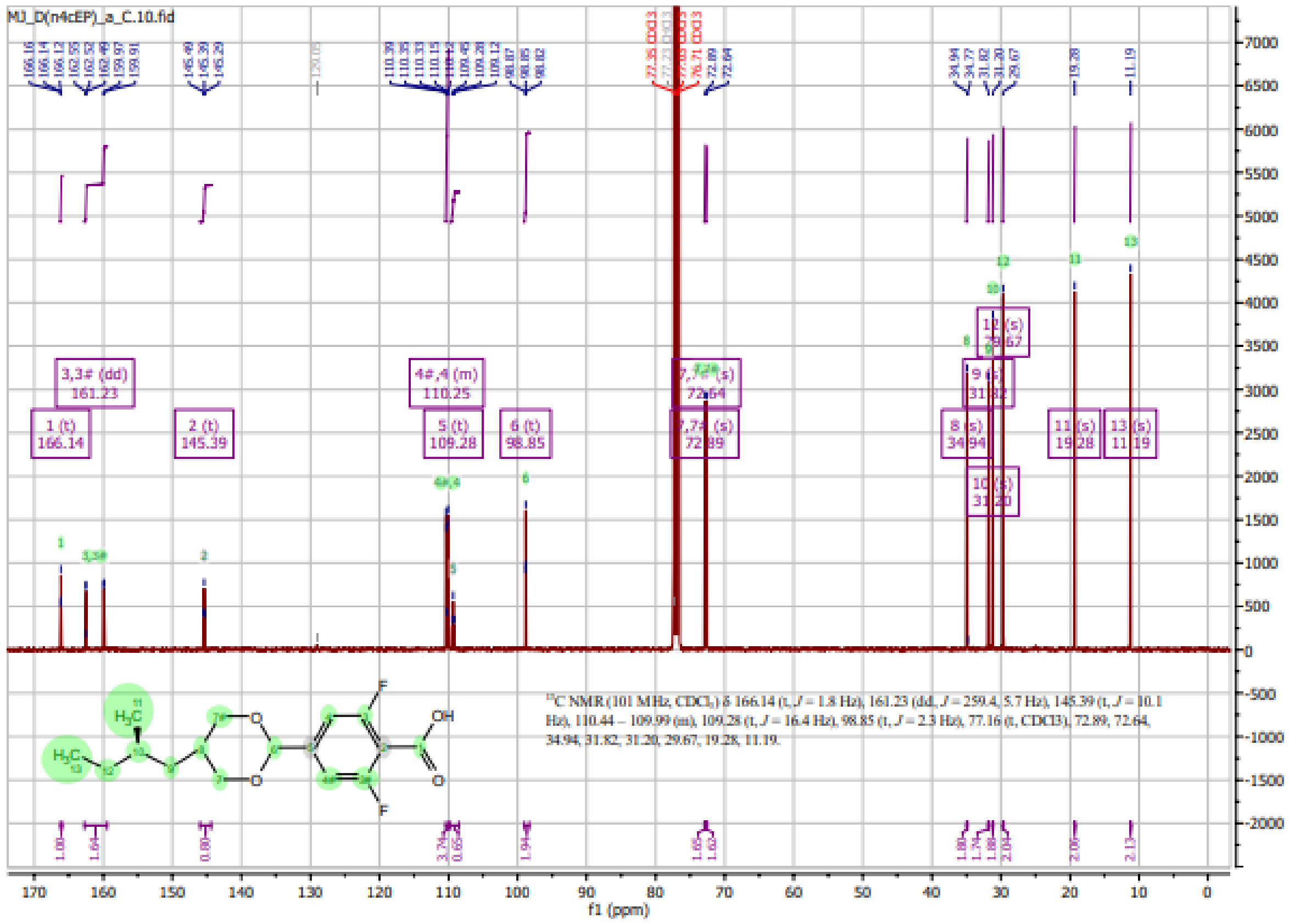

**Figure S20**: $^{13}C$ NMR spectrum of **4(S)** in $CDCl_3$.

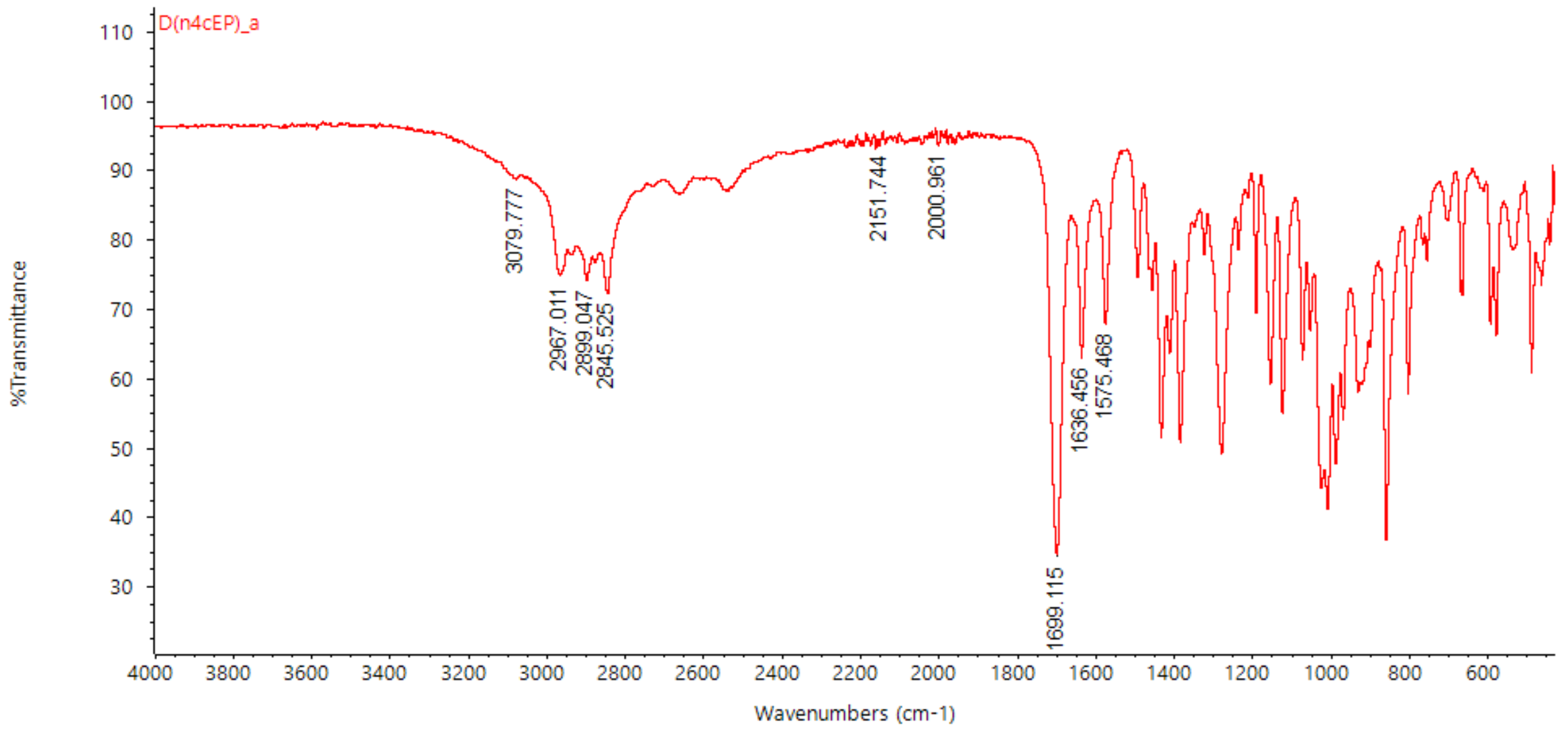

**Figure S21**: IR spectrum of **4(S)**.

**4(rac):**

**Yield:** 78.3 %

**m.p.** = 111.7 °C

**$^{1}$H NMR** (400 MHz, $CDCl_3$): δ ppm = 10.97 (1H, s, -COOH), 7.20 – 7.04 (2H, m, Ar-H), 5.38 (1H, s, Ar-CH-), 4.22 (2H, dddd, *J* 15.6 Hz, 11.2 Hz, 4.7 Hz, 2.3 Hz, -CH-$CH_2$-O-), 3.51 (2H, td, *J* 11.1 Hz, 7.1 Hz, -CH-$CH_2$-O-), 2.30 – 2.12 (1H, m, -$CH_2$-CH-($CH_2$-O-)$_2$), 1.41 – 1.28 (2H, m, -CH-CH2-CH(-$CH_3$)-), 1.23 – 1.12 (1H, m, -CH(-$CH_3$)-CH(-H)-$CH_3$), 1.10 – 0.99 (1H, m, -CH(-$CH_3$)-CH(-H)-$CH_3$), 0.96 – 0.75 (7H, m, -$CH_2$-$CH_3$, -$CH_2$-CH(-$CH_3$)-CH(-H)-).

**$^{13}$C NMR** (101 MHz, $CDCl_3$): δ ppm = 166.63, 161.36 (dd, $J_{C\text{-}F\text{-}i}$ = 259.4 Hz, $J_{C\text{-}F\text{-}m}$ = 5.7 Hz), 145.53 (t, $J_{C\text{-}F\text{-}o}$ = 10.0 Hz), 110.77 – 110.02 (m), 109.43 (t, $J_{C\text{-}F\text{-}m}$ = 16.3 Hz), 98.98 (t, $J_{C\text{-}F\text{-}l.r.c.}$ = 2.3 Hz), 73.01, 72.77, 35.06, 31.95, 31.33, 29.79, 19.40, 11.31.

(*i* – ipso coupling; *o* – ortho coupling; *m* – meta coupling; l.r.c. – long range coupling)

**IR** ($\nu_{max}$/cm$^{-1}$): 3400-2500 (OH stretch, carboxylic acid), 3087 (sp$^2$ hybridised -C-H stretch), 2960 (sp$^3$ hybridised -C-H stretch), 2926 (sp$^3$ hybridised -C-H stretch) 1988 (aromatic overtone), 1702 (C=O stretching, carboxylic acid), 1637 (C=C stretch, aromatic).

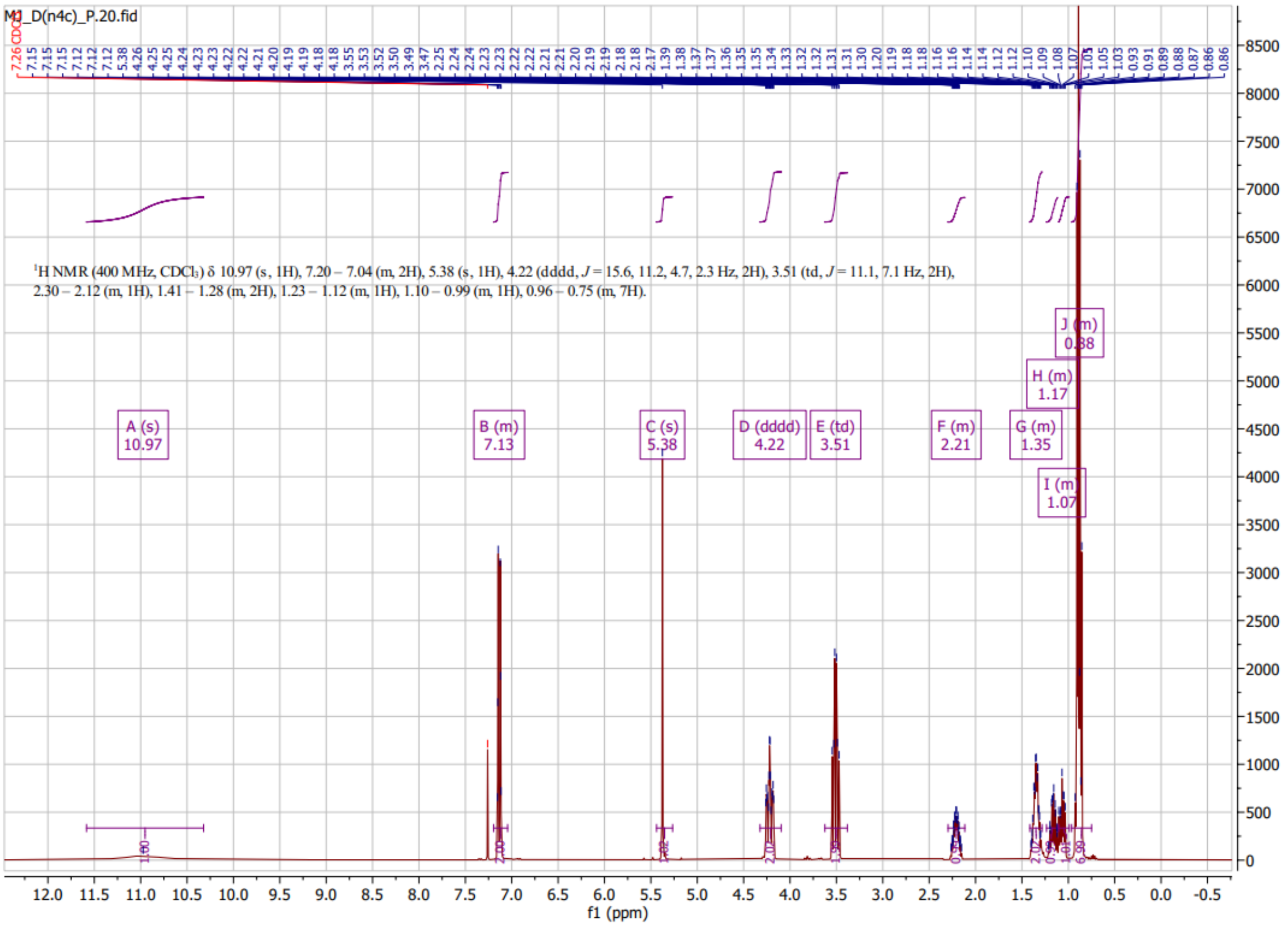

**Figure S22**: $^{1}$H NMR spectrum of **4(S)** in $CDCl_3$.

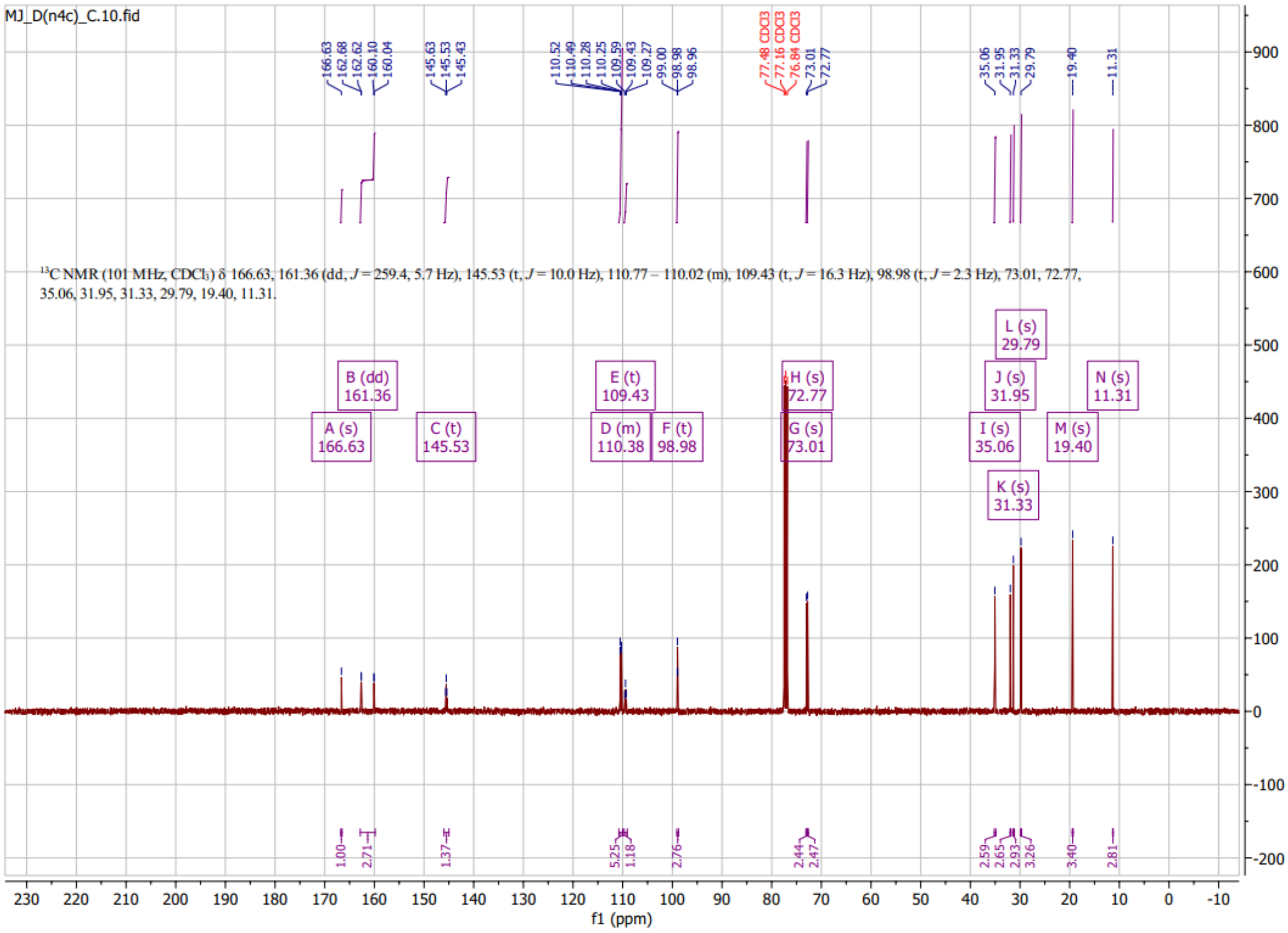

**Figure S23**: $^{13}C$ NMR spectrum of **4(S)** in $CDCl_3$.

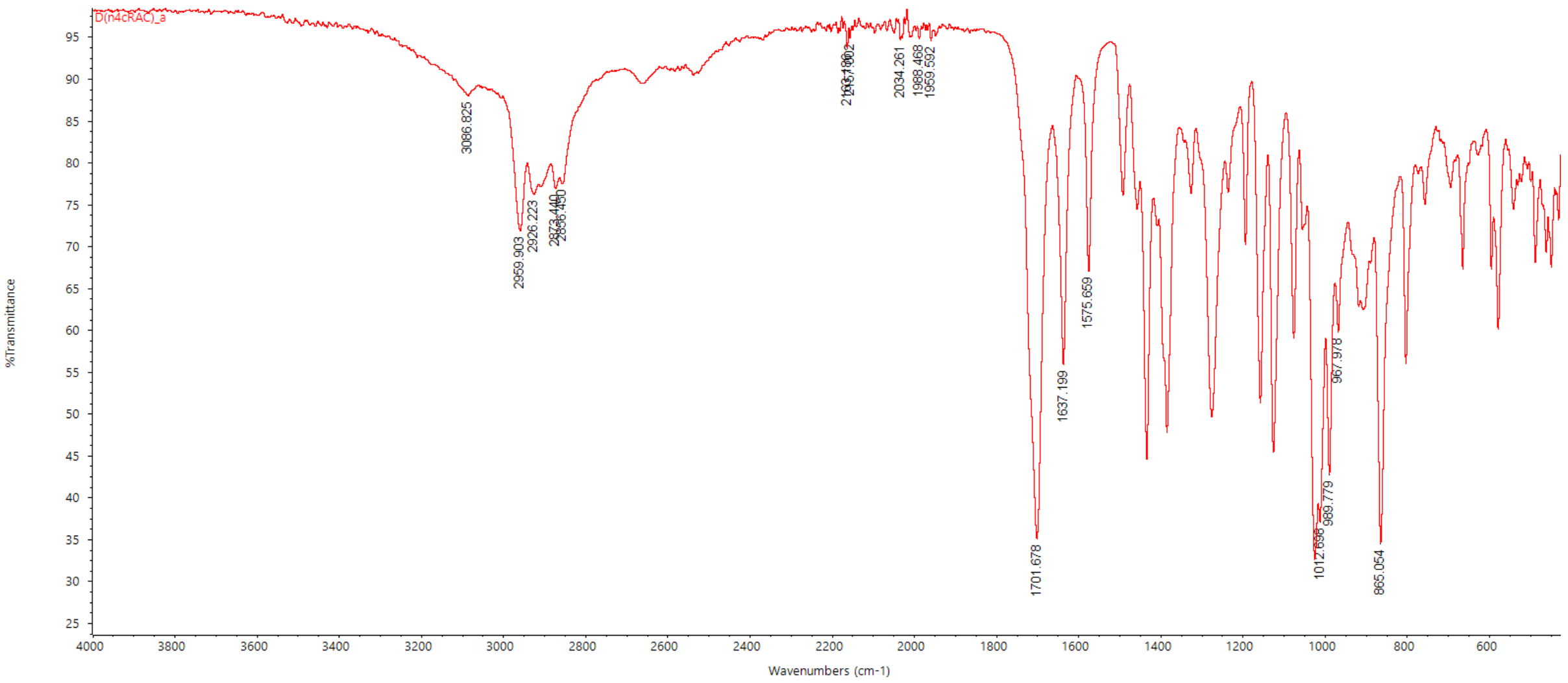

**Figure S24**: IR spectrum of **4(S)**.

**2-(4-(difluoro(3,4,5-trifluorophenoxy)methyl)-3,5-difluorophenyl)-4,4,5,5-tetramethyl-1,3,2-dioxaborolane (5)[3]**

5-Bromo-2-(difluoro(3,4,5-trifluorophenoxy)methyl)-1,3-difluorobenzene (4.95 g, 12.7 mmol, 1.00 eq.), bis(pinacolato)diboron (3.23 g, 12.7 mmol, 1.00 eq.) and dry potassium acetate (3.75 g, 38.2 mmol, 3.00 eq.) were suspended in a 1:1 mixture of dry toluene and dry 1,4-dioxane (200 mL) and sparged with argon for 1 hour. After adding the Pd-catalyst (1,1'-Bis(diphenylphosphino)ferrocene]palladium(II) dichloride, 278 mg, 381 µmol, 0.03 eq.) the mixture was stirred at 100°C for 20 h. The reaction mixture was allowed to cool down to rt and 1 M HCl was added to pH 4. The phases were separated, and the aqueous phase was extracted with toluene (3 x 50 mL). The combined organic extracts were washed with water (3 x 50 mL), while brine was added to enhance the separation. The combined organic extracts were dried over $MgSO_4$, and the solvent was removed under reduced pressure. The crude product was purified by column chromatography (DCM : *n*-hexane 1:1, Rf = 0.67) and recrystallisation in ethanol.

**Yield**: 3.81 g (75 %), white crystalline powder.

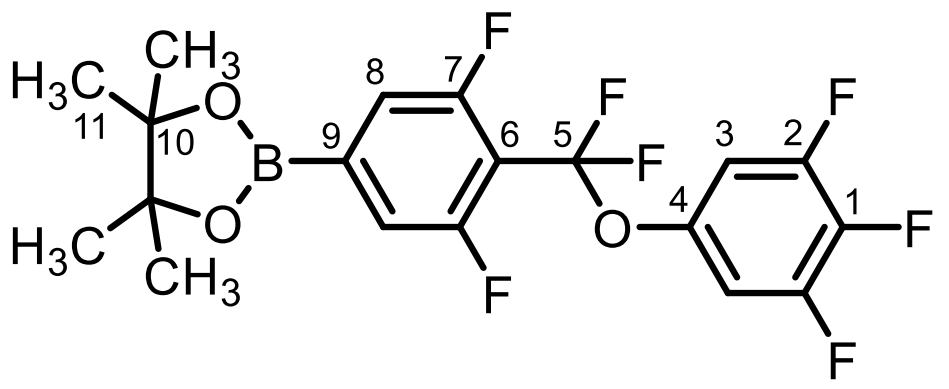

**$^1$H NMR** (400 MHz, $CDCl_3$): $\delta$ = 7.38 (d, $^3J_{HF}$ = 9.60 Hz, 2H, *H*-8), 6.96 (dd, $^3J_{HF}$ = 7.79 Hz, $^4J_{HF}$ = 6.00 Hz, 2H, *H*-3), 1.35 (s, 12H, *H*-11) ppm. **$^{19}$F NMR** (376 MHz, $CDCl_3$): $\delta$ = −61.8 (t, $^4J_{FF}$ = 27.1 Hz, 2F, *F*-5), −111.7 (td, $^4J_{FF}$ = 27.1 Hz, $^3J_{HF}$ = 9.84 Hz, 2F, *F*-7), −132.5 (dd, $^3J_{FF}$ = 20.9 Hz, $^3J_{HF}$ = 7.91 Hz, 2F, *F*-2), −163.5 (tt, $^3J_{FF}$ = 20.7 Hz, $^4J_{HF}$ = 5.62 Hz, 1F, *F*-1) ppm. **$^{13}$C NMR** (100 MHz, $CDCl_3$): $\delta$ = 159.6 (dt, $^1J_{CF}$ = 259 Hz, $^4J_{CF}$ = 2.20 Hz, *C*-7), 151.0 (app. dq, $^1J_{CF}$ = 251 Hz, $^2J_{CF}$ = 5.13 Hz, *C*-2), 144.7 (tt, $^2J_{CF}$ = 17.6 Hz, $^2J_{CF}$ = 2.20 Hz, *C*-6), 138.5 (dt, $^1J_{CF}$ = 250 Hz, $^2J_{CF}$ = 15.2 Hz, *C*-1), 135.8 (br, *C*-9), 120.3 (t, $^1J_{CF}$ = 266 Hz, *C*-5), 118.1 (m, *C*-3 or *C*-8), 111.7 (m, *C*-4), 107.5 (m, *C*-3 or *C*-8). 84.9 (s, *C*-10), 24.9 (s, *C*-11) ppm.

**IR** ($\nu_{max}$/cm$^{-1}$): 3117 (=C−H), 2981 (−C−H), 1632 (C=C), 1043 (C−F) cm$^{-1}$.

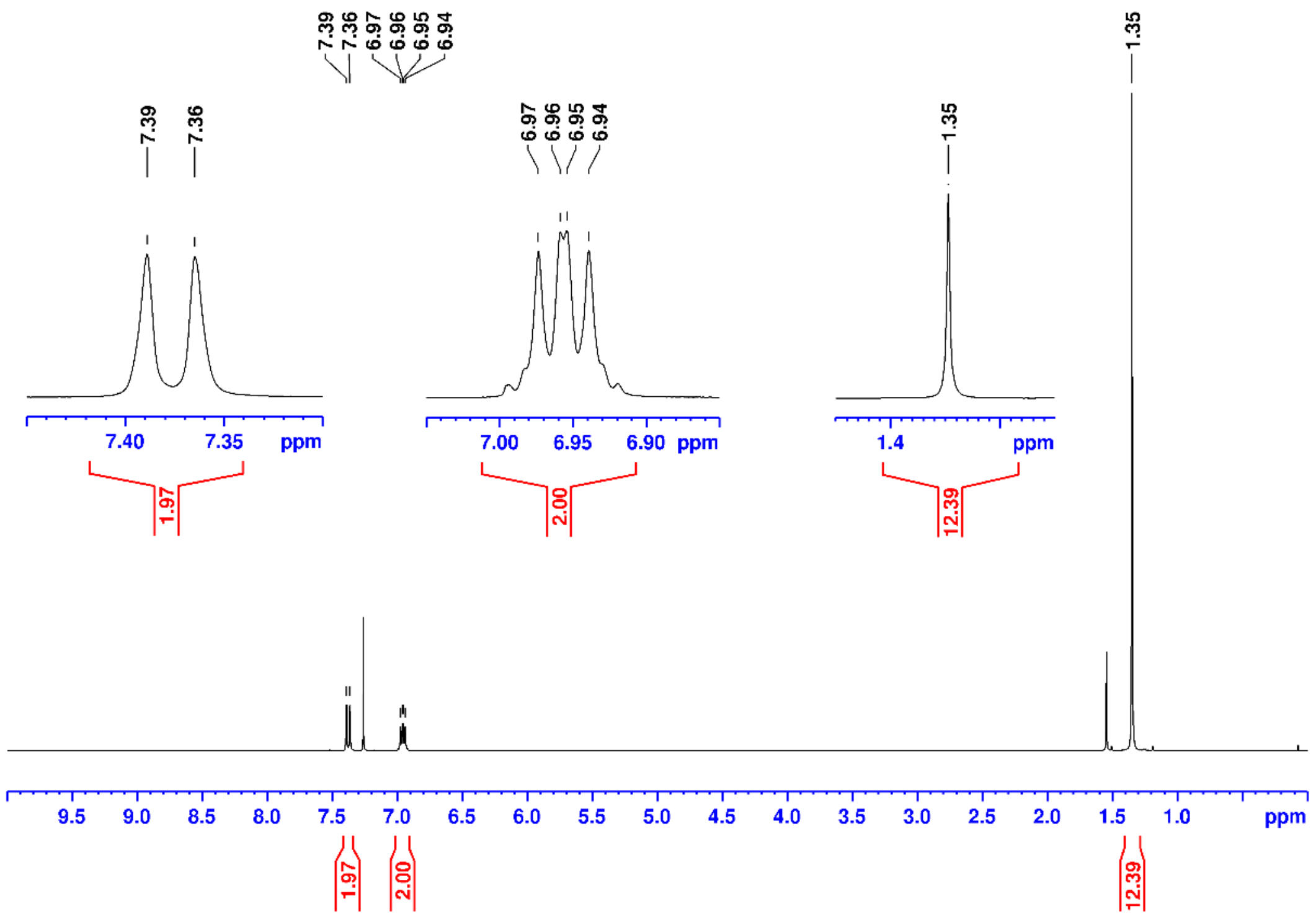

**Figure S25**: $^{1}$H NMR spectrum of **5** in $CDCl_3$.

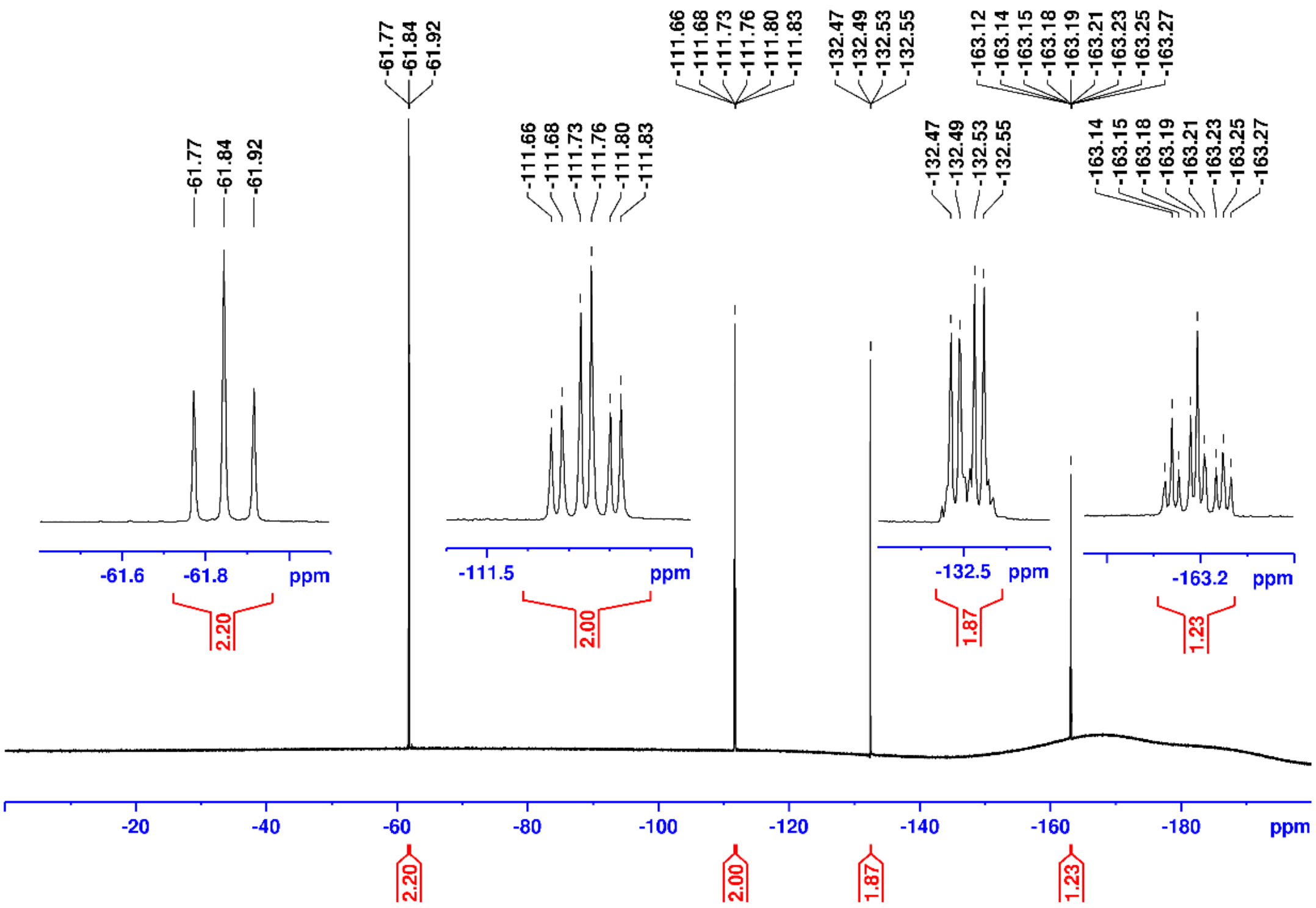

**Figure S26:** $^{19}$F NMR spectrum of **5** in $CDCl_3$.

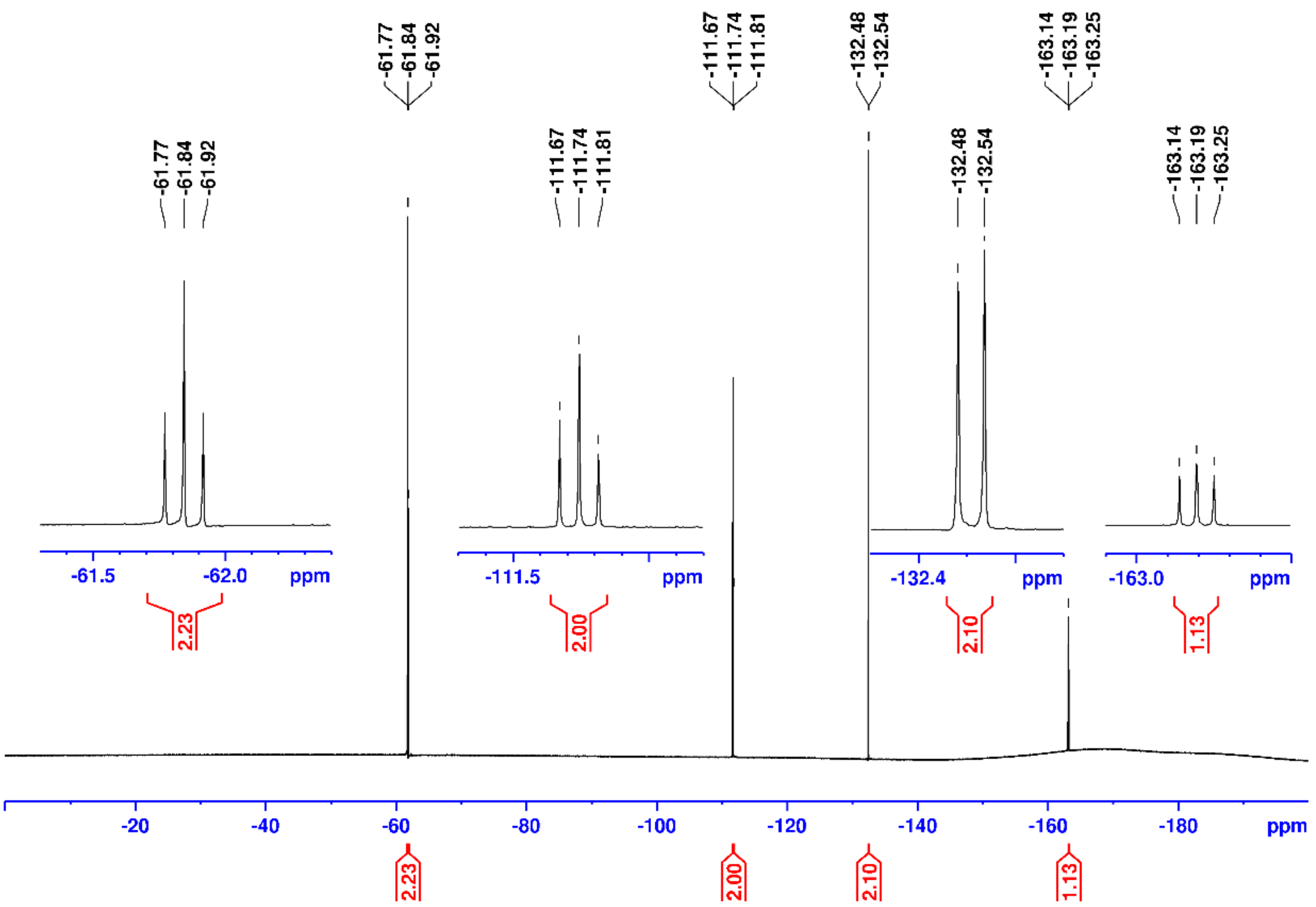

**Figure S27**: $^{19}F\{^1H\}$ NMR spectrum of **5** in $CDCl_3$.

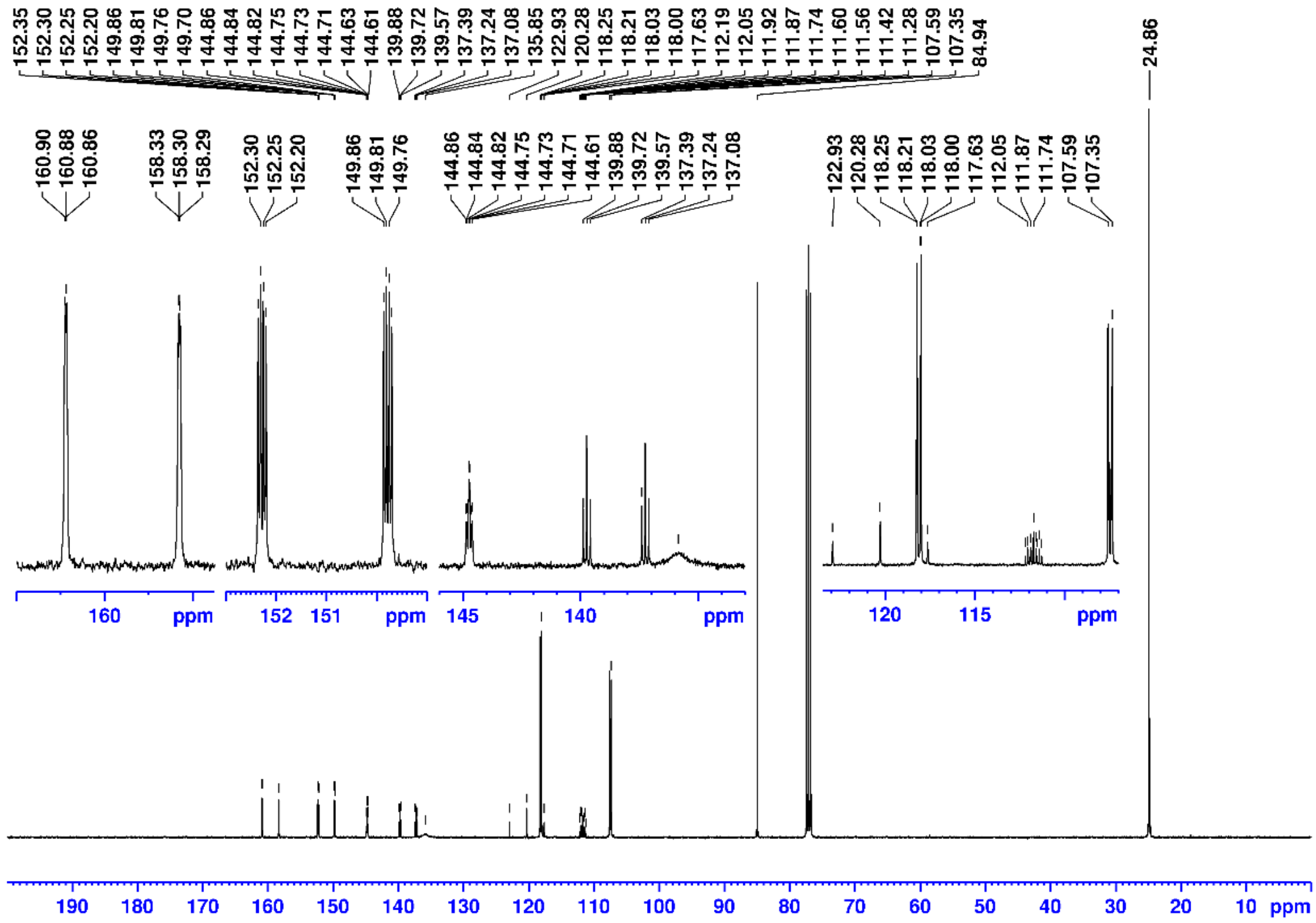

**Figure S28**: $^{13}C$ NMR spectra of **5** in $CDCl_3$.

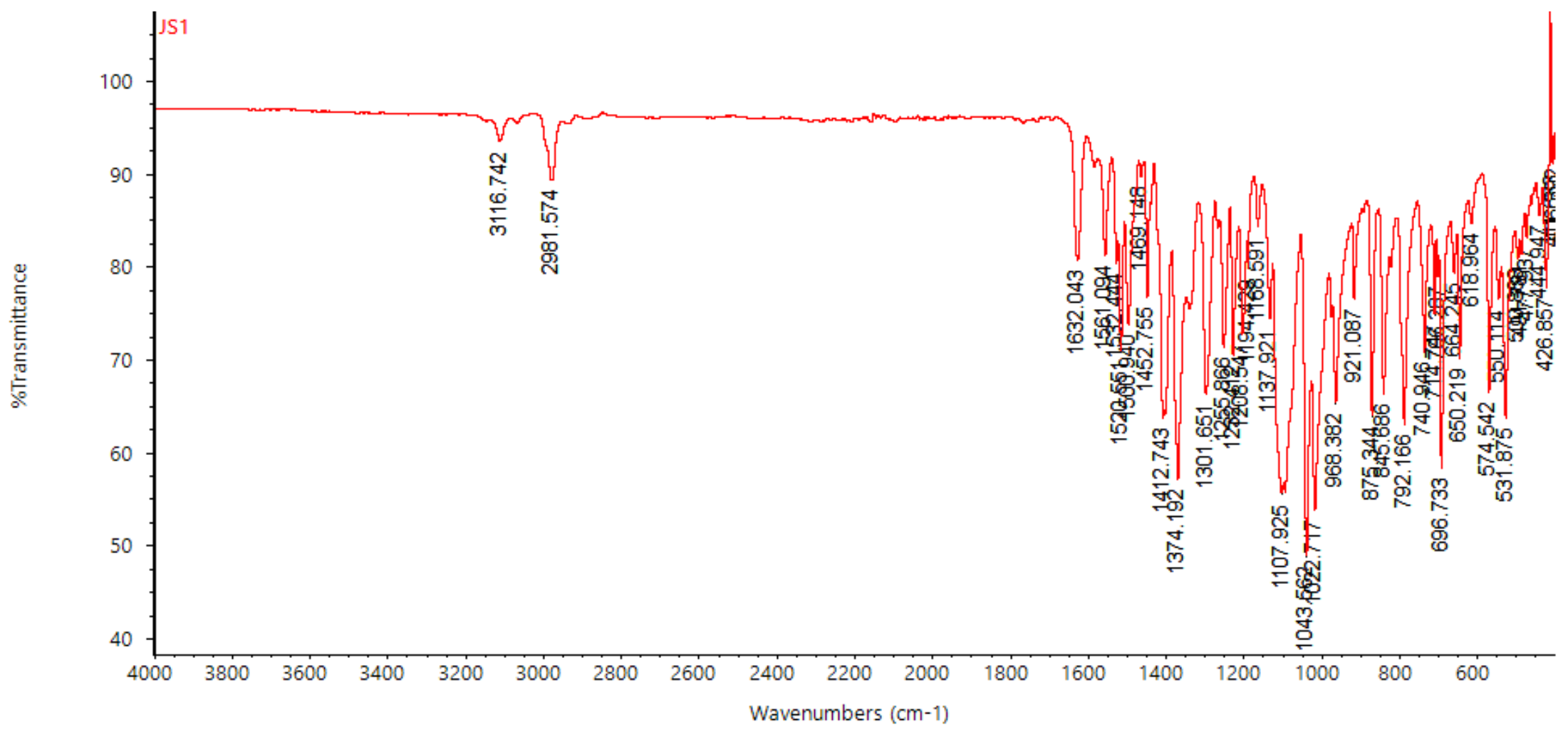

**Figure S29**: IR spectrum of **5**.

**4'-(difluoro(3,4,5-trifluorophenoxy)methyl)-2,3',5'-trifluoro-[1,1'-biphenyl]-4-ol (6)**[3]

Anhydrous potassium phosphate (3.41 g, 16.1 mmol, 3.50 eq.) was dissolved in 2.78 mL of water to form a solution of potassium phosphate trihydrate. **5** (2.00 g, 4.59 mmol, 1.00 eq.) and 4-bromo-3-fluorophenol (946 mg, 4.95 mmol, 1.08 eq.) were dissolved in 40 mL of THF and the potassium phosphate trihydrate solution was added. The mixture was sparged with argon for 30 min and then refluxed for 1 h. SPhos-ligand (188 mg, 459 µmol, 0.10 eq.) and palladium acetate (61.8 mg, 275 µmol, 0.06 eq.) were added and it was refluxed for 4 hours. After cooling down to rt, the mixture was acidified with 1 M HCl to pH = 4 and extracted with DCM (3 x 50 mL). The combined organic extracts were dried over $MgSO_4$ and the solvent was removed under reduced pressure. The crude product in form of an orange oil was further purified by column chromatography (DCM : *n*-hexane : ethyl acetate 2:7:1, Rf = 0.30) and recrystallisation in hexane.

**Yield**: 347 mg (18%), off-white crystalline powder.

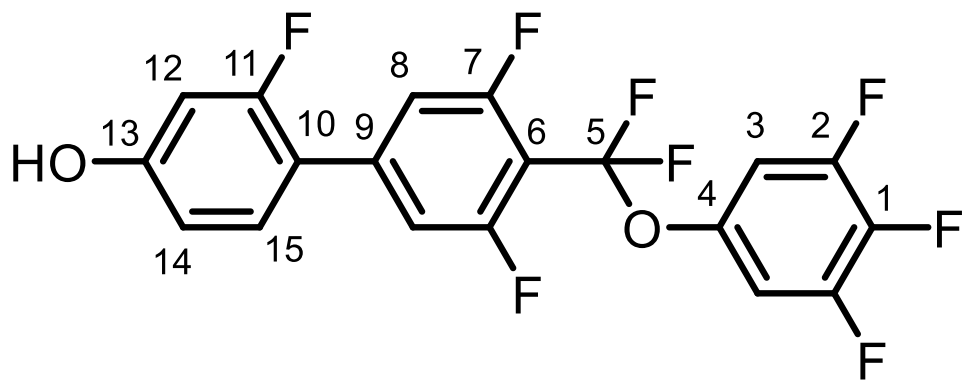

**$^1$H NMR** (400 MHz, $CDCl_3$): $\delta$ = 7.31 (app. t, $^3J_{HH}$ = 8.64 Hz, 1H, *H*-Ar), 7.17 (d, $^3J_{HF}$ = 11.2 Hz, 2H, *H*-8), 6.98 (dd, $^3J_{HF}$ = 7.88 Hz, $^4J_{HF}$ = 5.94 Hz, 2H, *H*-3), 6.76 – 6.68 (m, 2H, *H*-Ar) ppm. **$^{19}$F NMR** (376 MHz, 298 K, $CDCl_3$): $\delta$ = −61.7 (t, $^4J_{FF}$ = 26.2 Hz, 2F, *F*-5), −111.7 (td, $^4J_{FF}$ = 26.3 Hz, $^3J_{HF}$ = 10.5 Hz, 2F, *F*-7), −114.4 (m, 1F, *F*-11), −132.5 (dd, $^3J_{FF}$ = 20.6 Hz, $^3J_{HF}$ = 8.16 Hz, 2F, *F*-2), −163.5 (tt, $^3J_{FF}$ = 21.1 Hz, $^4J_{HF}$ = 5.99 Hz, 1F, *F*-1) ppm. **$^{13}$C NMR** (100 MHz, $CDCl_3$):

$\delta$ = 160.4 (d, ${}^{1}J_{CF}$ = 251 Hz, *C*-11), 159.9 (dm, ${}^{1}J_{CF}$ = 257 Hz, *C*-7), 157.8 (d, ${}^{2}J_{CF}$ = 11.7 Hz), 151.0 (app. dq, ${}^{1}J_{CF}$ = 251 Hz, ${}^{2}J_{CF}$ = 5.26 Hz, *C*-2), 144.7 (tt, ${}^{2}J_{CF}$ = 11.7 Hz, ${}^{2}J_{CF}$ = 2.20 Hz, *C*-6), 141.5 (t, ${}^{2}J_{CF}$ = 11.2 Hz), 138.5 (dt, ${}^{1}J_{CF}$ = 250 Hz, ${}^{2}J_{CF}$ = 15.2 Hz, *C*-1), 130.9 (d, ${}^{2}J_{CF}$ = 4.40 Hz), 120.3 (t, ${}^{1}J_{CF}$ = 266 Hz, *C*-5), 118.2 (dt, ${}^{2}J_{CF}$ = 12.5 Hz, ${}^{3}J_{CF}$ = 2.20 Hz), 112.6 (dt, ${}^{2}J_{CF}$ = 24.0 Hz, ${}^{3}J_{CF}$ = 23.35 Hz), 112.2 (d, ${}^{3}J_{CF}$ = 2.94 Hz), 108.1 (m, *C*-4), 107.5 (m), 104.2 (d, ${}^{2}J_{CF}$ = 25.8 Hz) ppm.

**IR** ($\nu_{max}$/cm$^{-1}$): 3432 (−O−H), 3098 (=C−H), 1621 (C=C), 1019 (C−F) cm$^{-1}$.

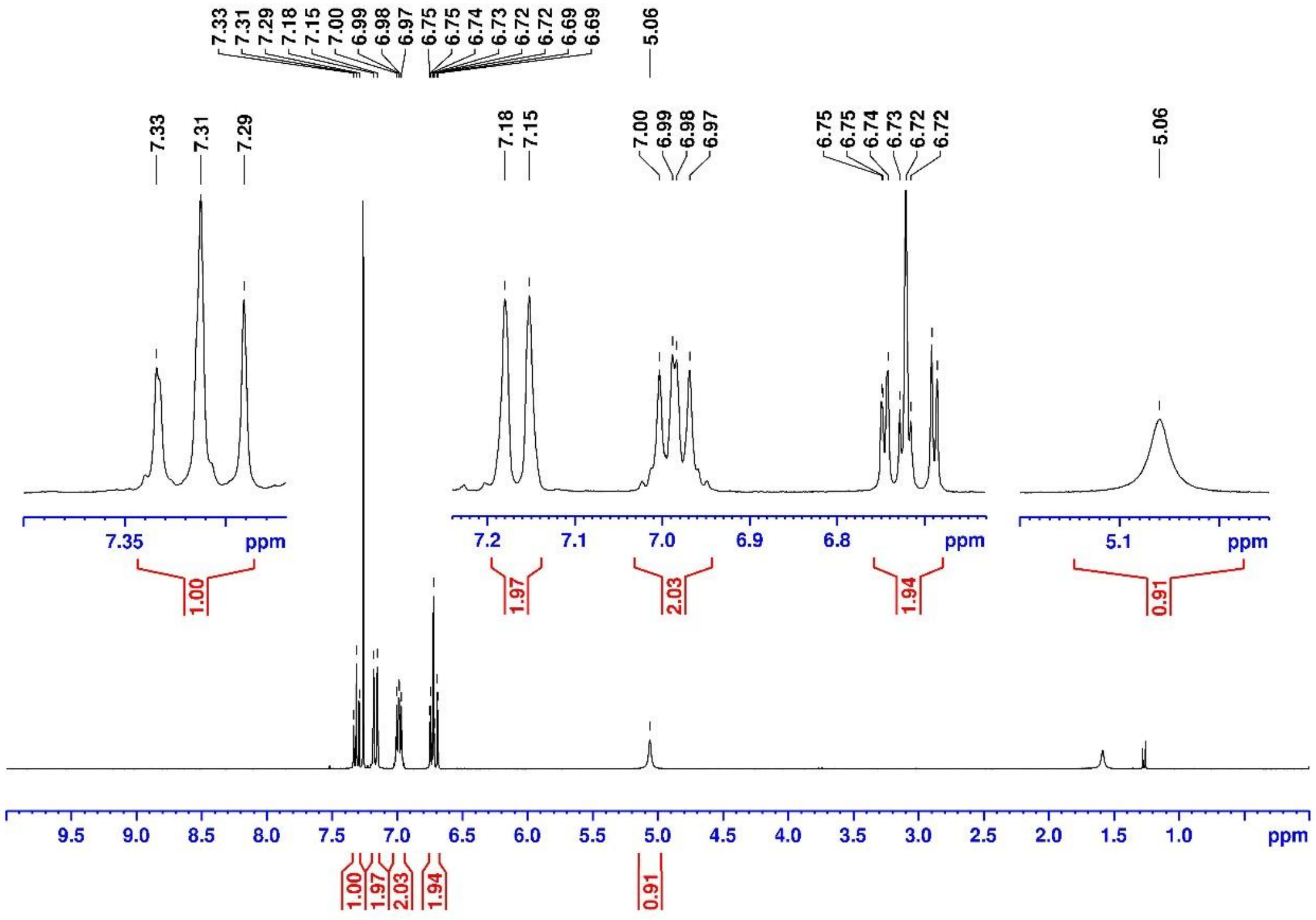

**Figure S30**: $^{1}$H NMR spectrum of **6** in $CDCl_3$.

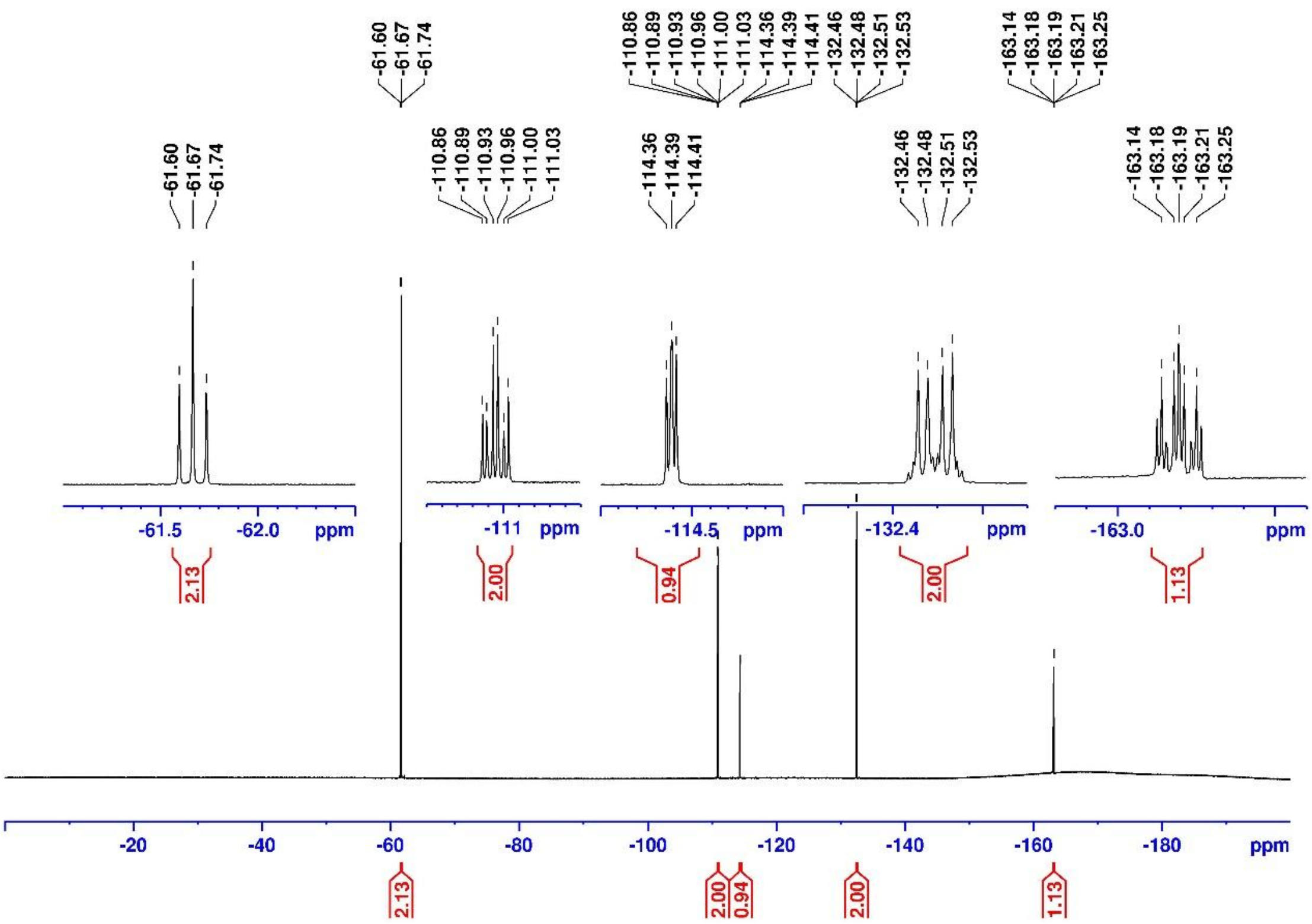

**Figure S31: $^{19}$F** *NMR spectrum of 6 in $CDCl_3$.*

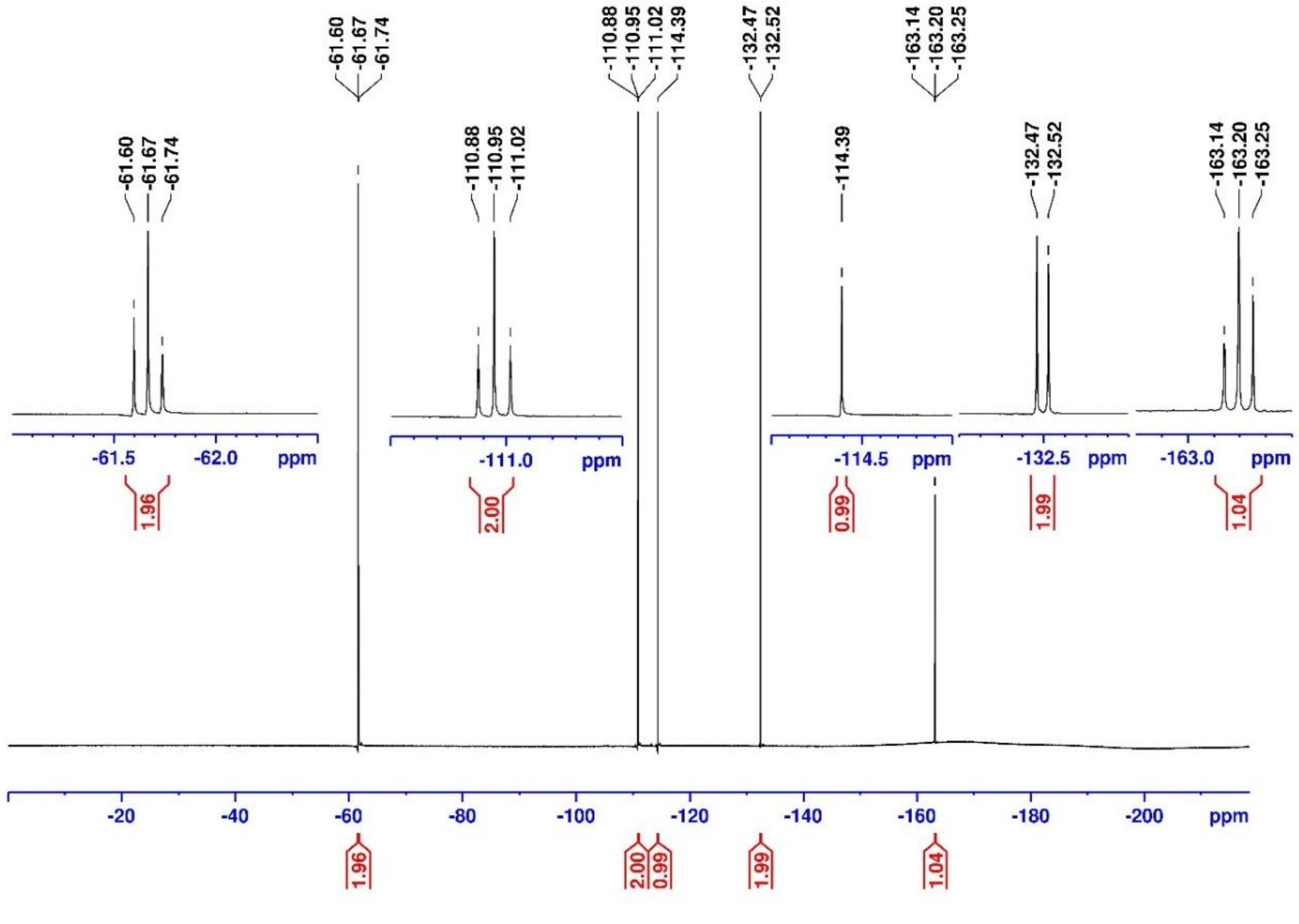

**Figure S32**: $^{19}F\{^1H\}$ NMR spectrum of **6** in $CDCl_3$.

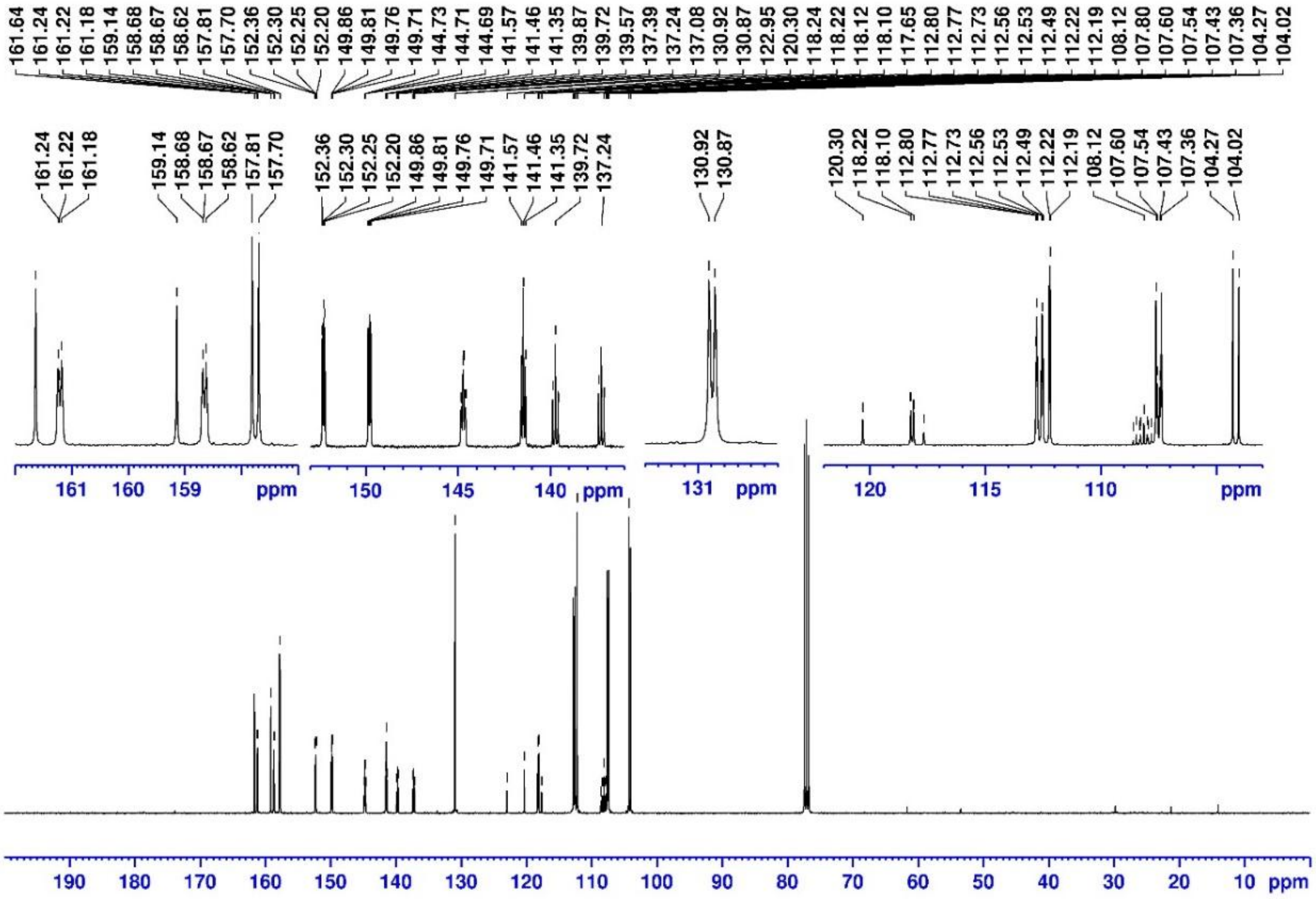

**Figure S33**: $^{13}C$ NMR spectrum of **6** in $CDCl_3$.

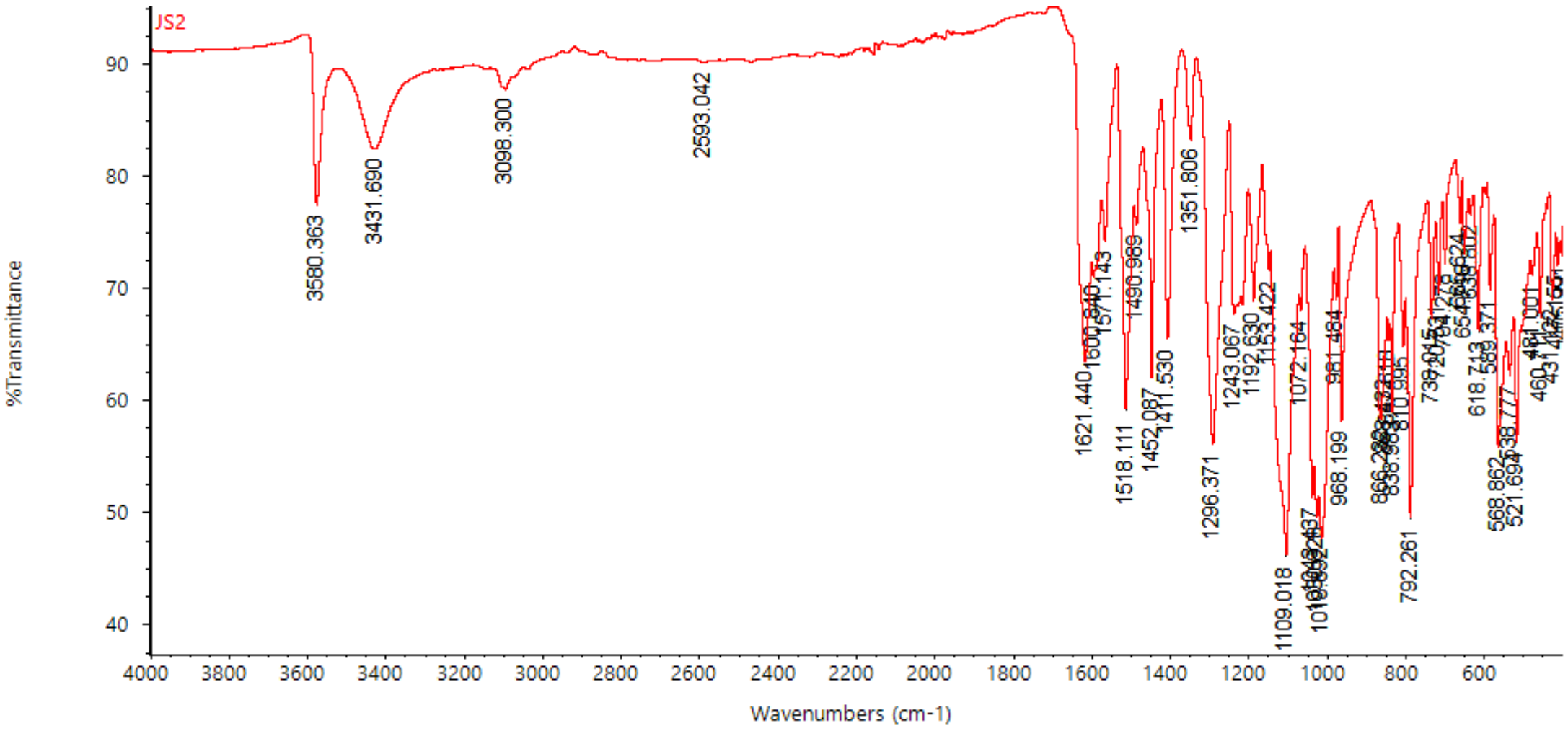

**Figure S34**: IR spectrum of **6.**

**4'-(difluoro(3,4,5-trifluorophenoxy)methyl)-2,3',5'-trifluoro-[1,1'-biphenyl]-4-yl (S)-2,6-difluoro-4-(5-(2-methylbutyl)-1,3-dioxan-2-yl)benzoate (7(S), S-RW4*)**

(S)-2,6-Difluoro-4-(5-(2-methylbutyl)-1,3-dioxan-2-yl)benzoic acid (**4**, 100 mg, 318 µmol, 1.00 eq.) was dissolved in 10 mL of DCM and DCC (85.3 mg, 414 µmol, 1.30 eq.) were added and the mixture was stirred until it changed to opaque (2 min). **6** (147 mg, 350 µmol, 1.10 eq.) and DMAP (5.05 mg, 41.4 µmol, 0.13 eq.) were added which created a clean mixture again. It was stirred for 22 h at rt. During the reaction a white precipitant formed, which was removed through filtration. The filtrate was dried in vacuo. The crude product was purified by column chromatography (DCM : petroleum ether 1:1, Rf = 0.62) and recrystallisation in ethanol.

**Yield**: 136 mg (60%), white solid.

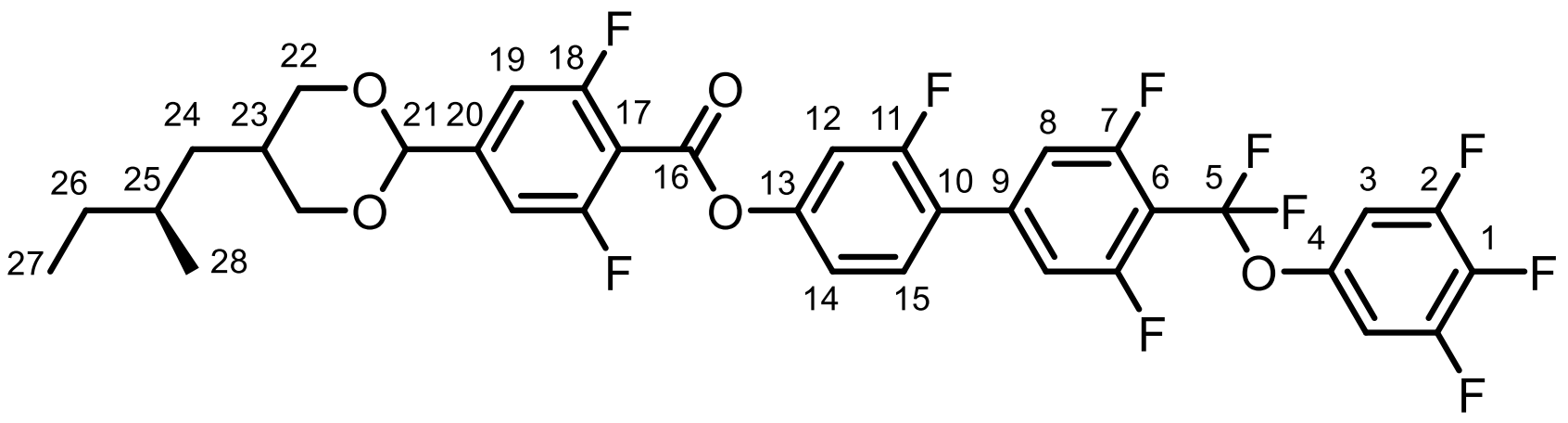

**$^{1}$H-NMR** (400 MHz, CDCl$_3$): $\delta$ = 7.49 (app. t, $^{3}J_{HF}$ = 8.57 Hz, 1H, *H*-Ar), 7.25 – 7.17 (m, 6H, *H*-Ar), 7.00 (dd, $^{3}J_{HF}$ = 7.67 Hz, $^{4}J_{HF}$ = 5.99 Hz, 2H, *H*-3), 5.41 (s, 1H, *H*-21), 4.29 - 4.19 (m, 2H, *H*-22), 3.58 – 3.48 (m, 2H, *H*-22), 2.32 – 2.16 (m, 1H, *H*-23), 1.43 – 1.29 (m, 2H, *H*-24), 1.27 – 1.04 (m, 2H, *H*-Al), 0.96 – 0.85 (m, 7H, *H*-Al) ppm. **$^{19}$F-NMR** (376 MHz, 298 K, CDCl$_3$): $\delta$ = −61.8 (t, $^{4}J_{FF}$ = 26.3 Hz, 2F, *F*-5), −108.4 (d, $^{3}J_{HF}$ = 8.79 Hz, 2F, *F*-18), −110.3 (td, $^{4}J_{FF}$ = 26.5 Hz, $^{3}J_{HF}$ = 10.2 Hz, 2F, *F*-7), −113.6 (app. t, $^{3}J_{HF}$ = 9.54 Hz 1F, *F*-11), −132.4 (dd, $^{3}J_{FF}$ = 20.7 Hz, $^{3}J_{HF}$ = 8.08 Hz, 2F, *F*-2), −163.5 (tt, $^{3}J_{FF}$ = 20.4 Hz, $^{4}J_{HF}$ = 5.45 Hz, 1F, *F*-1) ppm. **$^{13}$C-NMR** (101 MHz, CDCl$_3$) δ ppm: 161.09 (dd, $J_{C\text{-}F\text{-}i}$ = 258.5 Hz, $J_{C\text{-}F\text{-}m}$ = 5.7 Hz), 160.09 (dd, $J_{C\text{-}F\text{-}i}$ = 257.8 Hz, $J_{C\text{-}F\text{-}m}$ = 6.0 Hz), 159.65 (d, $J_{C\text{-}F\text{-}i}$ = 252.5 Hz), 159.33 (t, $J_{C\text{-}F\text{-}l.r.c.}$ = 1.9 Hz), 151.77 (d, $J_{C\text{-}F\text{-}m}$ = 11.1 Hz), 151.16 (ddd, $J_{C\text{-}F\text{-}i}$ = 251.0 Hz, $J_{C\text{-}F\text{-}o}$ = 10.7 Hz, $J_{C\text{-}F\text{-}m}$ = 5.3 Hz), 145.87 (t, $J_{C\text{-}F\text{-}m}$ = 9.9 Hz), 145.04 – 144.40 (m), 140.80 (app. t, $J_{C\text{-}F\text{-}m}$ = 11.1 Hz), 138.62 (dt, $J_{C\text{-}F\text{-}i}$ = 250.4 Hz, $J_{C\text{-}F\text{-}o}$ = 15.2 Hz), 130.75 (d, $J_{C\text{-}F\text{-}m}$ = 3.8 Hz), 123.99 – 123.48 (m), 120.27 (t, $J_{C\text{-}F\text{-}i}$ = 266.3 Hz), 118.48 (d, $J_{C\text{-}F\text{-}p}$ = 3.7 Hz), 113.27 (dt, $J_{C\text{-}F\text{-}o}$ = 24.4 Hz, $J_{C\text{-}F\text{-}m}$ = 3.5 Hz), 110.97 (d, $J_{C\text{-}F\text{-}o}$ = 26.0 Hz), 110.46 (dd, $J_{C\text{-}F\text{-}o}$ = 23.8, $J_{C\text{-}F\text{-}p}$ = 3.3 Hz), 109.46 (t, $J_{C\text{-}F\text{-}o}$ = 16.9 Hz), 109.17 – 108.63 (m), 107.81 – 107.33 (m), 98.92 (t, $J_{C\text{-}F\text{-}l.r.c.}$ = 2.2 Hz), 73.05, 72.81, 35.07, 31.98, 31.35, 29.80, 19.41, 11.33.

**IR** ($\nu_{max}$/cm$^{-1}$): 3100 (=C−H), 2963 (−C−H), 2919 (−C−H), 2876 (−C−H), 1748 (C=O), 1639 (C=C), 1033 (C−F) cm$^{-1}$.

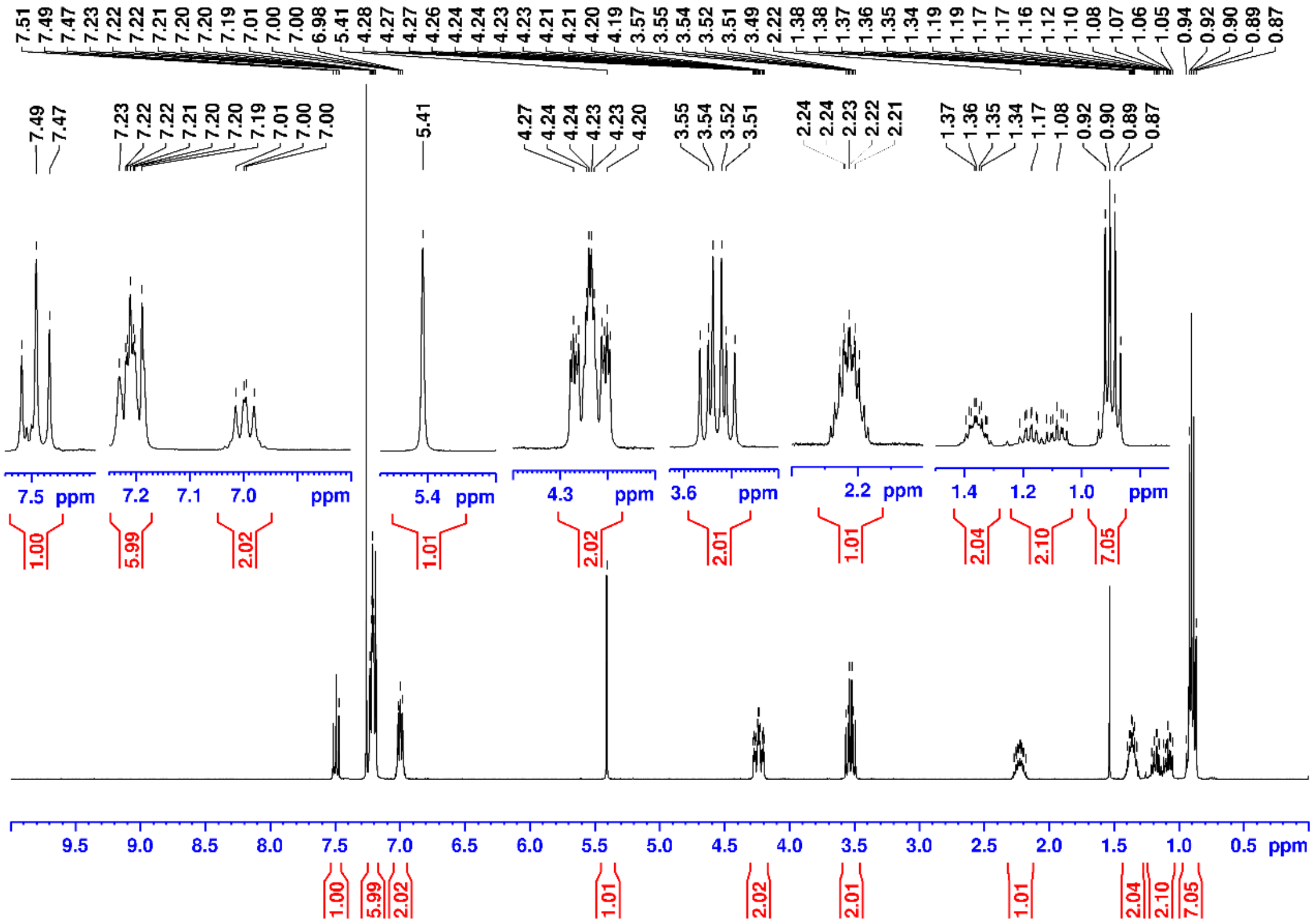

**Figure S35**: [1]H NMR spectrum of compound **S-RW4*** in $CDCl_3$.

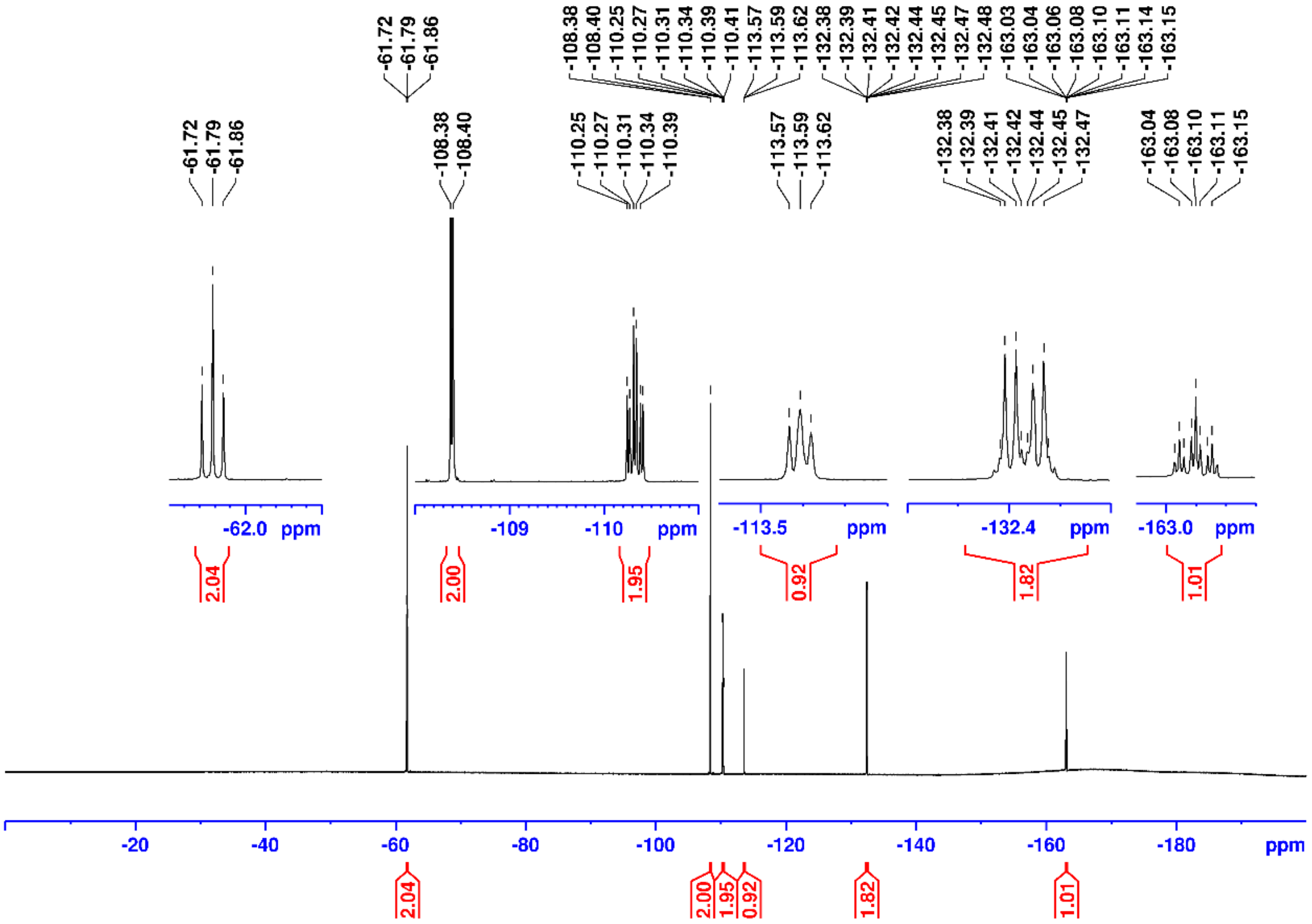

**Figure S36**: [19]F NMR spectrum of compound **S-RW4*** in $CDCl_3$.

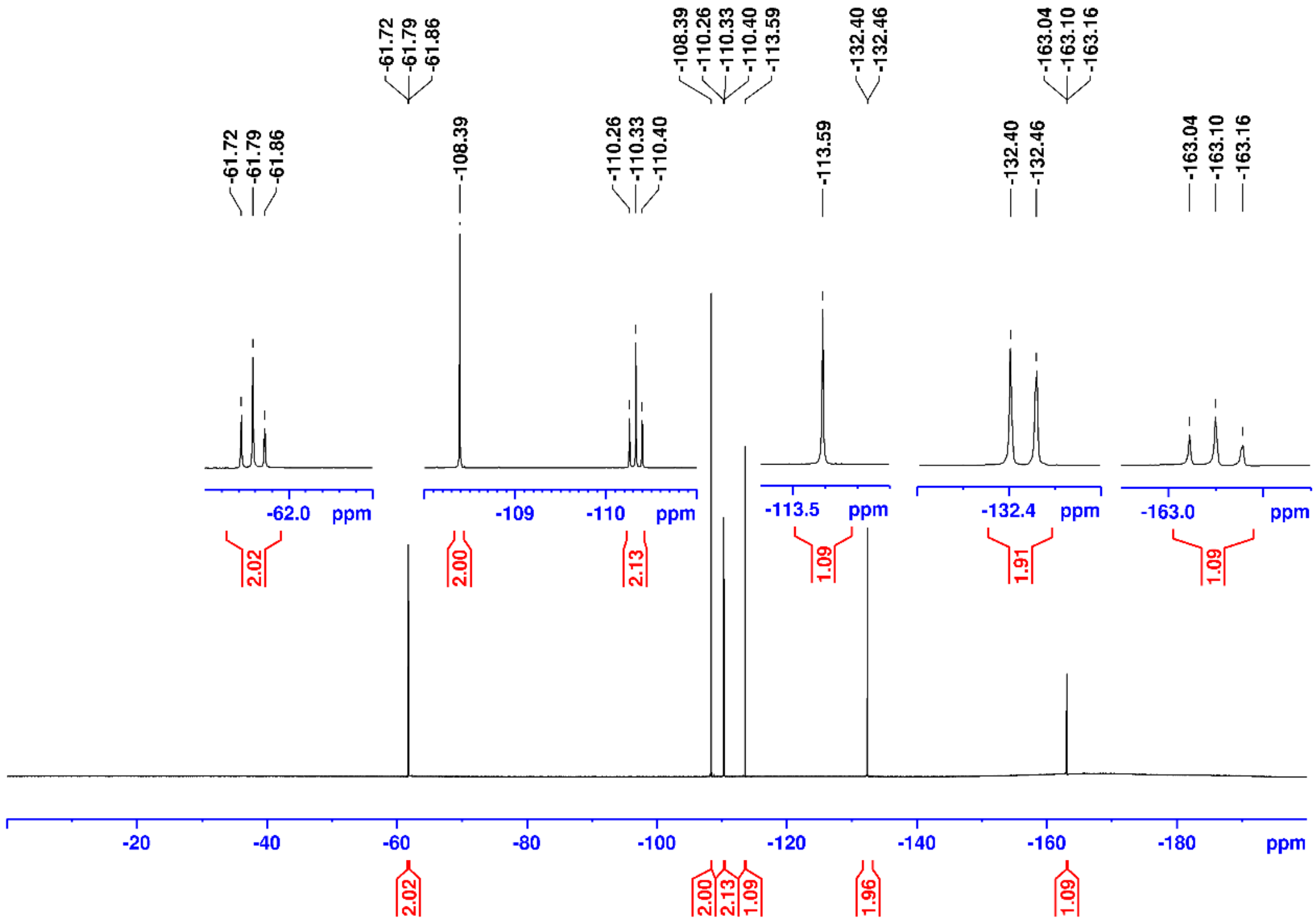

**Figure S37**: $^{19}F\{^{1}H\}$ NMR spectrum of **S-RW4***in $CDCl_3$.

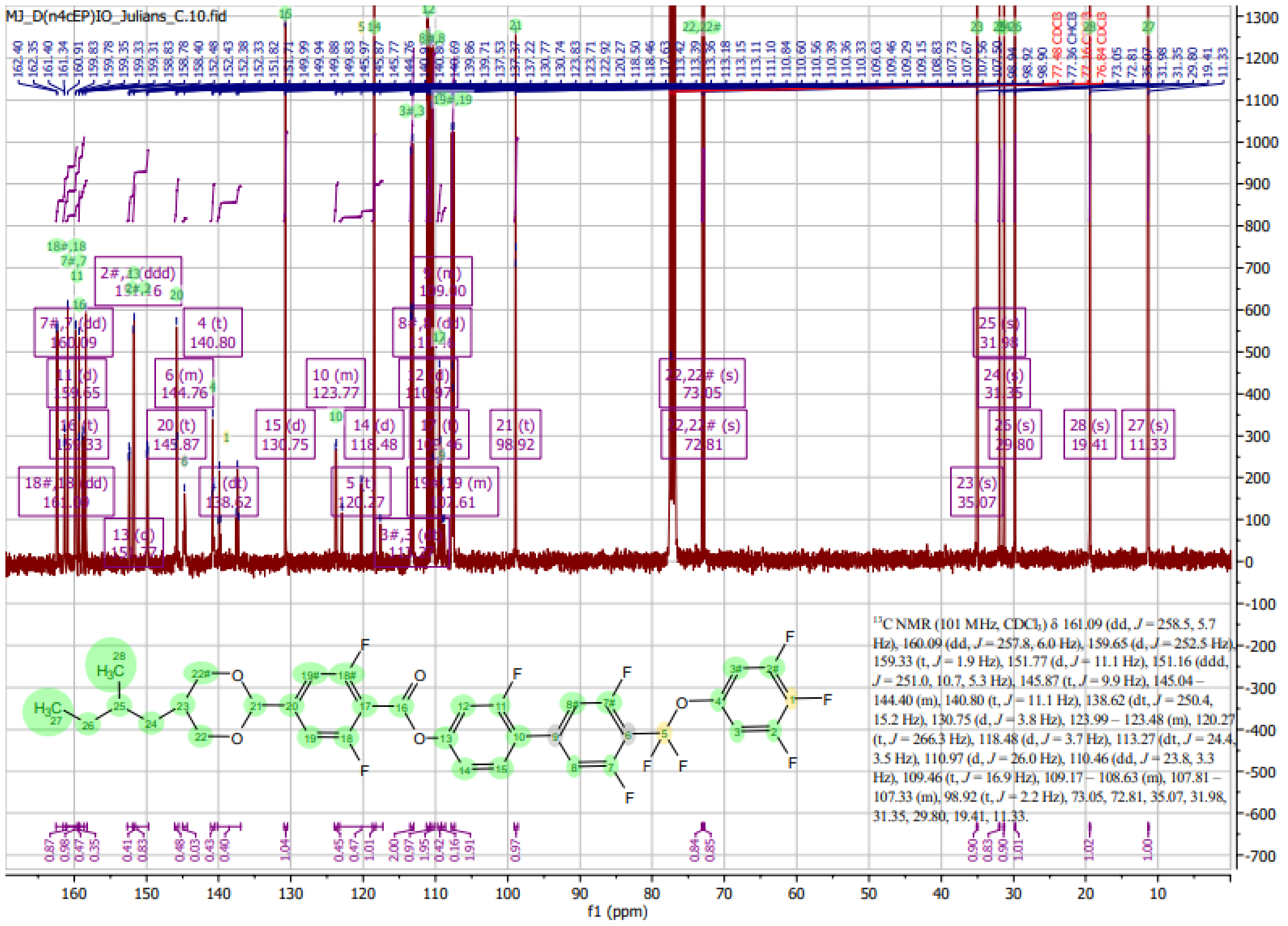

**Figure S38**: $^{13}C$ NMR spectrum of **S-RW4***in $CDCl_3$.

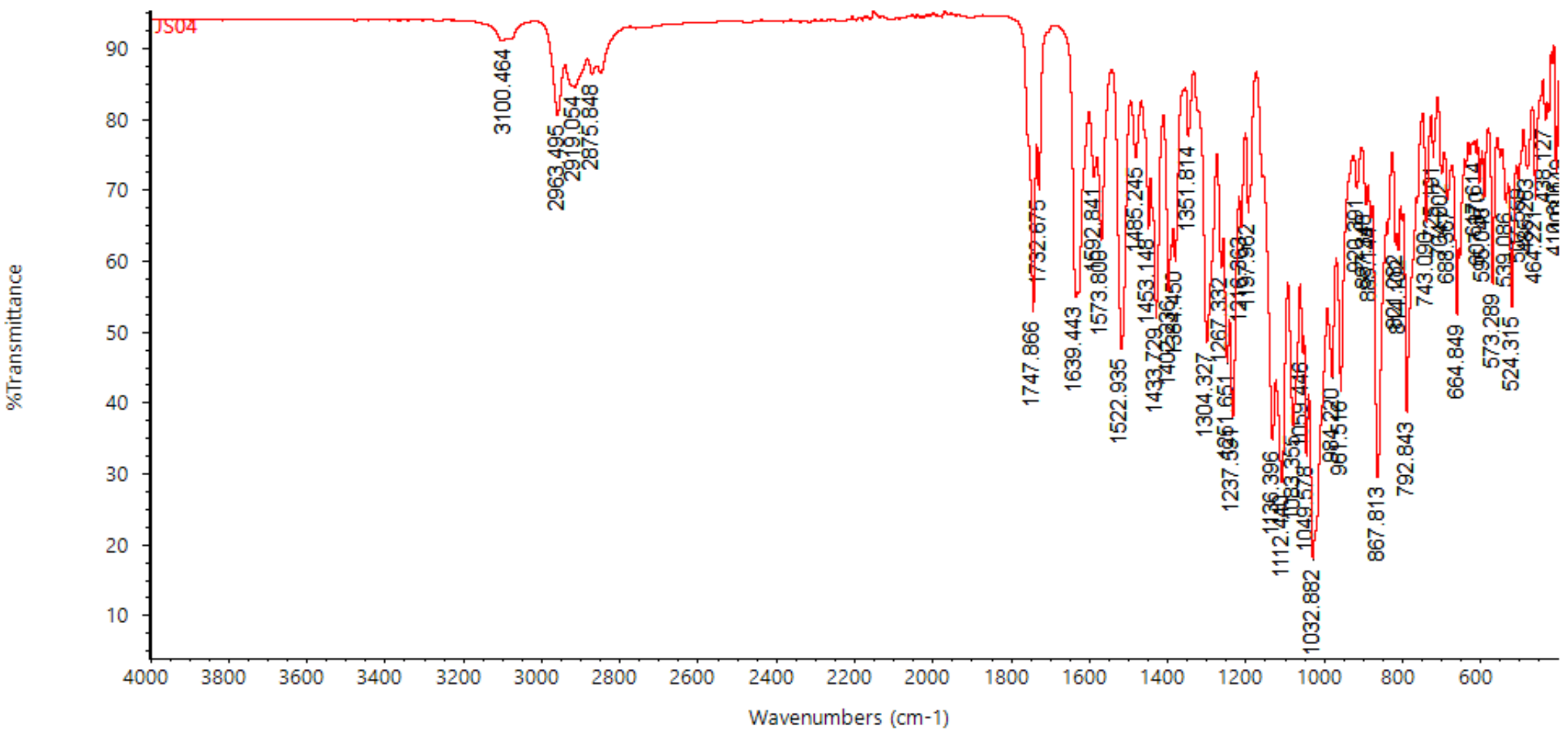

**Figure S39**: IR spectrum of compound **S-RW4***.

**4'-(difluoro(3,4,5-trifluorophenoxy)methyl)-2,3',5'-trifluoro-[1,1'-biphenyl]-4-yl (+/-)-2,6-difluoro-4-(5-(2-methylbutyl)-1,3-dioxan-2-yl)benzoate (7(rac), rac-RW4*)**

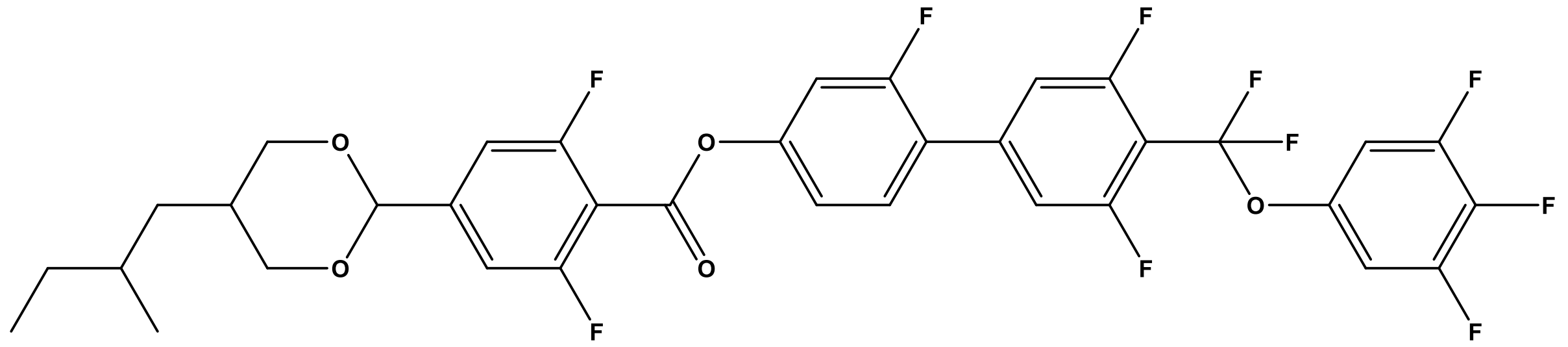

Prepared using the same method as the compound 7(S)

**Yield**: 37 mg (46%), white solid.

**$^{1}$H-NMR** (400 MHz, $CDCl_3$) δ ppm: 7.57 – 7.39 (1H, m, Ar-H), 7.26 – 7.11 (6H, m, Ar-H), 7.06 – 6.90 (2H, m, Ar-H), 5.41 (1H, s, Ar-CH), 4.31 – 4.16 (2H, m, -CH-$\underline{CH_2}$-O-), 3.62 – 3.44 (2H, m, -CH-$\underline{CH_2}$-O-), 2.32 – 2.09 (1H, m, , -$CH_2$-$\underline{CH}$-($CH_2$-O-)$_2$), 1.46 – 1.26 (2H,-CH-$\underline{CH_2}$-CH- m,), 1.24 – 1.14 (1H, m,-CH-$\underline{CH}$(-H)-$CH_3$), 1.13 – 1.03 (1H, m, -CH-CH($\underline{-H}$)-$CH_3$), 0.97 – 0.80 (7H, m, -($CH_2$)$_2$-$\underline{CH\text{-}CH_3}$; -$CH_2$-$\underline{CH_3}$). **$^{19}$F-NMR** (376 MHz, CDCl3) δ ppm: -61.78 (t, $J$ = 26.4 Hz), -108.39, -110.33 (t, $J$ = 26.3 Hz), -113.60, -132.44 (d, $J$ = 21.0 Hz), -163.11 (t, $J$ = 20.7 Hz). **$^{13}$C-NMR** (101 MHz, $CDCl_3$) δ ppm: 161.09 (dd, $J_{C\text{-}F\text{-}i}$ = 258.5 Hz, $J_{C\text{-}F\text{-}m}$ = 5.7 Hz), 160.09 (dd, $J_{C\text{-}F\text{-}i}$ = 257.9 Hz, $J_{C\text{-}F\text{-}m}$ = 6.2 Hz), 159.65 (d, $J_{C\text{-}F\text{-}i}$ = 252.4 Hz), 159.33 (m), 151.77 (d, $J_{C\text{-}F\text{-}m}$ = 11.0 Hz), 151.16 (ddd, $J_{C\text{-}F\text{-}i}$ = 251.2 Hz, $J_{C\text{-}F\text{-}o}$ = 10.7 Hz, $J_{C\text{-}F\text{-}m}$ = 5.4 Hz), 145.87 (t, $J_{C\text{-}F\text{-}m}$ = 9.9 Hz), 145.04 – 144.40 (m), 140.80 (app. t, $J_{C\text{-}F\text{-}m}$ = 10.9 Hz), 138.62 (dt, $J_{C\text{-}F\text{-}i}$ = 250.3 Hz, $J_{C\text{-}F\text{-}o}$ = 15.3 Hz), 130.76 (d, $J_{C\text{-}F\text{-}m}$ = 3.7 Hz), 123.78 (d, J = 12.6 Hz), 120.27 (t, $J_{C\text{-}F\text{-}i}$ = 267.3

Hz), 118.49 (d, $J_{C\text{-}F\text{-}p}$ = 3.8 Hz), 113.63 – 112.78 (m), 110.97 (d, $J_{C\text{-}F\text{-}o}$ = 25.9 Hz), 110.46 (dd, $J_{C\text{-}F\text{-}o}$ = 23.7, $J_{C\text{-}F\text{-}p}$ = 3.3 Hz), 109.46 (t, $J_{C\text{-}F\text{-}o}$ = 16.9 Hz), 109.20 – 108.76 (m), 107.90 – 107.28 (m), 99.05 – 98.74 (m), 73.06, 72.81, 35.08, 31.99, 31.35, 29.81, 19.42, 11.33.

**IR** ($\nu_{max}$/cm$^{-1}$): 3107 (=C−H), 2964 (−C−H), 2931 (−C−H), 2853 (−C−H), 1758 (C=O), 1635 (C=C).

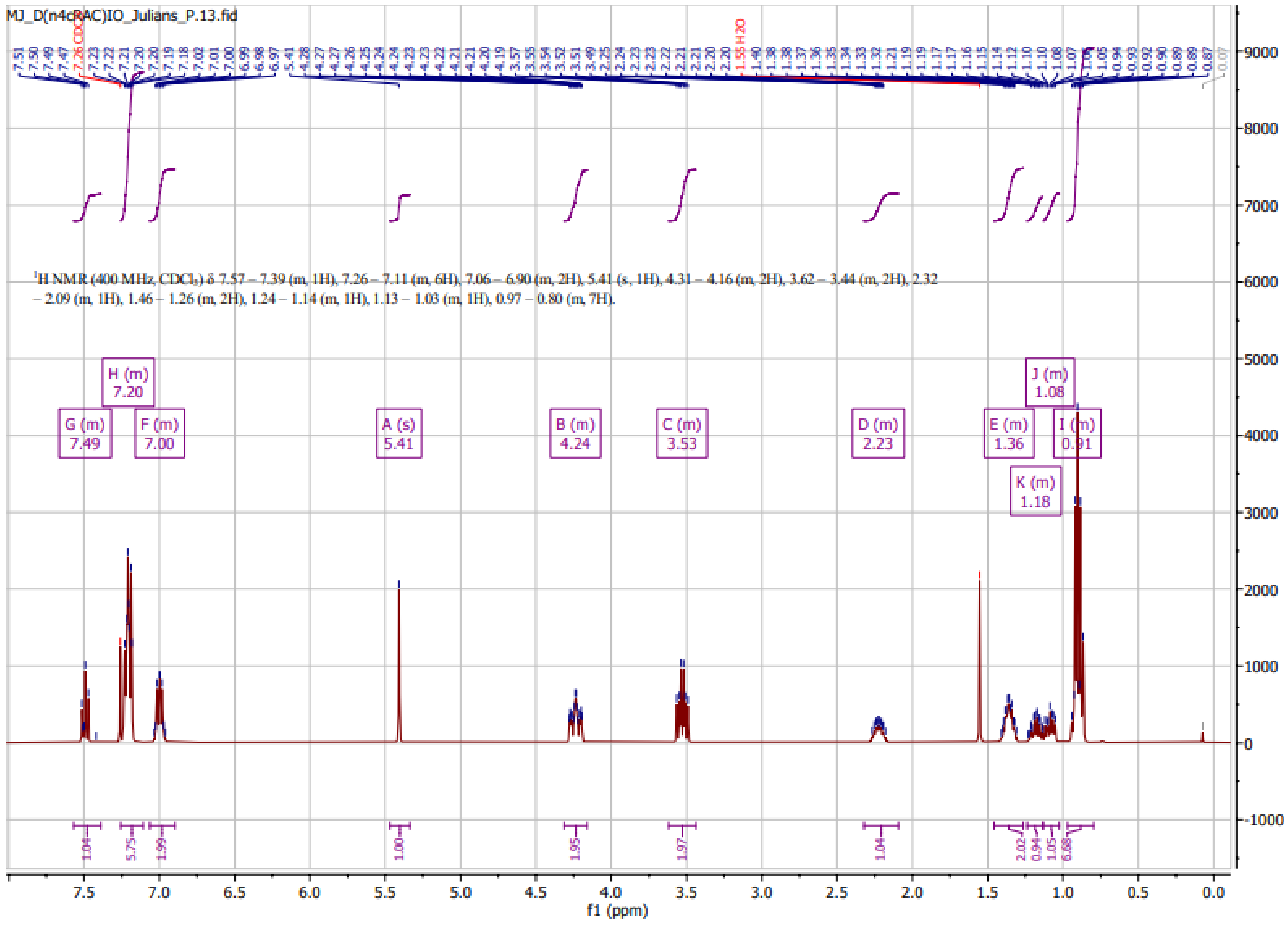

**Figure S40**: ¹H NMR spectrum of compound **rac-RW4*** in $CDCl_3$.

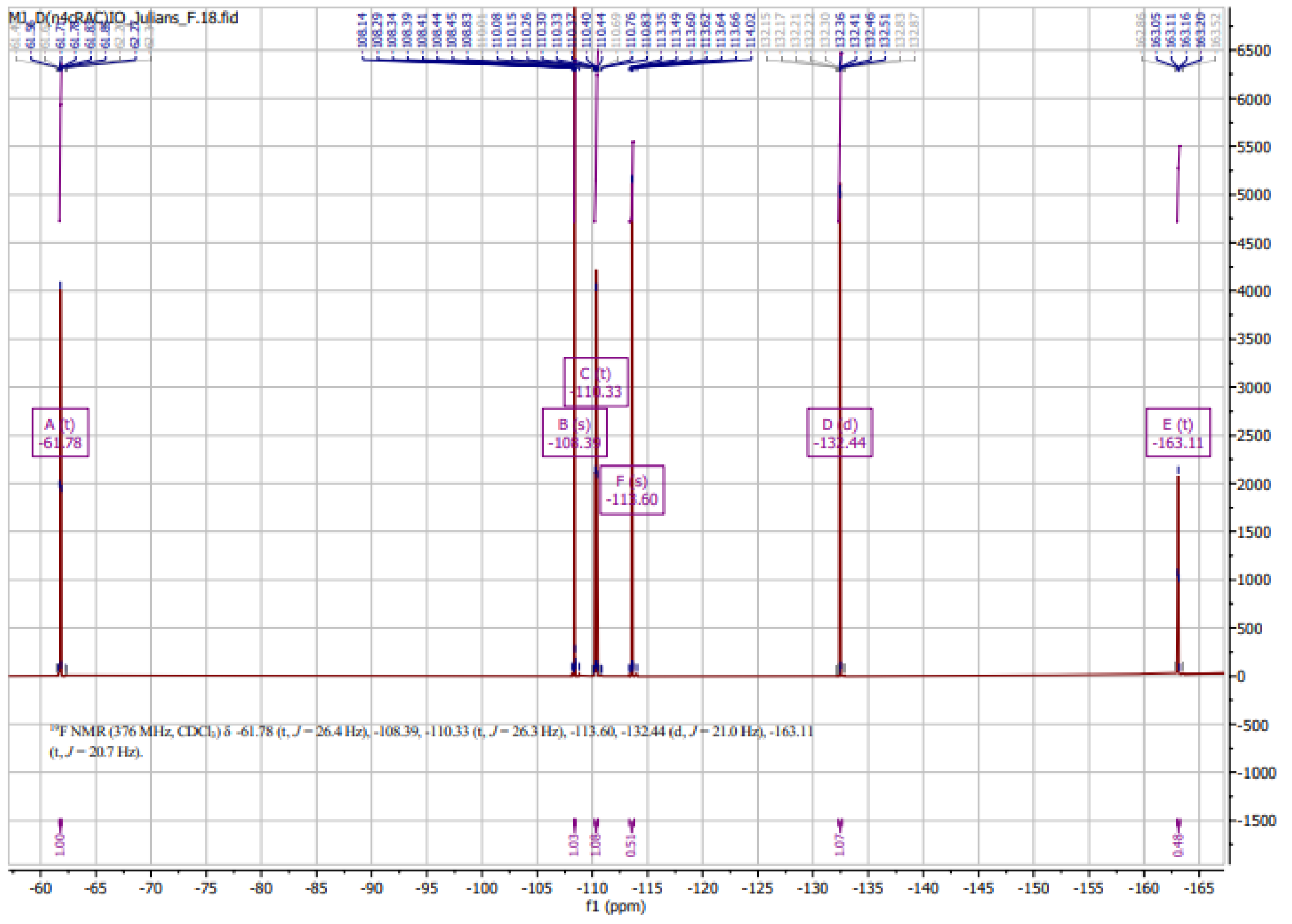

**Figure 41**: $^{19}$F NMR spectrum of compound **rac-RW4*** in $CDCl_3$.

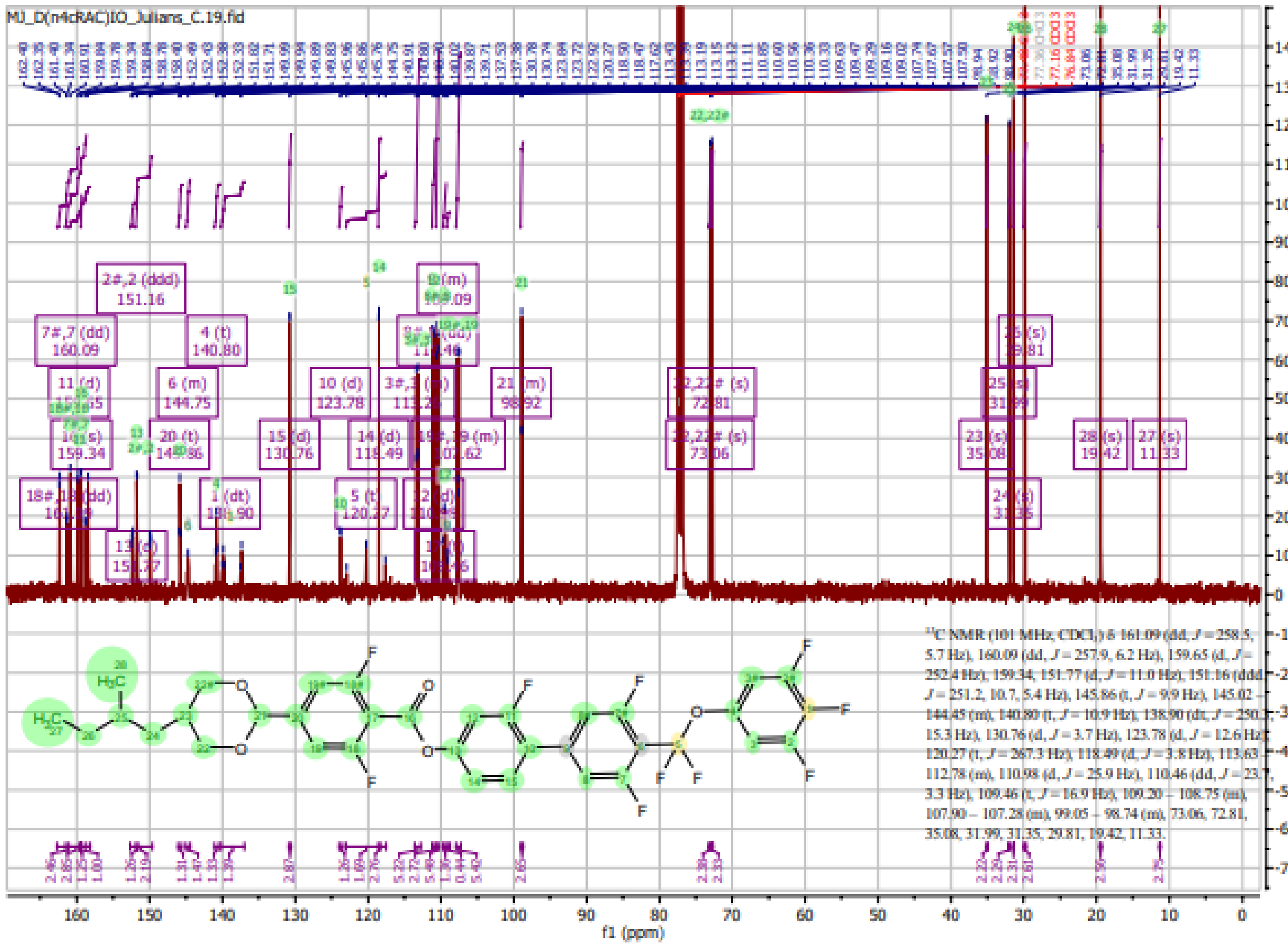

**Figure S42**: $^{13}$C NMR spectrum of compound **rac-RW4*** in $CDCl_3$.

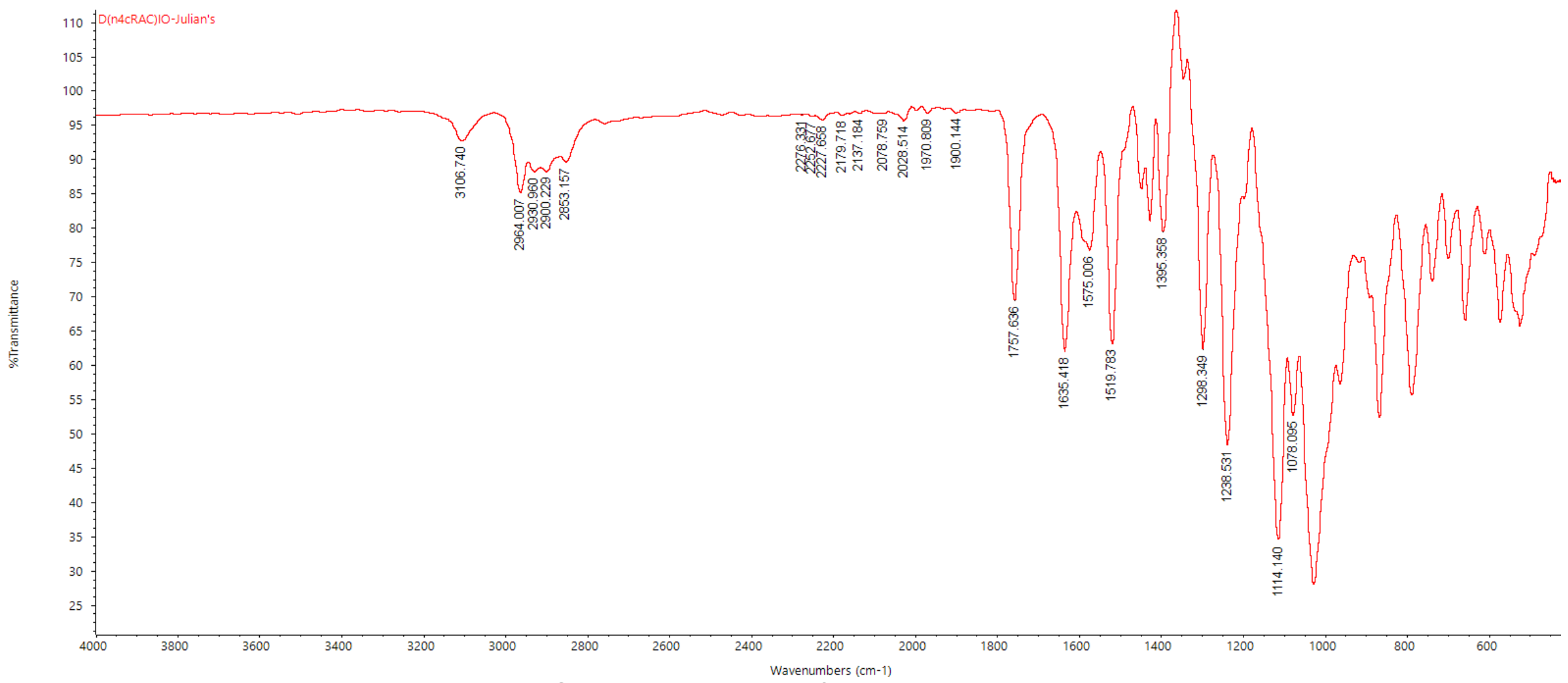

**Figure S43**: IR spectrum of compound **rac-RW4***.

# Experimental methods

**Calorimetric studies:** Differential scanning calorimetry was performed using a Mettler Toledo DSC3 instrument equipped with TSO 801RO sample robots and calibrated using indium and zinc standards. Heating and cooling rates were 10 K min$^{-1}$, between heating and cooling runs a 3 min isotherm steps were applied. Samples were measured under a nitrogen atmosphere.

**X-ray diffraction** (XRD) studies in broad diffraction angle range were performed with Bruker GADDS system equipped with micro-focus type X-ray tube with Cu anode and dedicated Montel optics, diffraction patterns were recorded with Vantec 2000 area detector. Samples were prepared in the form of small droplets placed on a heated surface and their temperature was controlled with a modified Linkam heating stage. For precise determination of smectic layer spacing temperature evolution small angle X-ray diffraction experiments were performed, using Bruker Nanostar system (micro-focus type X-ray tube with Cu anode and dedicated Montel optics, MRI TCPU-H heating stage, Vantec 2000 area detector). Samples were prepared in thin-walled glass capillaries, with 1.5 mm diameter.  The x-ray patterns was analyzed using TOPAS software.

**Microscopic studies:** Optical textures of LC phases were studied using a Zeiss Axio Imager A2m polarized light microscope, equipped with a Linkam LTS420 heating stage. Samples were prepared in commercial cells (AWAT) of various thicknesses (1.5–20 μm) with ITO electrodes and surfactant layers for planar or homeotropic alignment, in the case of planar cells parallel rubbing on both surfaces was applied. Cells for in-plane switching (~3-μm-thick, with planar alignment layers and 3 millimeter distance between electrodes) were provided by prof. O. Lavrentovich group at Kent State University.

**Selective light reflection and absorption** studies were carried out for material placed in the glass cell (3:10-μm-thick) with planar anchoring. The measurements were performed with a fiber-coupled spectrometer (Filmetrics F20-UV) mounted to the Zeiss Axio Imager A2m microscope working either in reflection or transmission mode. The tested sample area was confined to ~50 microns.

**Optical birefringence** measurements were measured with a setup based on a photoelastic modulator (PEM-90, Hinds) working at the base frequency f=50 kHz. As a light source, a halogen lamp (Hamamatsu LC8) equipped with a narrow bandpass filter (532±3 nm) was used. Samples were prepared in glass cells with a thickness of 1.5 μm, having surfactant layers for planar anchoring and parallel rubbing assuring uniform alignment of the optical axis in nematic phases. The sample and PEM were placed between crossed linear polarizers, with axes rotated ±45 deg with respect to the PEM axis, and the intensity of the light transmitted through this set-up was measured with a photodiode (FLC Electronics PIN-20). The registered signal was de-convoluted with a lock-in amplifier (EG&G 7265) into 1f and 2f components to yield a retardation induced by the sample.

**Dielectric spectroscopy:** The complex dielectric permittivity was measured in the 1 Hz–10 MHz frequency range using a Solartron 1260 impedance analyzer. The material was placed in 3:10-μm-thick glass cell with gold electrodes (without surfactant to avoid the influence of the high capacitance of a thin polymer layer). The amplitude of the applied ac voltage, 20 mV, was low enough to avoid Fréedericksz transition in nematic phases.

**Polarization current measurements:** electric polarization measurements were performed using cells with gold electrodes and no surfactant layers. The spontaneous polarization was calculated by analyzing the current flow through a resistor (500 Ω) connected in series with the cell, upon application of triangular-wave voltage, saturation of the current vs. applied electric field in few temperatures was checked. Siglent SDG2042X arbitrary waveform generator, FLC A200 amplifier and Siglent SDS2000X Plus oscilloscope were used.

**Second Harmonic Generation:** Second Harmonic Generation: The SHG response was investigated using a setup based on a solid-state IR laser EKSPLA NL202, λ=1064 nm. A series of collimated (~1 mm 1/e2 ), 9-ns laser pulses at a 10 Hz repetition rate and <2 mJ pulse energy were applied. The pulse energy was adjusted to avoid material decomposition. The IR beam was incident onto an LC cell of thickness 10 μm. An IR pass filter was placed at the entrance to the sample and a green pass filter at the exit of the sample. The emitted SHG radiation was detected using a photon counting head (Hamamatsu H7421) with a power supply unit (C8137). The SHG signal intensity was monitored with an oscilloscope (Agilent Technologies DSO6034A) using a custom-written Python script. The samples with in-plane electrodes were used. Optical SHG images were recording with homemade microscopic setup.

# Additional results

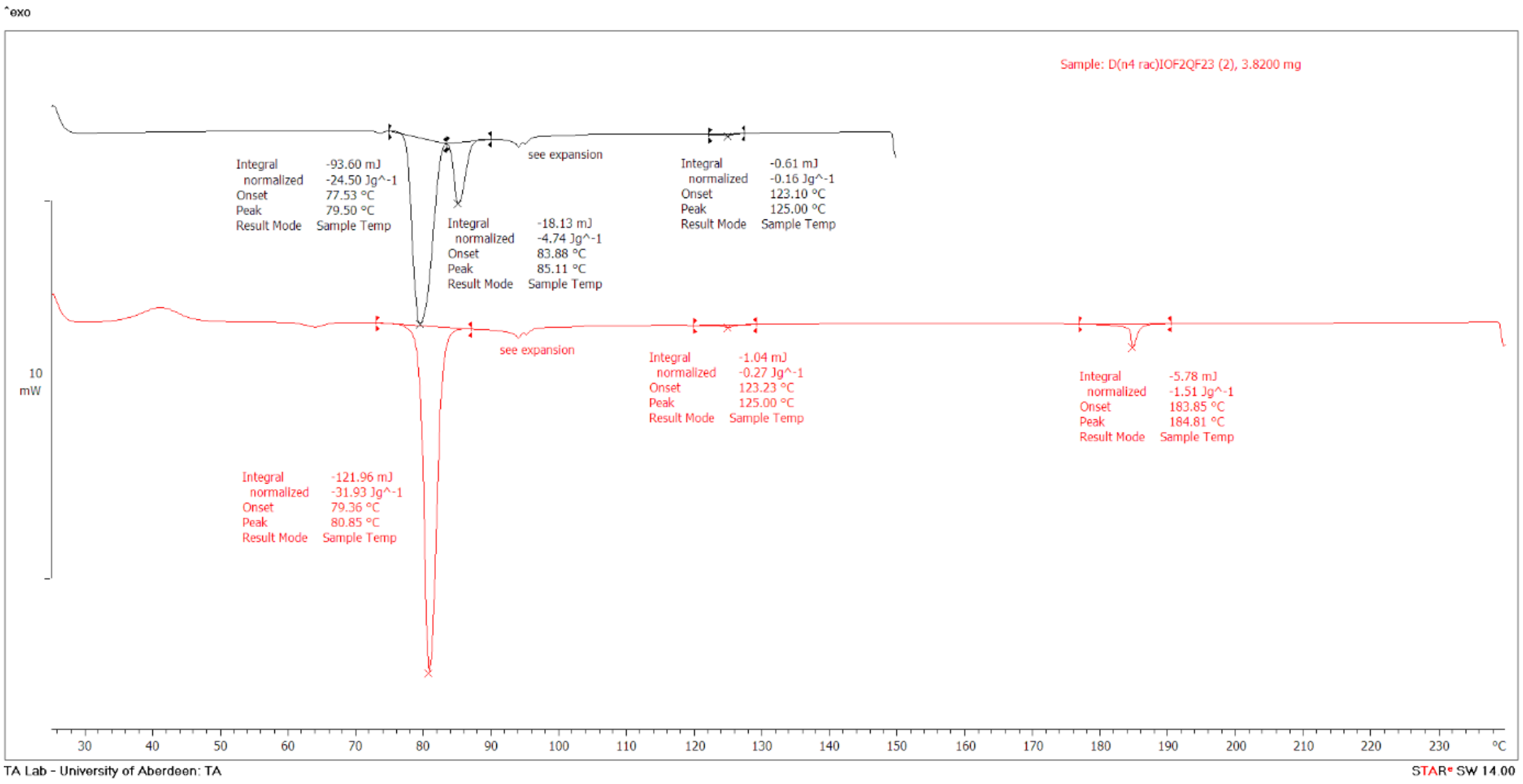

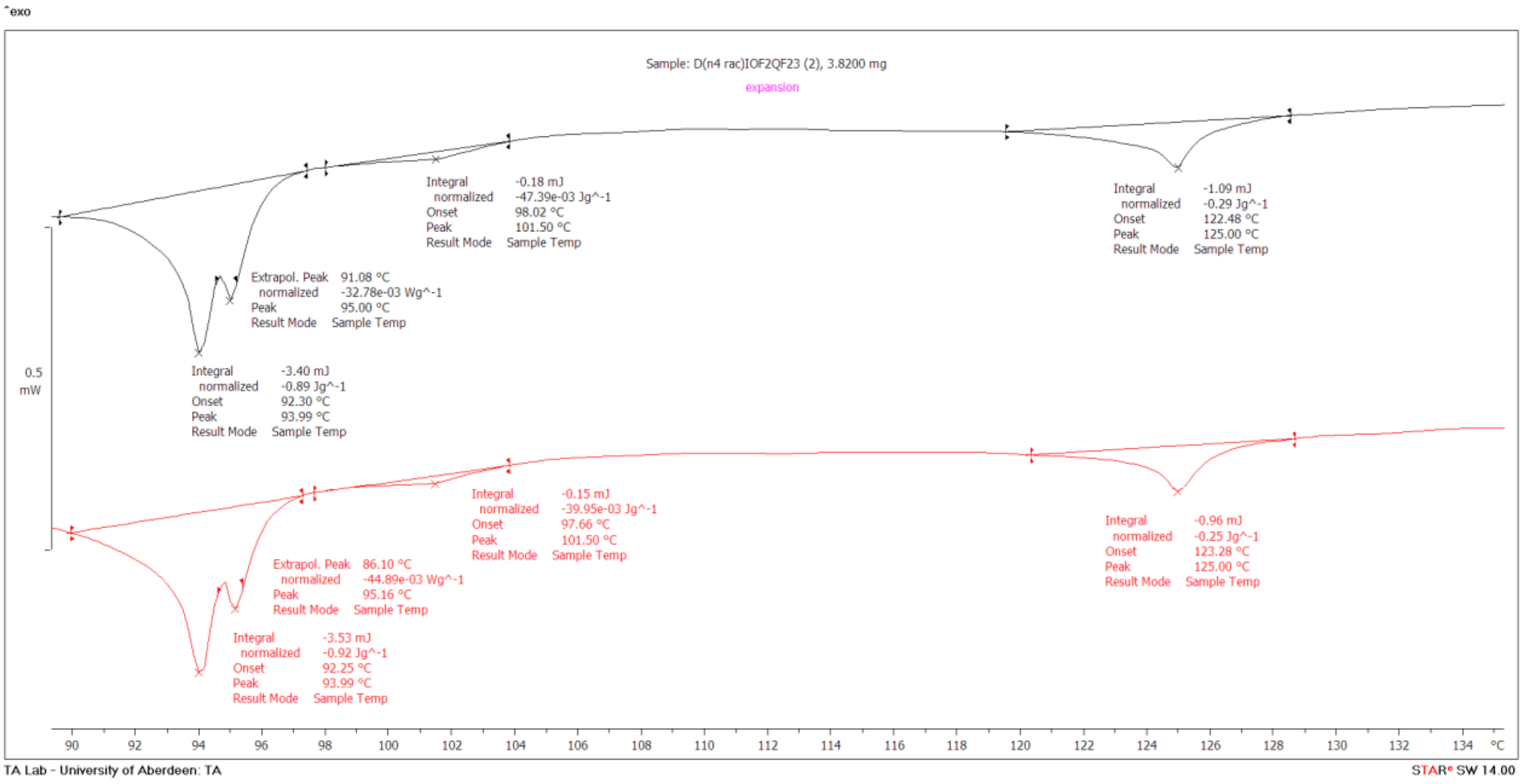

**Figure S45**: Heating DSC trace of **rac-RW4*,** full temperature range (top) and expanded region (bottom).

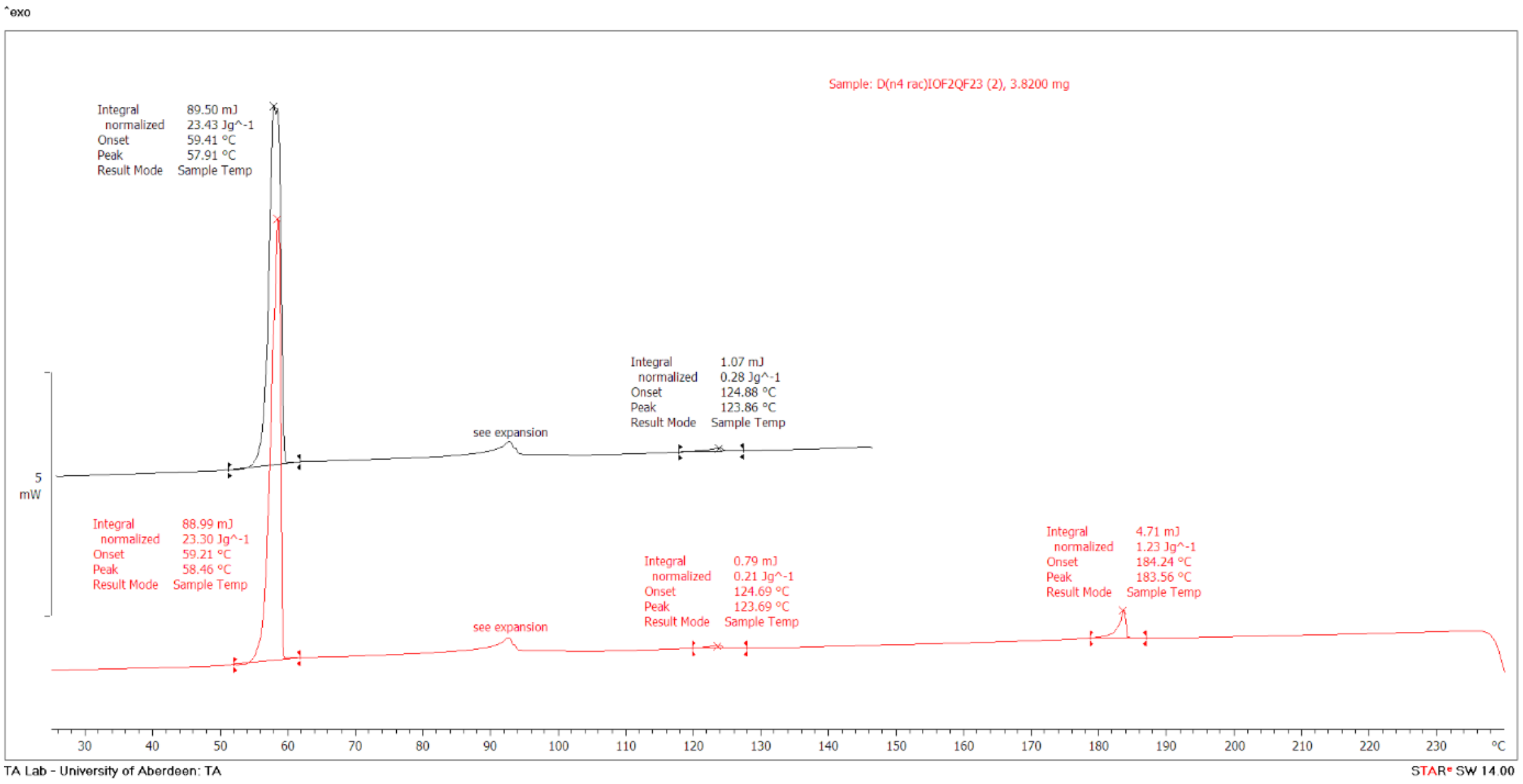

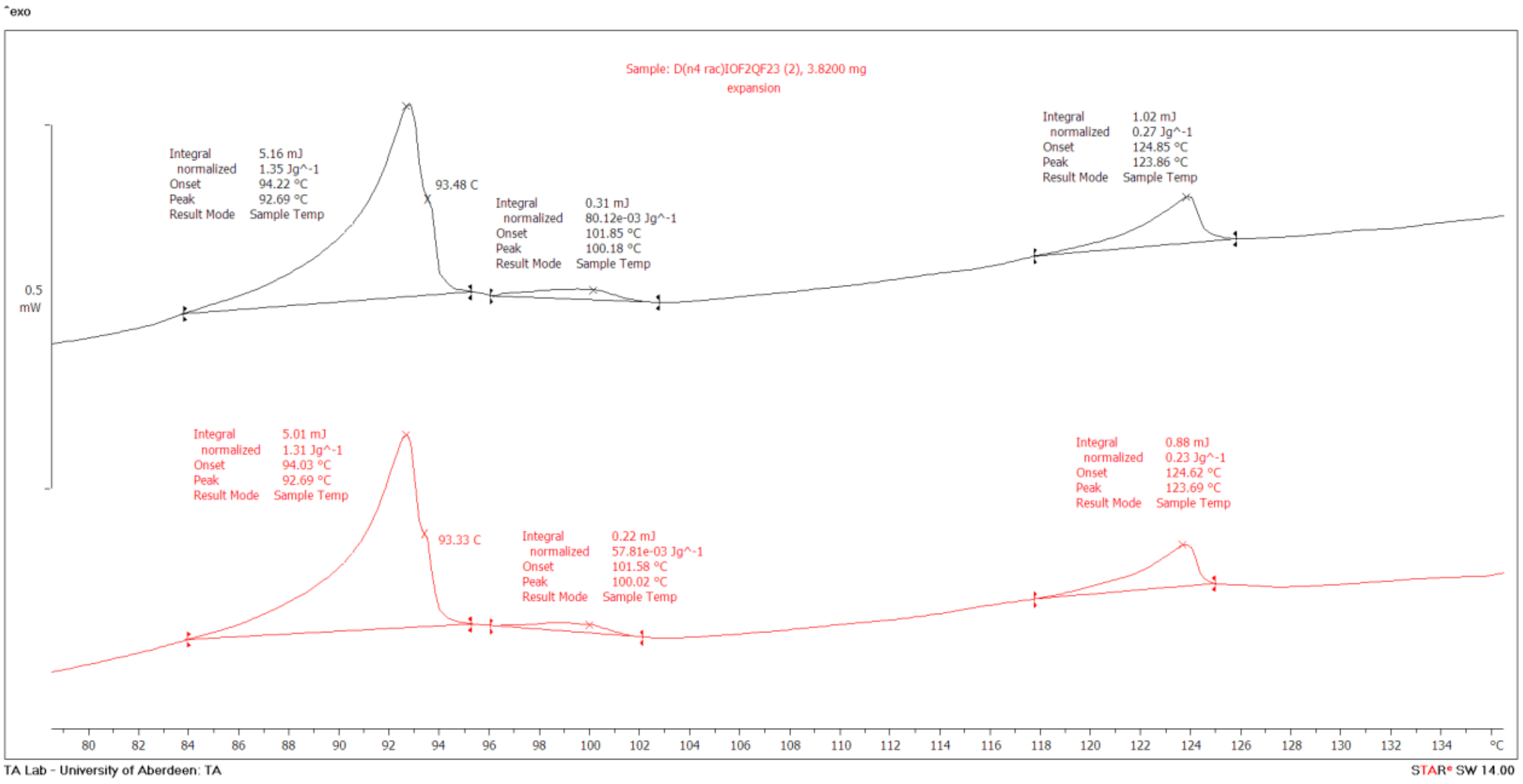

**Figure S46**: Cooling DSC trace of **rac-RW4*,** full temperature range (top) and expanded region (bottom).

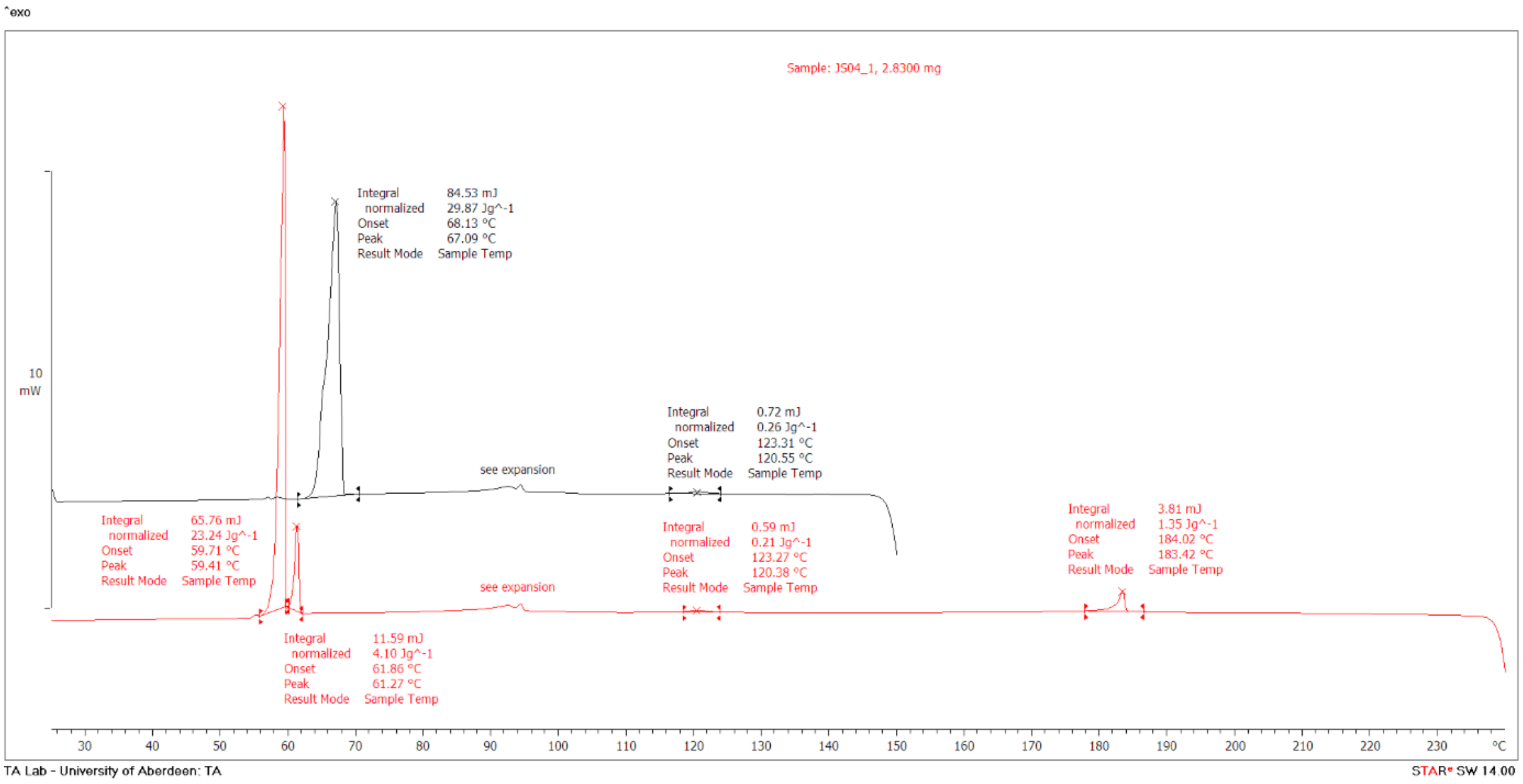

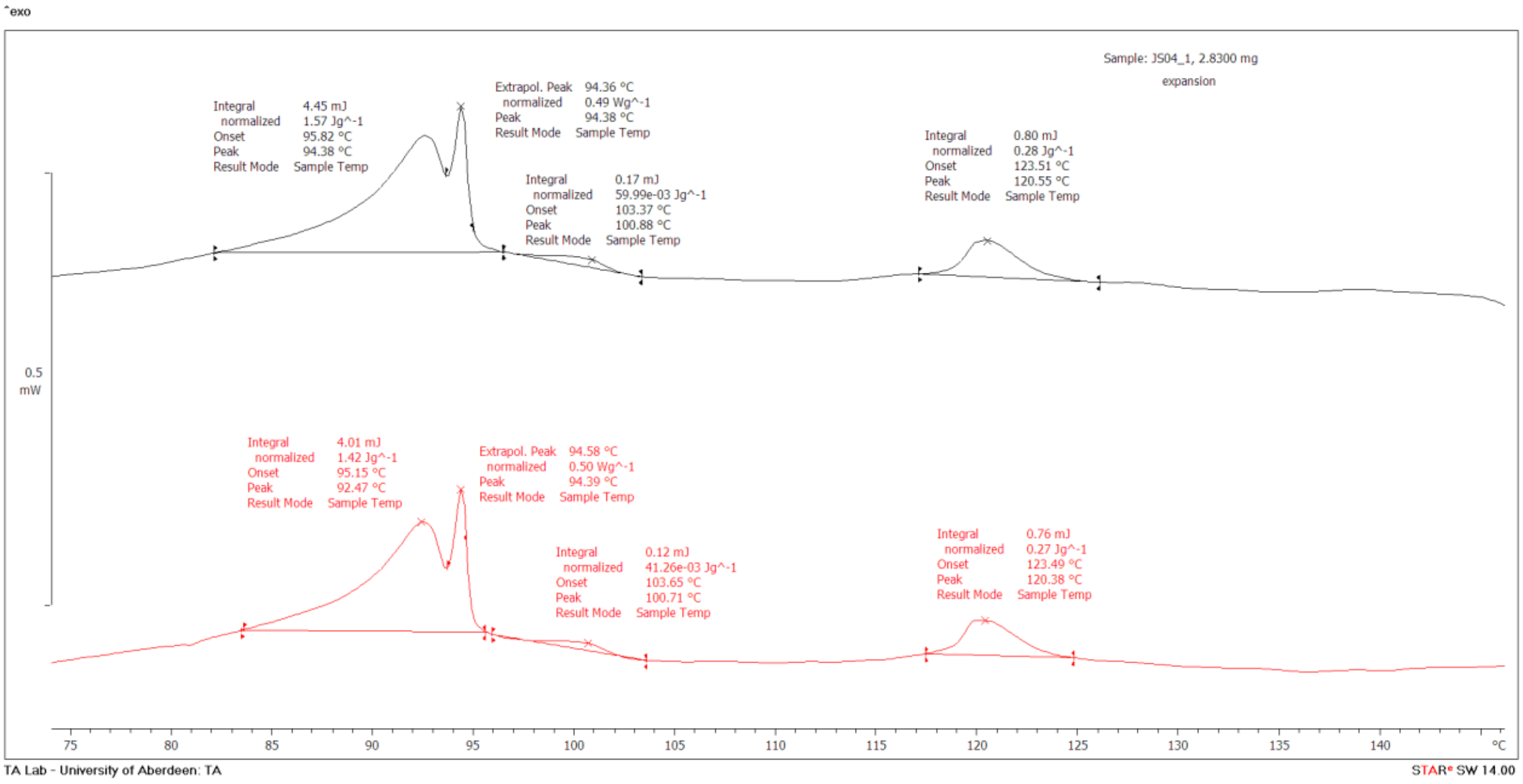

**Figure S47**: Cooling DSC trace of **S-RW4*,** full temperature range (top) and expanded region (bottom).

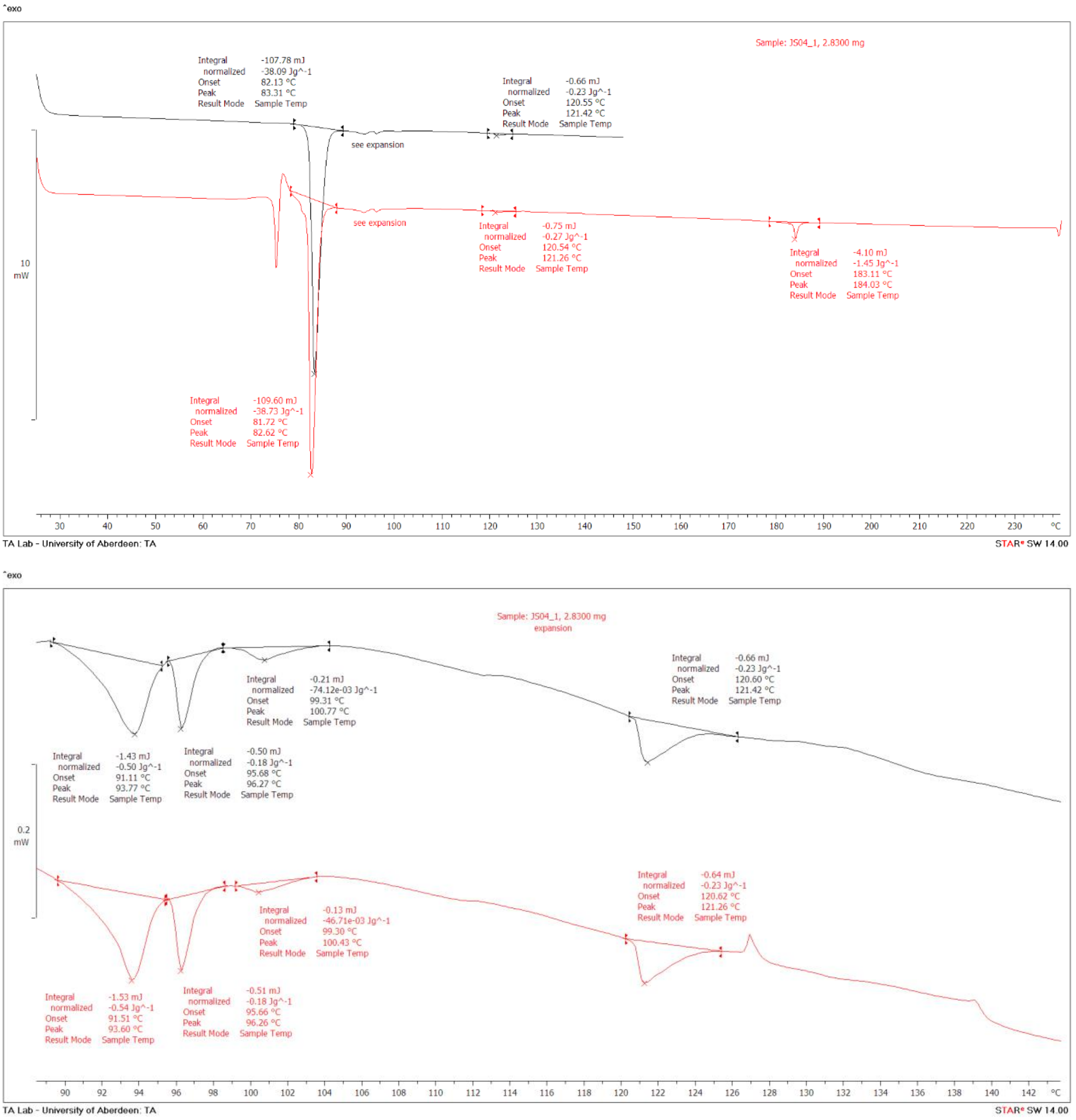

**Figure S48**: Heating DSC trace of **S-RW4*,** full temperature range (top) and expanded region (bottom).

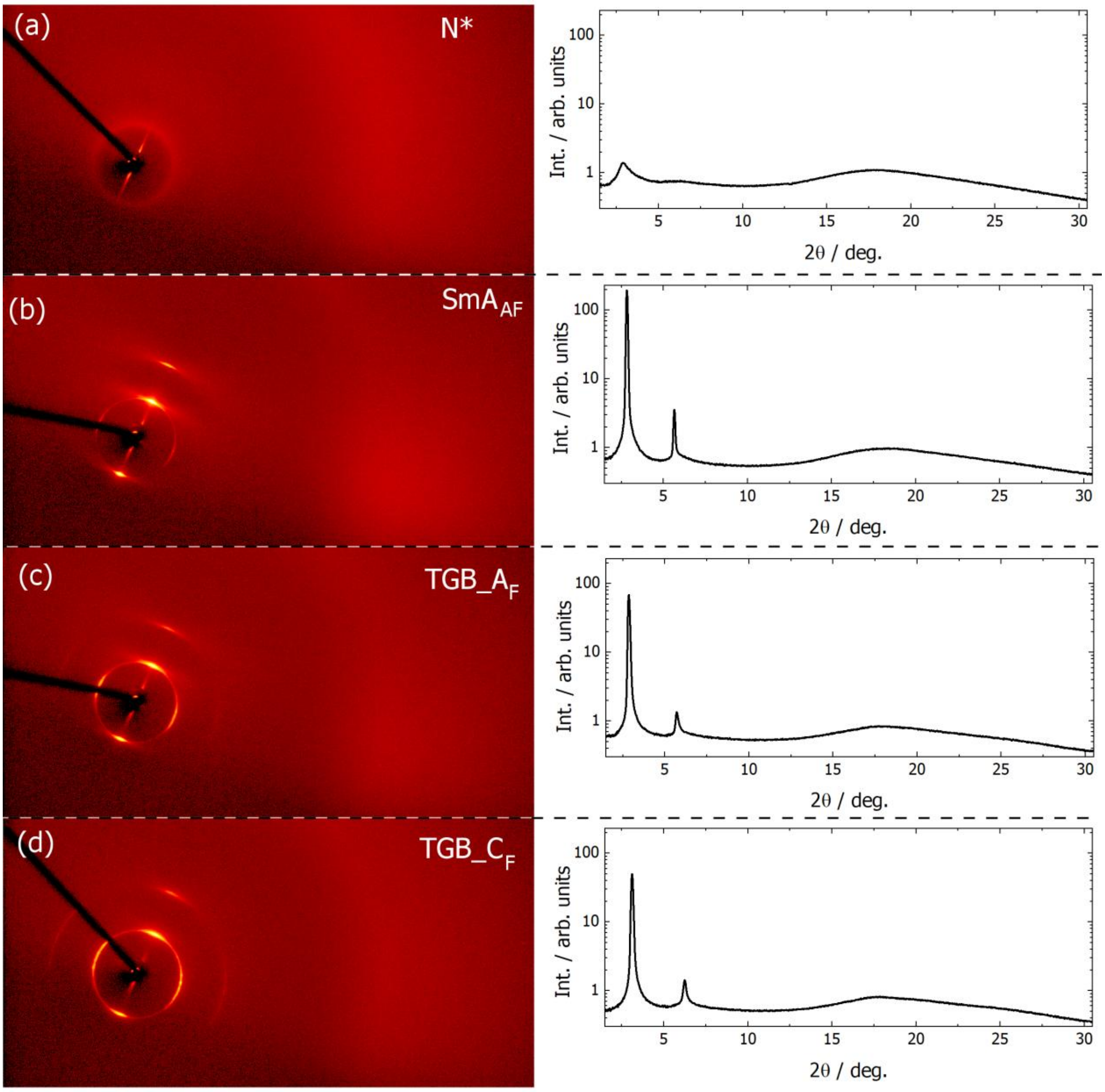

**Figure S49**: Wide angle X-ray diffraction patterns recorded in various phases of S-**RW4*** compound, diffractograms on the right were obtained by integration of 2D patterns presented on left, by integration over azimuthal angle.

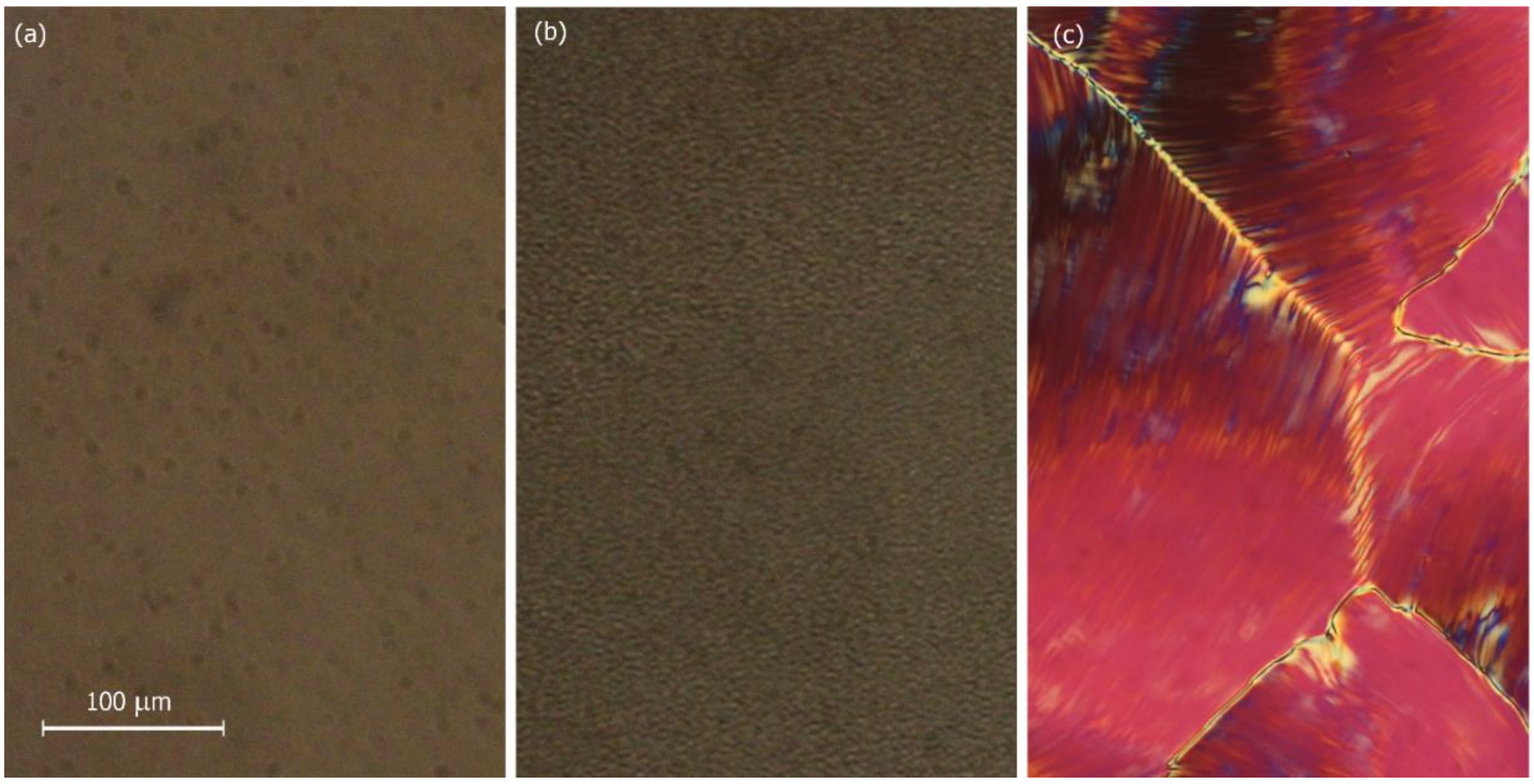

**Figure S50**: Optical texture of rac-**RW4*** compound observed in 3-μm-thick cell with homeotropic anchoring: in $SmA_{AF}$ phase perfect homeotropic texture is obtained( a), in $SmA_F$ phase small wrinkle-like defects develop (b) and the texture gradually rebuilds into planar one (c). Note, that the photos presented in panels (a) and (b) were overexposed.

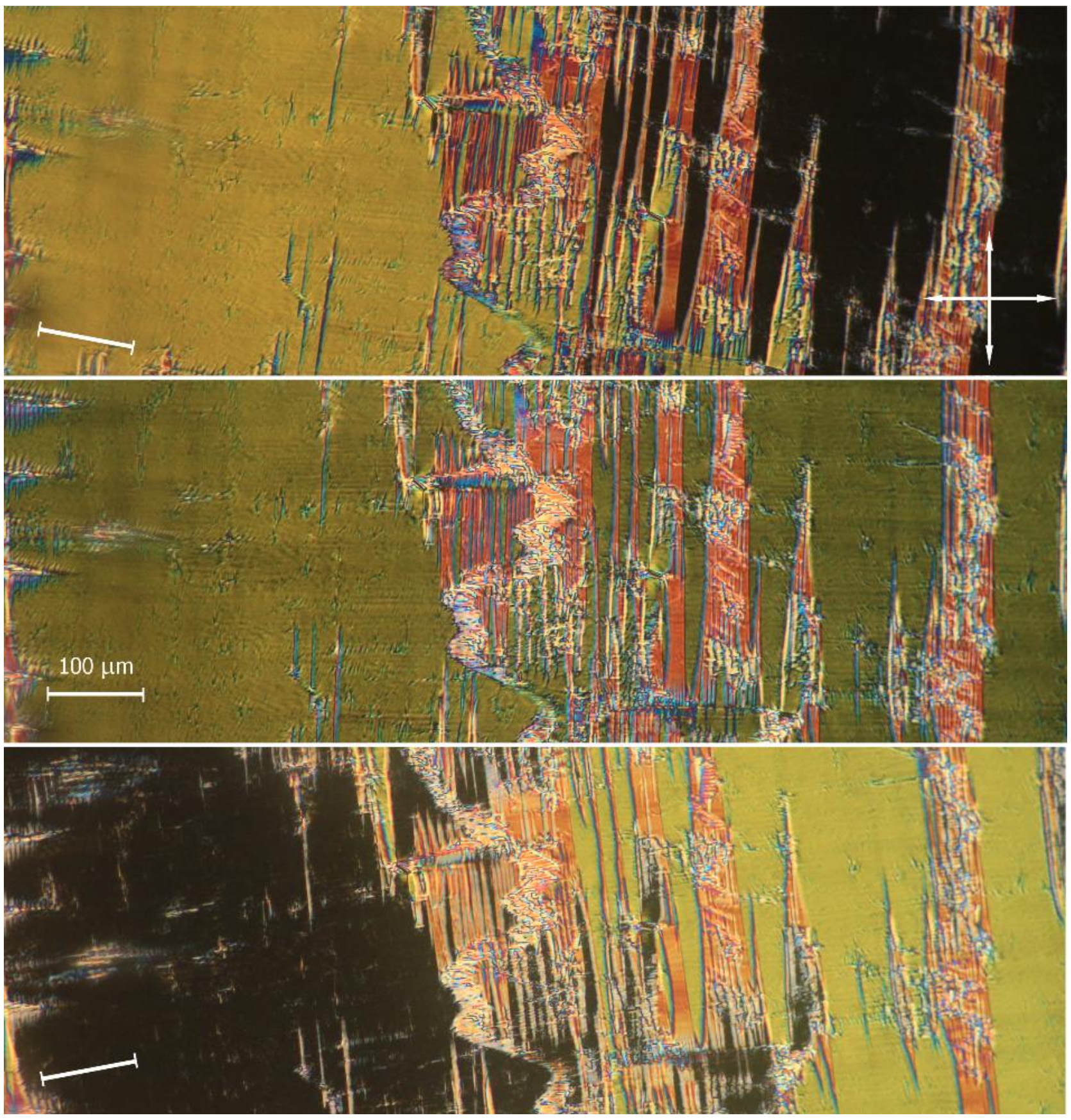

**Figure S51**: Optical texture of $SmC_F$ phase rac-**RW4*** compound in 5-μm-thick cell with planar anchoring, showing tilted domains which can be brought into extinction by rotating the sample with respect to crossed polarizers (arrows). Scale bar shows the rubbing direction.

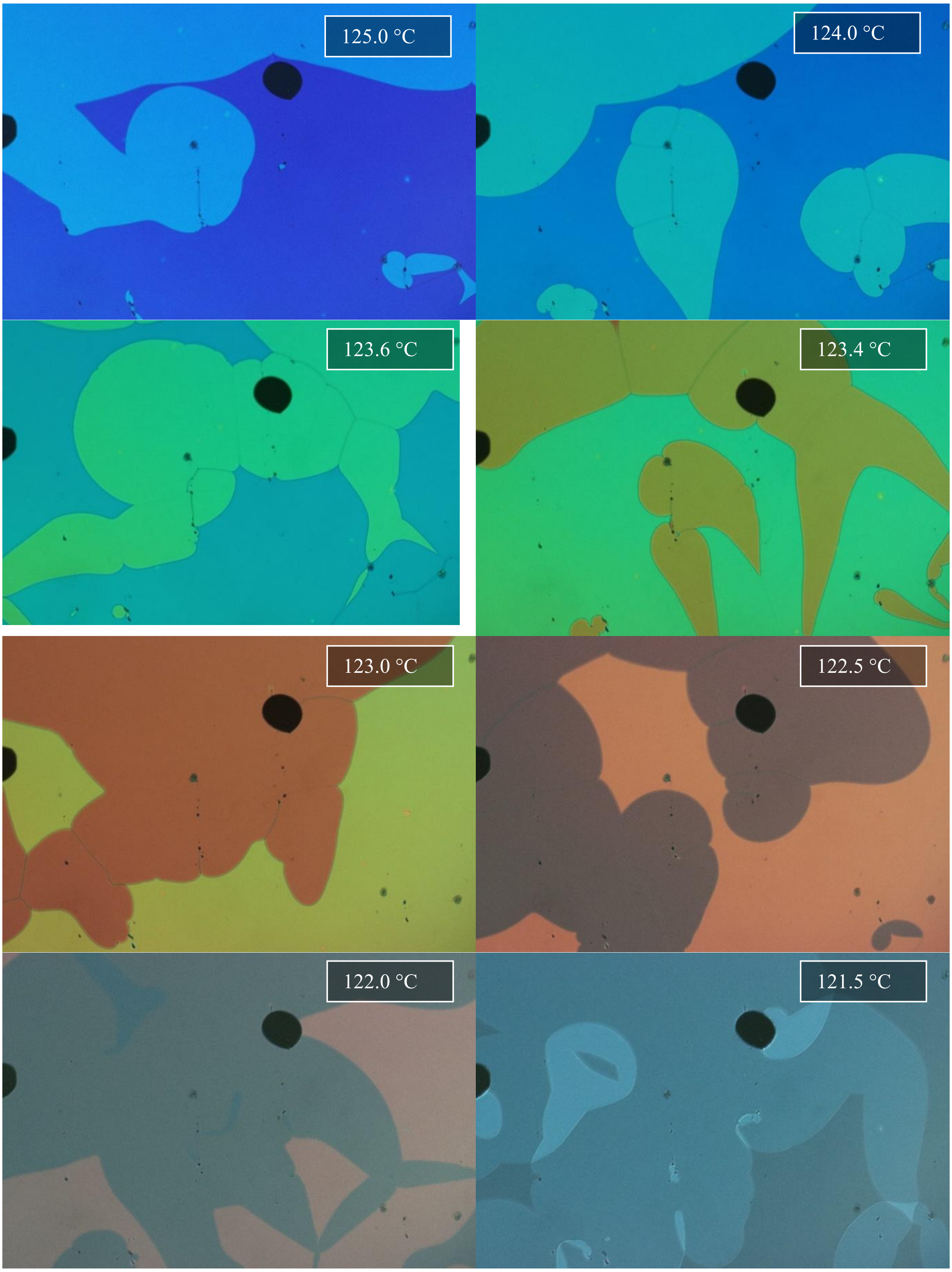

**Figure S52**: Grandjean textures of N* phase of S-**RW4***, changes of the colours are due to unwinding of the helix.

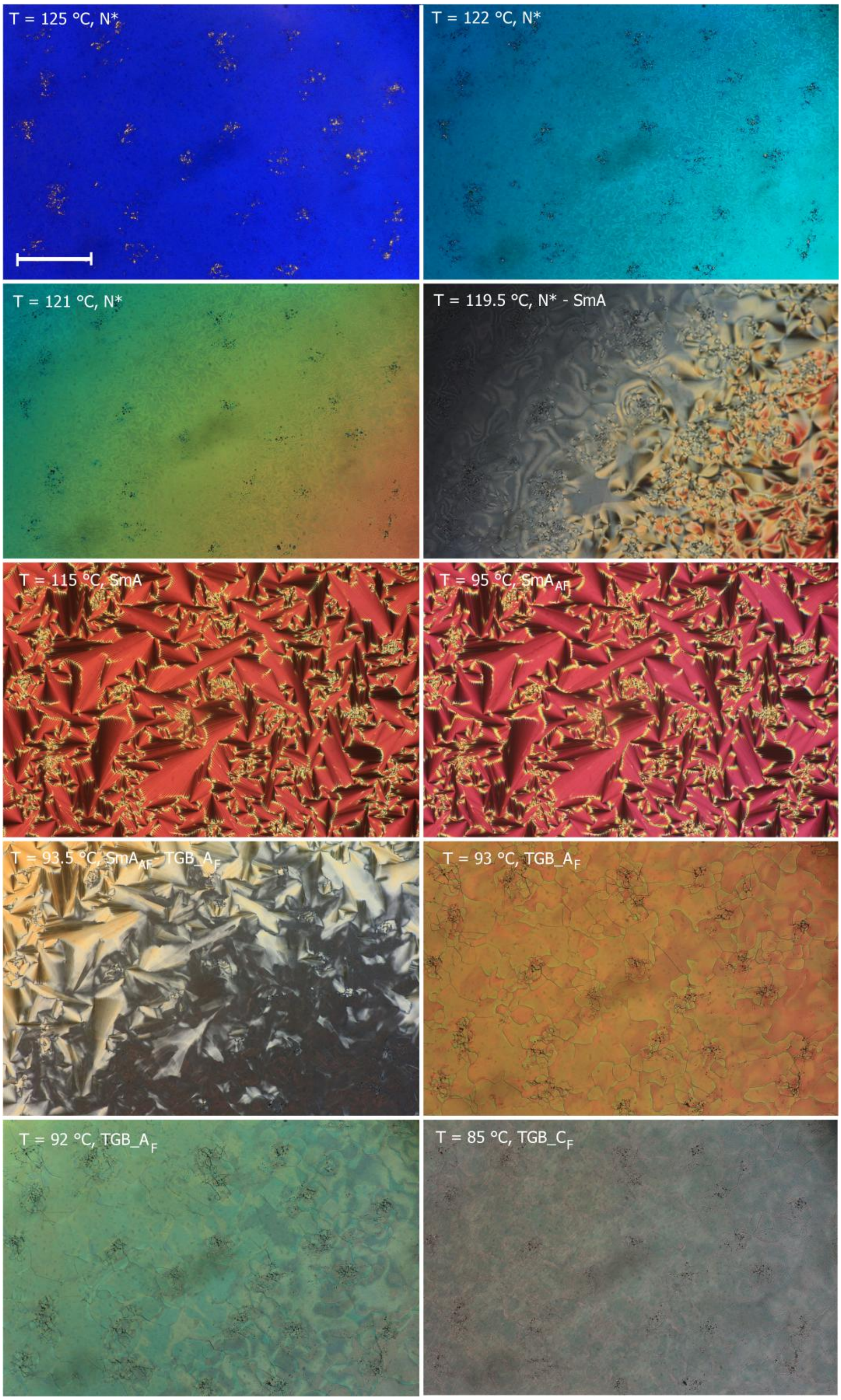

**Figure S53**: Optical textures of LC phases of S-**RW4*** observed in 3-μm-thick cell with degenerate planar anchoring.

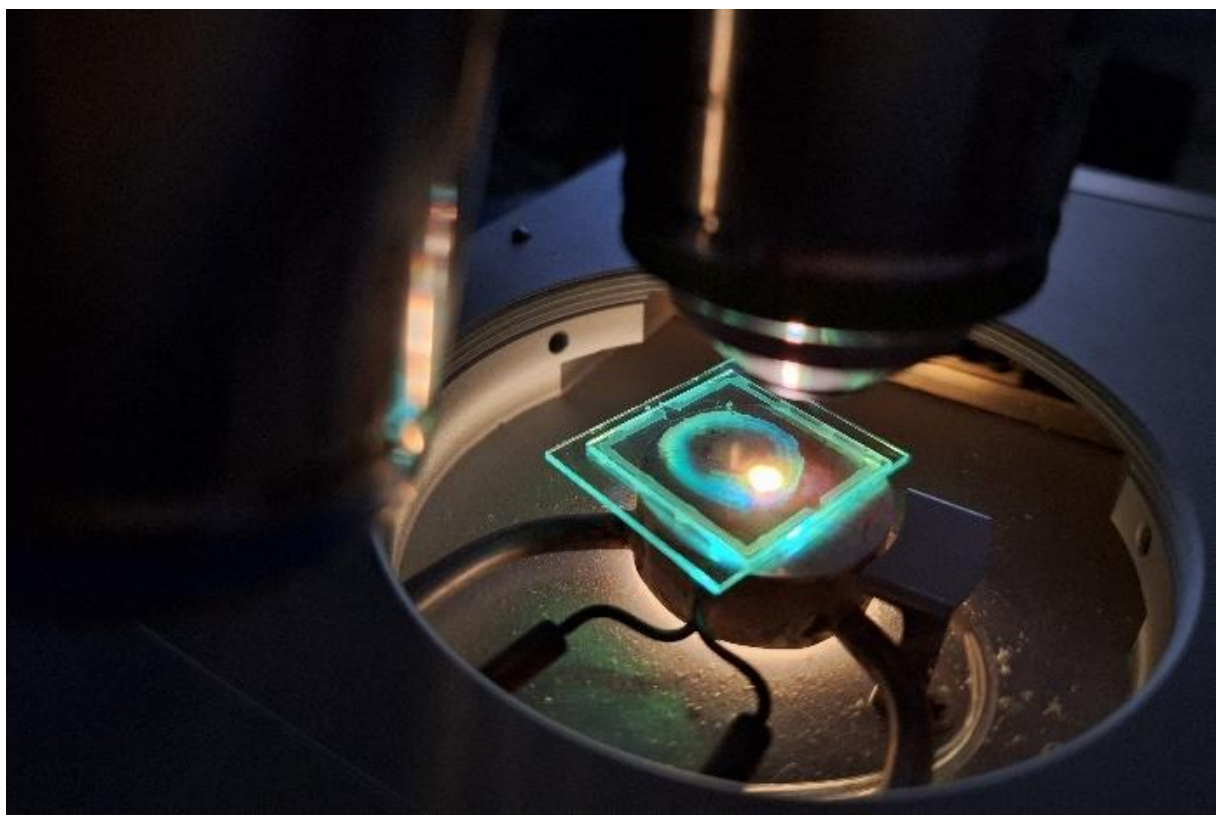

**Figure S54**: Selective reflection from the sample in TGB_$A_F$ phase.

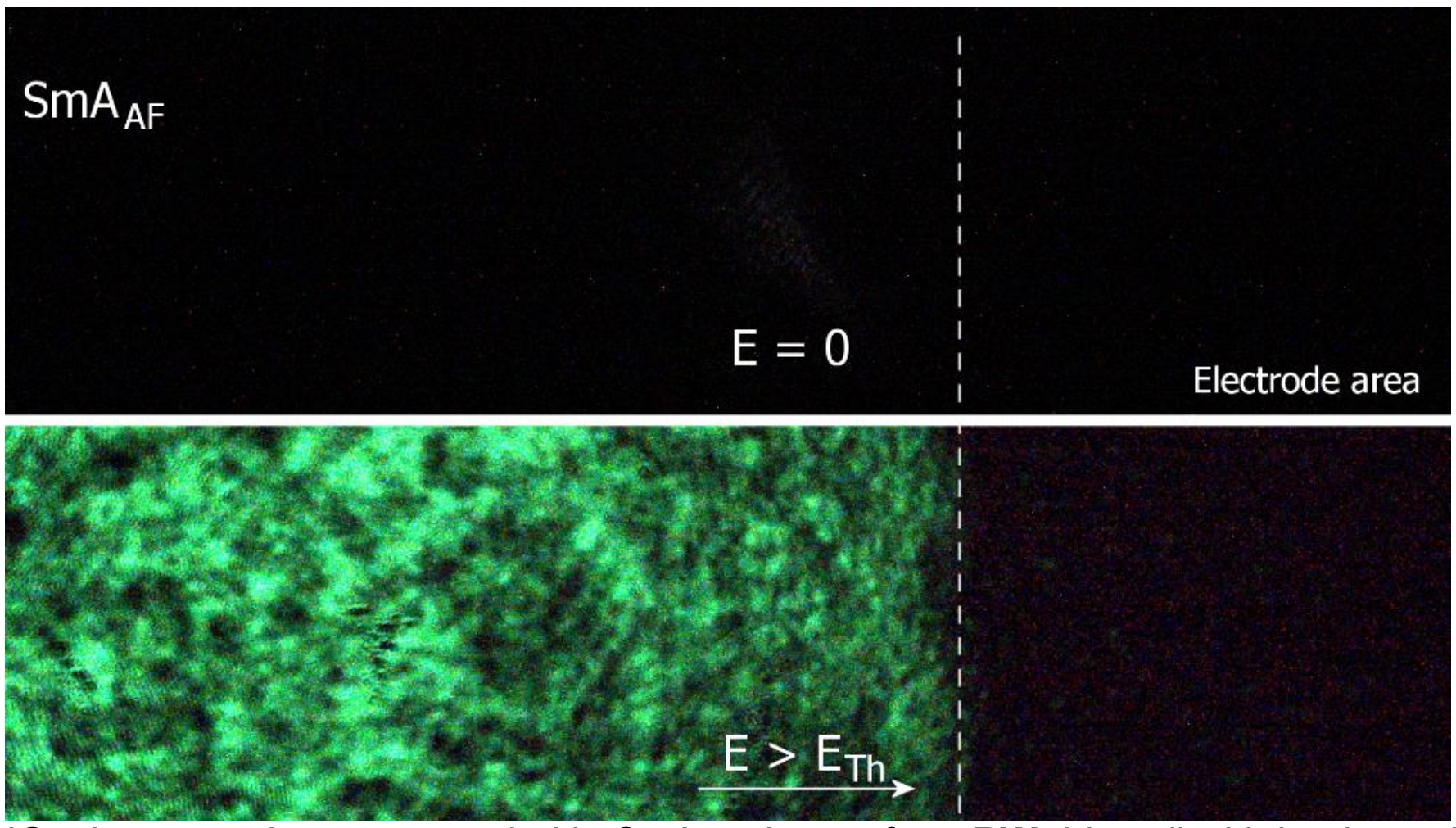

**Figure S55**: SHG microscopy images recorded in SmA$_{AF}$ phase of rac-**RW4*** in cell with in-plane electrodes. Ground state is clearly SHG silent, while under applied electric field (area on the left from dashed line) strong SHG activity is detected.

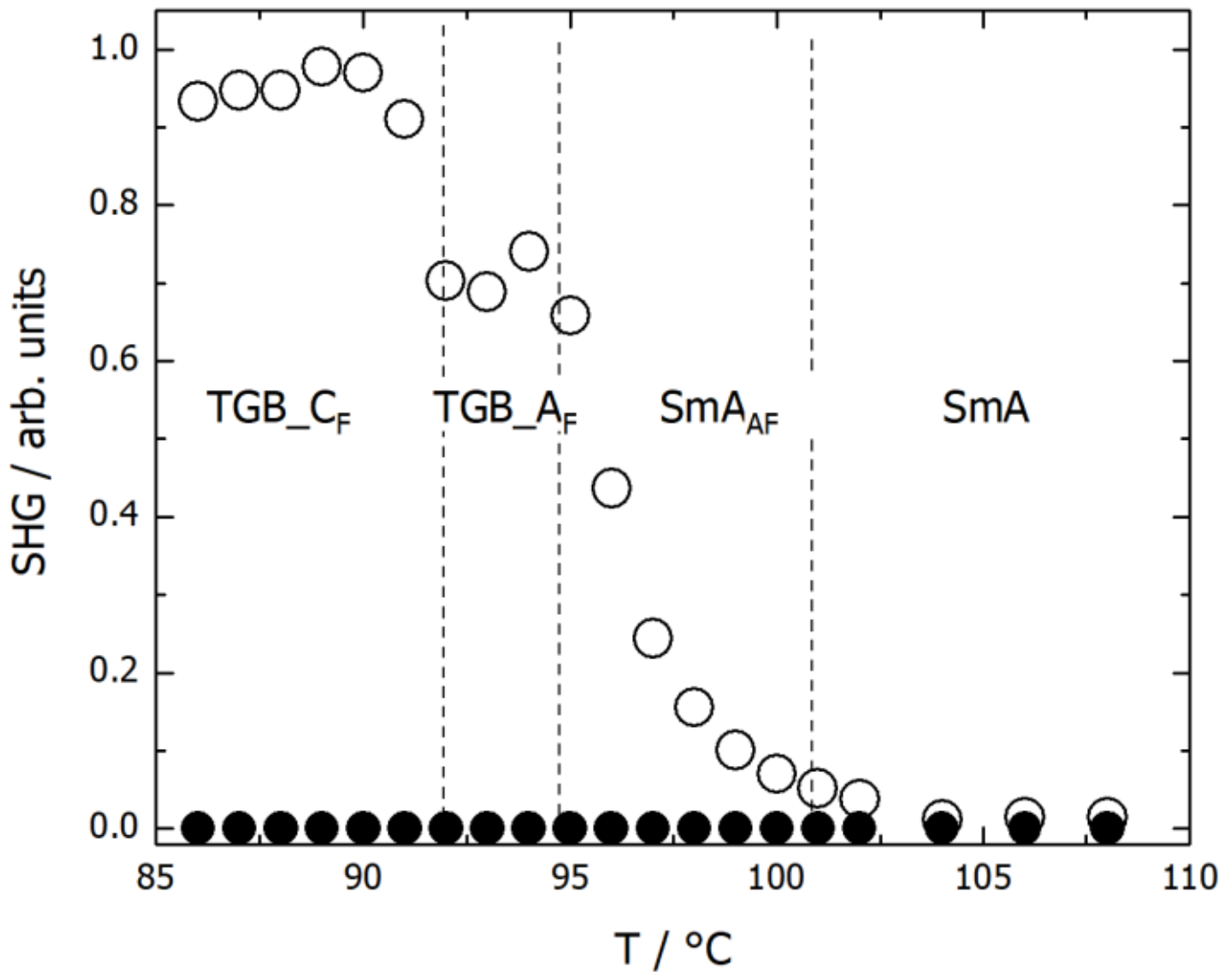

**Figure S56**: SHG intensity recorded on cooling for S-**RW4*** under applied electric field (open circles) and without electric field (solid circles). Note that the ground state of all the phases is SHG silent, evidencing full compensation of electric polarization.

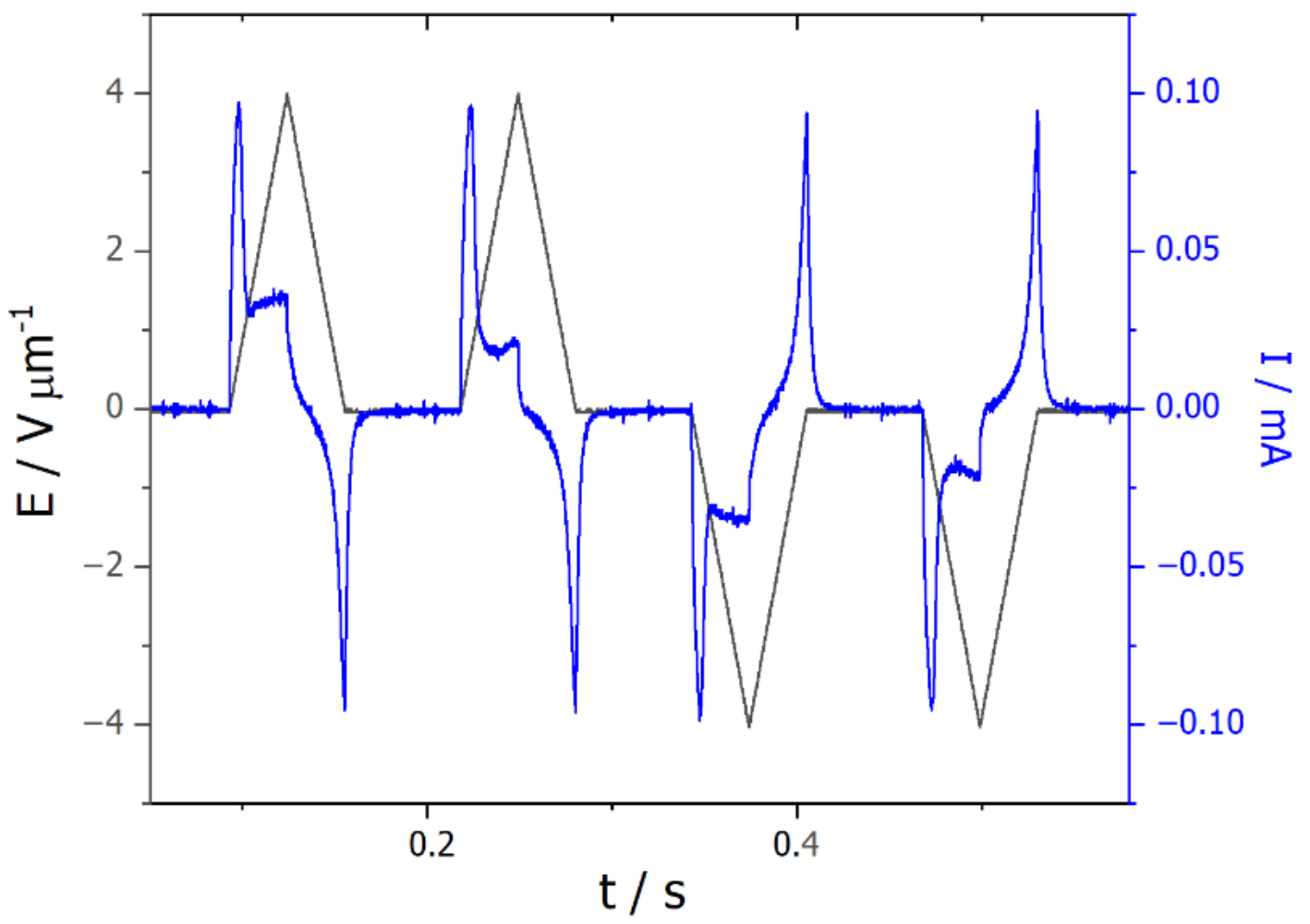

**Figure S57**: Switching current (blue line) recorded under application of modified triangular-wave voltage (black line) for SmA$_{AF}$ phase of S-**RW4*** compound. Repolarization current peaks appearing at each rise/fall of electric field confirm that the antiferroelectric ground state is restored at zero electric field.

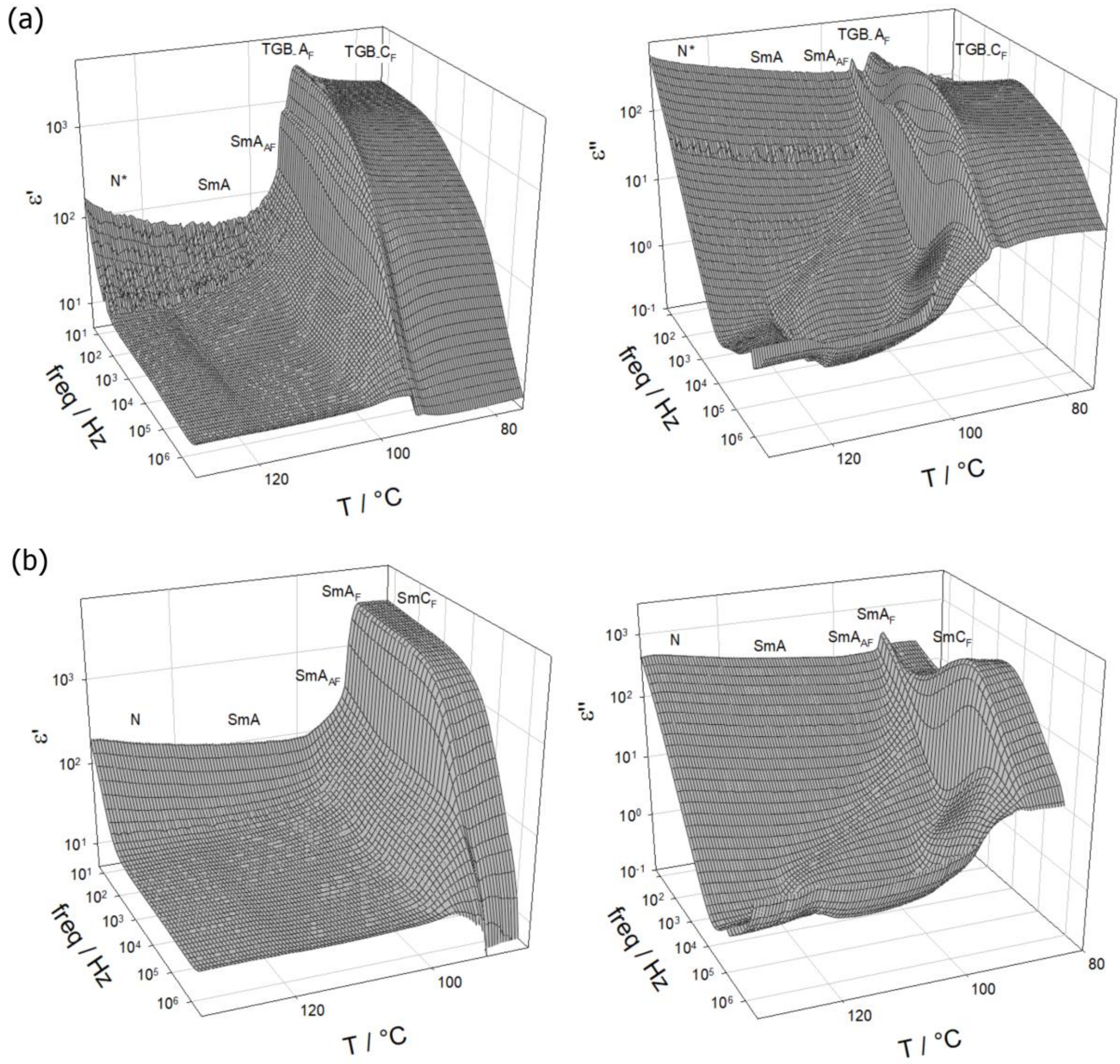

**Figure S58**: Real and imaginary parts of dielectric permittivity measured in cells with Au electrodes for (a) S-**RW4*** and (b) rac-**RW4***.

## Phenomenological Model

To consider in more detail observed phase sequences, we employ a phenomenological model. The bulk free energy density is written as a sum of a nematic ($f_n$), smectic ($f_s$) and polar ($f_P$) contributions, assuming a constant value of the nematic order parameter. The nematic contribution to the free energy density can be expressed as:

$$f_n = \frac{1}{2}K_1\big(\vec{n}(\nabla\cdot\vec{n}) - \vec{S}\big)^2 + \frac{1}{2}K_2(\vec{n}\cdot(\nabla\times\vec{n}) - q_0)^2 + \frac{1}{2}K_3\big(\vec{n}\times(\nabla\times\vec{n})\big)^2 \ , \qquad (1)$$

where $\vec{n}$ is the nematic director and $K_1$, $K_2$ and $K_3$ are the splay, twist and bend elastic constants. $\vec{S} = \gamma\vec{P}/K_1$ is a preferred spontaneous splay due to the flexoelectric effect [4], $\vec{P}$ being a spontaneous polarization and $\gamma$ a parameter related to the flexoelectric coefficient. If material is chiral, then $q_0$ is related to the pitch $p_0$ of the spontaneous twist ($q_0 = 2\pi/p_0$). In the case of a racemic mixture $q_0$ should be set to zero.

The smectic contribution can be expressed as [5]:

$$f_s = \frac{1}{2}a(T)|\psi|^2 + \frac{1}{4}b|\psi|^4 + c_{\parallel}(T)|(\vec{n}\cdot\nabla - iq_s)\psi|^2 + c_{\perp}|(\vec{n}\times\nabla)\psi|^2 \ , \qquad (2)$$

where $\psi$ is the smectic order parameter. If the smectic layer normal is along the $z$-direction, then $\psi = \psi_0 e^{iq_s z}$, $q_s = 2\pi/d_s$ and $d_s$ is a smectic layer thickness. The temperature dependent Landau parameter $a(T)$ is negative at $T < T_{NS}$. At $T < T_{NS}$ a smectic phase is stable and the term with Landau parameter $b$ stabilizes a finite magnitude ($\psi_0$) of the smectic order parameter. The sign of the parameter $c_{\parallel}(T)$ defines the type of the smectic phase. In the SmA phase, $c_{\parallel}(T)$ is positive, and it is negative in the SmC phase, where a finite tilt of the director with respect to smectic layer normal in stabilized by the last term in $f_s$, with parameter $c_{\perp}$ being positive. At $T = T_{AC}$, $c_{\parallel} = 0$. If $T_{AC} < T_{NS}$, the expression for $f_s$ describes a material with a phase sequence N → SmA → SmC.

Several terms have already been proposed for the free energy density due to the polar ordering [6–9]. Here, we consider only the most relevant ones needed to describe the observed phase sequences. To simplify the estimates, we assume a second order phase transition to the polar phase, thus

$$f_P = \frac{1}{2}\mu(T)P^2 + \frac{1}{4}\nu P^4 + \frac{1}{2}\kappa\big|\nabla\vec{P}\big|^2, \qquad (3)$$

where $\mu(T)$ and $\nu$ are Landau parameters defining the transition temperature to the polar state and the term with parameter $\kappa$ gives the energy penalty for spatial variation in direction and/or magnitude of $\vec{P}$. As pointed out in [9], the splay term in eq. (1) is larger than the sum of the splay elastic and flexoelectric energy contributions by $K_1S_0^2/2$, where $S_0 = \gamma P_0/K_1$ and $P_0$ is the equilibrium value of polarization. This should be accounted for when minimising the free energy for $P_0$, but it simply leads to a shift of the temperature at which the polar order becomes stable, and we can include this in the temperature dependent Landau term in eq. (3). The equilibrium value of $P_0$ is found by minimizing the following expression:

$$\frac{1}{2}\mu_0(T - T_P)P_0^2 + \frac{1}{4}\nu P_0^4 = min, \qquad (4)$$

where we used $\mu(T) = \mu_0(T - T_P)$ with $\mu_0 > 0$. At $T < T_P$ the equilibrium magnitude of polarization is

$$P_0^2 = \frac{1}{\nu}\ \mu_0(T_P - T) \ . \qquad (5)$$

When we plug $P_0$ from eq. (5) into eq. (3), we obtain $f_P = -\nu P_0^4/4$ if $\big|\nabla\vec{P}\big| = 0$.

If we assume that $T_P > T_{NS}$, then, at $T < T_P$, the material will first experience the phase transition from the nematic to the ferroelectric nematic phase and by further lowering the temperature the transition to the polar SmA phase and eventually to the polar SmC phase. However, if $T_P < T_{NS}$, the following phase sequence is expected: N → SmA → $SmA_F$ → $SmC_F$. Because polar phases favour spontaneous splay of polarization, and because the splay cannot be made favourable everywhere, one can also expect smectic structures that will be antiferroelectric but in the sense of antiferroelectricity along the smectic layer (and not perpendicular to it), as shown in Figure S59.

Let us first focus on the racemic mixture, in which the following phase sequence is observed: N → SmA → $\mathrm{SmA_{AF}}$ → $\mathrm{SmA_F}$ → $\mathrm{SmC_F}$. At temperatures below $T_{P\gamma}$ the polar order becomes stable. However, uniform polar order along the smectic layer in the SmA phase would require the splay of polarization to be zero, which would increase the nematic component of the free energy. Splay of polarization in the smectic phase leads to the undulation of smectic layers, but in the SmA phase this comes at no energy cost, because the smectic layer thickness is preserved. It is also more convenient for the material to be macroscopically apolar, which is obtained by the interchange of domains with the up and down polarization as shown in Figure S59. Such arrangement is easily obtained if the polar order between the two domains is melted. If the width of the melted region is $l_w$, the energy penalty will be ($F_w$):

$$F_w = \frac{1}{4}\nu P_0^4\, l_w \ . \tag{6}$$

On the other hand, the energy will be gained in the regions of the favourable splay. If there were no splay, the energy in the region of the length $l_b$ would be larger due to the lack of splay but lower because there would be no spatial variation of polarization. The net result which should be compared with the energy of the wall is

$$F_b = \frac{1}{2}K_1 S_0^2 l_b - \frac{1}{2}\kappa\left(\frac{\Delta P_x}{l_b}\right)^2 l_b \ , \tag{7}$$

where $\Delta P_x = 2P_0 \sin\theta_0$ and $\theta_0$ is the maximum splay angle (see Figure S59). By equating energies given by eqs. (6) and (7), we find

$$\frac{l_b}{l_w} = \frac{\nu K_1 P_0^2}{2\gamma^2}\left(1 - \frac{4\kappa K_1}{\gamma^2 l_b^2}\sin^2\theta_0\right)^{-1} \ . \tag{8}$$

In the case of the equilibrium splay $\nabla\cdot\vec{n} = \vec{n}\cdot\vec{P}_0$, from where it follows that $2\sin\theta_0 / l_b = \gamma P_0 / K_1$, so

$$l_b = \frac{2K_1}{\gamma P_0}\sin\theta_0 \ . \tag{9}$$

By plugging $l_b$ from eq. (9) into the right-hand-side of eq. (8) we obtain

$$\frac{l_b}{l_w} = \frac{\frac{\nu K_1 P_0^2}{2\gamma^2}}{1 - \frac{\kappa P_0^2}{K_1}} \ . \tag{10}$$

From eq. (10) we see that the ratio between the width of the block and width of the wall increases as temperature decreases, because $P_0$ increases by decreasing temperature, so the numerator in eq. (10) increases, and the denominator decreases. Also, when temperature decreases, the energy of the wall (eq. (5)) increases with $(T_P - T)^2$, while the energy penalty of no or unfavourable splay increases with $(T_P - T)$. As a result, at some temperature the formation of the wall with no polar order will not be favourable anymore. This will lead to the transition to the $\mathrm{SmA_F}$ phase, where the energy of the wall is related to the unfavourable splay (see Figure S59b) and to the spatial variation in polarization as

$$F_w = \frac{1}{2}K_1\left(\frac{\Delta n_x}{l_w} + \frac{\gamma P_0}{K_1}\right)^2 l_w + \frac{1}{2}\kappa\left(\frac{\Delta P_x}{l_w}\right)^2 l_w \ , \tag{11}$$

where $\Delta n_x = 2\sin\theta_0$. The transition temperature $T_F$ from the $\mathrm{SmA_{AF}}$ to the $\mathrm{SmA_F}$ phase is obtained by equating eqs. (6) and (11). If we assume that the main temperature dependence is in the magnitude of polarization and that the width of the wall is very weakly dependent on temperature, we obtain a fourth order polynomial equation for $P_{0,T_F}$ – polarization value at temperature $T = T_F$:

$$\frac{1}{2}K_1\left(\frac{2\sin\theta_0}{l_w}+\frac{\gamma P_{0,T_F}}{K_1}\right)^2+\frac{1}{2}\kappa\left(\frac{2P_{0,T_F}\sin\theta_0}{l_w}\right)^2=\frac{1}{4}\nu P_{0,T_F}^4\,. \qquad (12)$$

From eq. (12) one can find $P_{0,T_F}$ and then estimate $T_F$ from eq. (6), if other parameters are known.

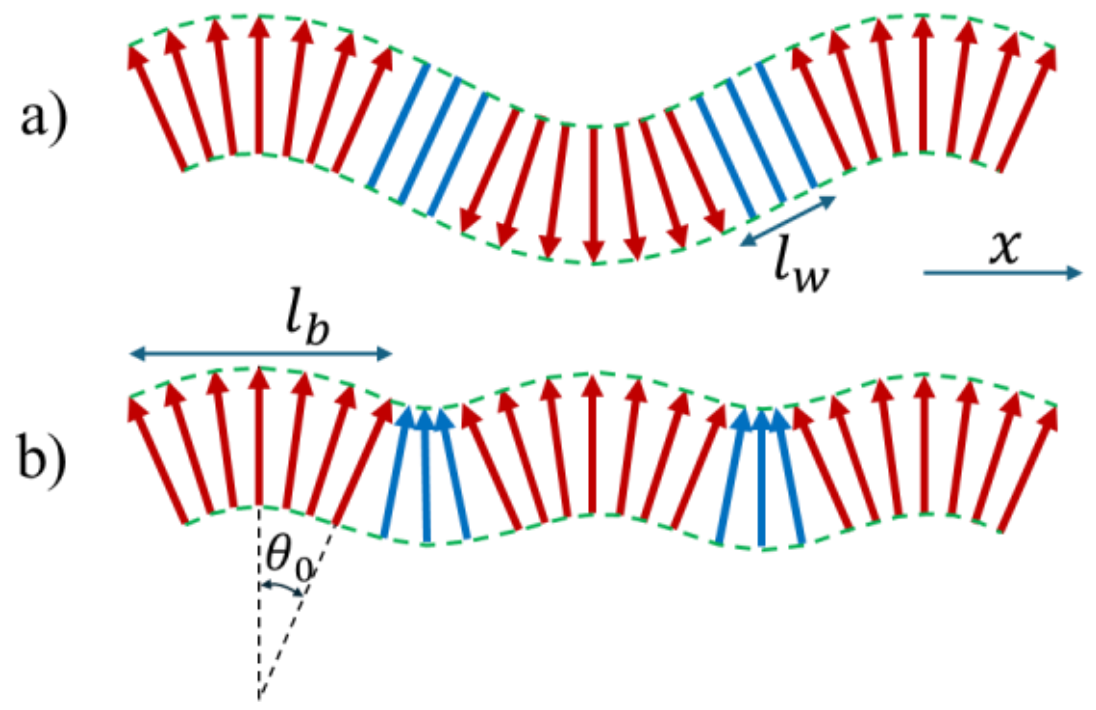

**Figure S59**. Tendency for the polarization splay leads to undulation of smectic layers in the a) $SmA_{AF}$ and b) $SmA_F$ phase observed in a racemic mixture. Red arrows present the local polarization direction with a favourable splay, blue arrows regions with unfavourable splay, blue rods present the direction of the average molecular long axes in regions with no polar order. The width of the blocks with a favourable splay is $l_b$, the width of the walls with an unfavourable splay or no polar order is $l_w$, $\theta_0$ is the angle of the maximum splay along the $x$ – axis.

Now, let us focus on a chiral material. In this case, the material prefers also a spontaneous twist. Spontaneous twist, however, is impossible to accommodate along the smectic layer without changing the thickness of the layer. Upon transition to the polar phase, we first obtain the $\mathrm{SmA_{AF}}$ phase, because the cost of melting the polar order is low at low values of $T_P - T$. In the SmA and $\mathrm{SmA_{AF}}$ phase, the energy density price for no spontaneous twist is $K_2 q_0^2/2$ , for $\mathrm{SmA_{AF}}$, both in the block of favourable splay and in the walls with no polar order. When temperature is further reduced below $T_F$, the system can finally achieve spontaneous twist as well, by a phase transition to the polar TGB_$A_F$ phase instead of the $\mathrm{SmA_F}$ phase, where the wall between two regions of favourable splay contains only the unfavourable splay. In a chiral material, the energy of the wall can be reduced by forming a series of screw dislocation in a TGB boundary. Such a boundary contains favourable twist and unfavourable splay, the energy price of the latter being compensated by the energy gain due to the former.

Finally, we estimate the width of the block ($l_b$). We assume a typical value for the elastic constant $K_1 \approx 2\times10^{-11}\mathrm{N}$ and take the value of $P_0$ at $T \approx 100°\mathrm{C}$, $P_0 \approx 10^{-2}\mathrm{Cm^{-2}}$ (see Figure 4a in the main text). From the measurements of the flexoelectric coefficient $e^*/K \sim 1\ \mathrm{Cm^{-1}N^{-1}}$ in the nematic phase of material exhibiting the ferroelectric nematic phase [10] and relating it to the “energy” flexoelectric coefficient $\gamma$ as $\gamma = e^*/(\varepsilon\varepsilon_0)$, with $\varepsilon \sim 100$, we find $\gamma \sim 0.01\ \mathrm{V}$, which agrees with the value chosen for $\gamma$ in [9]. From the birefringence measurements it can be deduced (see the next section in SI) that the splay amplitude is $\theta_0 \approx 0.025$. The value of $l_b$ is then estimated from eq. (9) and found to be approximately 10 nm. The obtained order of magnitude is consistent with the widths of the blocks measured in the ordinary, non-polar TGB_A phase [11].

## Estimation of the maximum splay angle

To estimate the maximum splay angle from the measurements of birefringence, we use the following procedure. The dielectric tensor at optical frequencies ($\underline{\varepsilon}$) of a uniaxial material with the optic axis along the $z$-direction (direction of the smectic layer normal in the case of planar smectic layers in Figure S59) is

$$\underline{\varepsilon} = \begin{pmatrix} \varepsilon_\perp & 0 & 0 \\ 0 & \varepsilon_\perp & 0 \\ 0 & 0 & \varepsilon_\parallel \end{pmatrix} . \tag{13}$$

If the optic axis rotates by $\theta$ around the $y$-axis, the dielectric tensor transforms to

$$\underline{\varepsilon}(\theta) = \begin{pmatrix} \varepsilon_\perp \cos^2\theta + \varepsilon_\parallel \sin^2\theta & 0 & \frac{1}{2}\Delta\varepsilon \sin(2\theta) \\ 0 & \varepsilon_\perp & 0 \\ \frac{1}{2}\Delta\varepsilon \sin(2\theta) & 0 & \varepsilon_\perp \sin^2\theta + \varepsilon_\parallel \cos^2\theta \end{pmatrix} , \tag{14}$$

where $\Delta\varepsilon = \varepsilon_\parallel - \varepsilon_\perp$ is an anisotropy of a uniform phase, in our case the SmA phase. To find anisotropy of a modulated phase, we average $\underline{\varepsilon}(\theta)$ over all $\theta$ between $\pm\theta_0$ and find the dielectric tensor ($\underline{\varepsilon}_{splay}$) of the splayed structure as

$$\underline{\varepsilon}_{splay} = \frac{1}{2\theta_0}\int_{-\theta_0}^{\theta_0} \underline{\varepsilon}(\theta) d\theta \ , \tag{15}$$

from where it follows

$$\underline{\varepsilon}_{splay} = \begin{pmatrix} \frac{\varepsilon_\perp + \varepsilon_\parallel}{2} + \frac{\Delta\varepsilon}{4\theta_0}\sin(2\theta_0) & 0 & 0 \\ 0 & \varepsilon_\perp & 0 \\ 0 & 0 & \frac{\varepsilon_\perp + \varepsilon_\parallel}{2} - \frac{\Delta\varepsilon}{4\theta_0}\sin(2\theta_0) \end{pmatrix} . \tag{16}$$

The anisotropy of the splayed structure ($\Delta\varepsilon_{splay}$) is thus

$$\Delta\varepsilon_{splay} = \frac{\Delta\varepsilon}{2\theta_0}\sin(2\theta_0) \ . \tag{17}$$

In the limit of very small splay angle, $\Delta\varepsilon_{splay} = \Delta\varepsilon$, but we are interested in the first nonzero correction to this limit:

$$\Delta\varepsilon_{splay} = \Delta\varepsilon\left(1 - \frac{4}{3}\theta_0^2\right) \ . \tag{18}$$

By using $\Delta\varepsilon = (n_e + n_o)(n_e - n_o) = (n_e + n_o)\Delta n,$ where $n_e$ and $n_o$ are the extraordinary and ordinary index of refraction, respectively, and $\Delta n$ is birefringence, we find that the difference in the birefringence between the splayed and uniform structure is

$$\Delta n_{splay} - \Delta n = -\frac{4}{3}\Delta n\, \theta_0^2 \ . \tag{19}$$

The maximum splay angle is thus

$$\theta_0^2 = -\frac{3}{4}\frac{\Delta n_{splay} - \Delta n}{\Delta n} \ . \tag{20}$$

The birefringence is $\Delta n \approx 0.17$ close to the SmA – SmA$_{AF}$ phase transition and $\Delta n_{splay} - \Delta n \approx -1.4 \times 10^{-4}$ (both values can be deduced from Figure 1e in the main text). By plugging these values into eq. (20), we find $\theta_0 \approx 0.025\,\text{rad} \approx 1.4\,\text{deg}$. The total splay is thus approximately 3 deg (from $-\theta_0$ to $\theta_0$).

## References

[1] H. Maebayashi, T. Fuchigami, Y. Gotoh, M. Inoue, Stereoselective Acetalization for the Synthesis of Liquid-Crystal Compounds Possessing a trans-2,5-Disubstituted 1,3-Dioxane Ring with Saturated Aqueous Solutions of Inorganic Salts, *Org. Process Res. Dev.*, **23**, 477, (2019).
[2] S. Brown, E. Cruickshank, J. M. D. Storey, C. T. Imrie, D. Pociecha, M. Majewska, A. Makal, E. Gorecka, Multiple Polar and Non-polar Nematic Phases, *ChemPhysChem*, **22**, 2506, (2021).
[3] G. J. Strachan, E. Górecka, J. Szydłowska, A. Makal, D. Pociecha, Nematic and Smectic Phases with Proper Ferroelectric Order, *Adv. Sci.*, **12**, 2409754 (2025).
[4] N. Sebastián et al., Polarization patterning in ferroelectric nematic liquids via flexoelectric coupling, *Nat. Commun*. **14**, 3029, (2023).
[5] S. R. Renn and T. C. Lubensky, Abrikosov dislocation lattice in a model of the cholesteric – to – smectic- *A* transition, *Phys. Rev. A* **38**, 2132, (1988).
[6] E. I. Kats, Stability of the uniform ferroelectric nematic phase, *Phys. Rev. E* **103**, 012704, (2021).
[7] N. Vaupotič, D. Pociecha, P. Rybak, J. Matraszek, M. Čepič, J. M. Wolska, and E. Gorecka, Dielectric response of a ferroelectric nematic liquid crystalline phase in thin cells, *Liq. Cryst.* **50**, 584, (2023).
[8] L. Paik and J. V. Selinger, Flexoelectricity versus electrostatics in polar nematic liquid crystals, *Phys. Rev. E* **111**, L053402, (2025).
[9] P. Medle Rupnik, E. Hanžel, M. Lovšin, N. Osterman, C. J. Gibb, R. J. Mandle, N. Sebastián, and A. Mertelj, Antiferroelectric Order in Nematic Liquids: Flexoelectricity Versus Electrostatics, *Adv. Science* **12**, 2414818, (2025).
[10] A. Barthakur, J. Karcz, P. Kula, and S. Dhara, Critical splay fluctuations and colossal flexoelectric effect above the nonpolar to polar nematic phase transition, *Phys. Rev. Mater.* **7**, 035603, (2023).
[11] L. Navailles, B. Pansu, L. Gorre-Talini, and H. T. Nguyen, Structural Study of a Commensurate TGB A Phase and of a Presumed Chiral Line Liquid Phase, *Phys. Rev. Lett*. **81**, 4168, (1998).